\documentclass[aps,prd,times,graphicx,amssymb,tighten,showpacs,groupedaddress,superscriptaddress,floatfix]{revtex4-1}

\usepackage{graphicx,amssymb,amsmath}
\usepackage[dvips]{xcolor}
\usepackage{hyperref}
\usepackage{amsmath}
\usepackage{amssymb}
\newcommand{\be}{\begin{equation}}
\newcommand{\ee}{\end{equation}}

\begin{document}
\title{New evaluation of axial nucleon form factor from electron- and neutrino-scattering data and impact on neutrino-nucleus cross-section}
 
 \author{G.D.~Megias}
  \affiliation{University of Tokyo,  Institute for Cosmic Ray Research,  Research Center for Cosmic Neutrinos,  Kashiwa,  Japan}
 \affiliation{Departamento de F\'{i}sica At\'omica, Molecular y Nuclear, Universidad de Sevilla, 41080 Sevilla, Spain}

\author{S.~Bolognesi}
\affiliation{IRFU, CEA, Universit\'e Paris-Saclay, 91191 Gif-sur-Yvette, France}

\author{M.B.~Barbaro}
 \affiliation{Dipartimento di Fisica, Universit\`{a} di Torino and INFN, Sezione di Torino, Via P. Giuria 1, 10125 Torino, Italy}

\author{E.~Tomasi-Gustafsson}
\affiliation{IRFU, CEA, Universit\'e Paris-Saclay, 91191 Gif-sur-Yvette, France}

\date{\today}
\begin{abstract}
A joint fit to neutrino-nucleon scattering and pion electroproduction data is performed to evaluate the nucleon axial form factor in the two-component model consisting of a three-quark intrinsic structure surrounded by a meson cloud. Further constrains on the model are obtained by re-evaluating the electromagnetic form factor using electron scattering data. The results of the axial form factor show sizable differences with respect to the widely used dipole model. The impact of such changes on the Charged-Current Quasi-Elastic neutrino-nucleus cross-section is evaluated in the SuSAv2 nuclear model, based on the Relativistic Mean Field and including the contribution of two-body currents. How the different parametrizations of the axial form factor affect the cross-section prediction is assessed in full details and comparisons to recent T2K and MINERvA data are presented. 
\end{abstract}


\maketitle

\section{Introduction}
\label{sec:Intro}

The accurate knowledge of the nucleon's form factors and a good control of nuclear effects in neutrino-nucleus scattering in the GeV region are mandatory requirements for the analysis and interpretation of ongoing and planned neutrino oscillation experiments~\cite{Acero:2019ksn,Abe:2018wpn,Abe:2018hyperk,Abi:2018dune}, which aim at measuring neutrino properties with unprecedented precision and search for CP violation in the leptonic sector.

While the weak vector form factors of the nucleon are related to the electromagnetic ones through conservation of vector current (CVC) and are relatively well under control in the kinematical region of interest for these experiments, an important source of uncertainty arises from the poor knowledge of the axial form factor $G_A$. This occurrence gave rise to the so-called "$M_A$ puzzle" when the first neutrino-carbon cross-sections were published by the MiniBooNE collaboration~\cite{AguilarArevalo:2010zc} and found to be largely underestimated by the theoretical prediction, unless a value of the axial cutoff $M_A$=1.35 GeV was used in place of the standard value $M_A\approxeq$ 1 GeV. It was soon realized in Refs.~\cite{Martini:2009uj,Martini:2010ex,Amaro:2010sd,Amaro:2011aa,Nieves:2011yp} that the discrepancy was mainly due to the incorrect treatment of nuclear effects, and in particular those going beyond the Impulse Approximation (IA), corresponding to the excitation of two-particle two-hole (2p2h) states. While this can solve the puzzle at a qualitative level, yet it is important to quantify the errors related to the nucleon structure input and their interplay with nuclear corrections.

After a general introduction on the axial nucleon form factor and the neutrino-nucleus Charged-Current Quasi-Elastic cross-section (Section \ref{sec:Intro}), in Section \ref{sec:model} we describe the main features of the two-component model for the axial form factor of the nucleon (\ref{sec:IJL}), we re-evaluate the electromagnetic form factors in such model (\ref{sec:EMfit}) and we perform the first evaluation of two-component model axial form factor with a joint fit to neutrino scattering and pion electroproduction data (\ref{sec:fitaxial}). Section \ref{sec:neutrinoNucleus} focuses on the impact of the nucleon axial form factor on the neutrino-nucleus cross-section:  the superscaling model (SuSAv2) used to describe the nuclear dynamics is introduced in Sec.\ref{sec:SuSAv2}, then in Sec.\ref{sec:results} we show the results obtained within the SuSAv2 model for neutrino and antineutrino scattering on a carbon target using different prescriptions for the axial form factor and compare with experimental data in different kinematical conditions. Conclusions are drawn in Sec.\ref{sec:concl}.

\subsection{The nucleon axial form factor}
The axial form factor $G_A$ has been measured directly by (anti)neutrino scattering on nucleon  $\nu_{\mu}+n\to \mu^-+p$ ($\bar\nu_{\mu}+p\to \mu^++n$), or nuclei, or indirectly, by near threshold charged pion electroproduction in space-like region. It is function of one kinematical variable, the momentum transfer squared $Q^2$
(for a review see \cite{Bernard:2001rs}).

At the nucleon level, the effective Lagrangian for (anti)neutrino-hadron ($\nu h$) elastic 
neutral current (NC) $\nu h \to \nu h$  or 
charged current (CC) $\nu h\to  \ell h$ scattering ($\ell$ is a lepton) can be written as:

\begin{equation}
{\cal L}^{\nu h} =-\frac{G_F}{\sqrt{2}} 
 J^{\mu\lambda} J_\mu^h. 
\label{eq:Lnuh}
\end{equation}
where the the leptonic current is given by:
\begin{equation}
J^{\mu\lambda}= \bar \ell\gamma^\mu(1-\gamma_5)\nu .
\label{eq:Lnuh}
\end{equation}
The hadronic weak current is:
\begin{equation}
J_\mu^h=(1-2\sin^2\theta_W)V_\mu^3+A^3_\mu-\frac{2}{3}\sin^2\theta_WV_\mu^0 + a_0A^0_\mu .
\label{eq:Jmuh}
\end{equation}
In Eq. \eqref{eq:Jmuh},  the light quark contributions are combined to form the vector currents, isoscalar and isovector : ($V^0_\mu$, $V^3_\mu$) and axial-vector currents ($A^0_\mu$ , $A^3_\mu$ ); $\theta_W$ is the weak mixing angle, and $a_0$ vanishes.
Assuming isospin invariance and time reversal invariance, the hadronic current matrix element between nucleon states can be written as:
\begin{equation}
<N(p')|J_{\mu}^h|N(p)>=\bar u (p')\left [\gamma_\mu F_1(Q^2)+i\displaystyle\frac{\sigma_{\mu\nu}q^\nu}{2m_N} F_2(Q^2) + \gamma_\mu \gamma_5 G_A(Q^2)+q_\mu\gamma_5\frac{G_P(Q^2)}{2m_N}\right ] u(p),
\label{eq:mat}
\end{equation}
where $m_N$ is the nucleon mass, $p (p')$ is the four momentum of the initial(final) hadron and $Q^2=-(p'-p)^2$ is the transferred four momentum squared. Vector current conservation requires that the Dirac and Pauli form factors, $F_1$ and $F_2$, are the same as the ones entering the hadron electromagnetic current and measured by electron scattering. $G_A$ is the axial form factor 
and it is extracted from the $Q^2$ dependence of the cross-section for 
CC elastic scattering of (anti)neutrino on nucleons and nuclei. 
Experiments in the period 1969-1990 using bubble chambers~\cite{Baker:1981su,Kitagaki:1990vs,Mann:1973pr,Kitagaki:1983px,Barish:1977qk,Allasia:1990uy,Miller:1982qi} assumed a dipole form for $G_A$:
\begin{equation}
G_A^D(Q^2)=\displaystyle\frac{G_A(0)}{(1+Q^2/M_A^2)^2} ,
\label{eq:dipole}
\end{equation}
where $G_A(0)$ is the axial-vector coupling constant $G_A(0)=1.2695 \pm 0.0029$~\cite{Tanabashi:2018oca}.  
The extracted values for $M_A$  were recently reanalyzed at the light of more recent and precise electromagnetic form factors data in Ref.~\cite{Bodek:2007ym}. 
The pseudoscalar form factor, $G_P$, can be connected to the axial one making use of the PCAC (partially conserved axial current) hypothesis (see~\cite{Thomas1} for details), through the Goldberger-Treiman relation,
\begin{equation}\label{gp}
 G_P(Q^2)=\frac{4M_N^2}{|Q^2|+m_\pi^2}G_A(Q^2) \quad;\quad m_\pi:\mbox{pion mass}\hspace{0.1cm}.
\end{equation}


The $Q^2$ dependence of the pion electroproduction cross-section at threshold gives information on  $G_A(Q^2)$, but the numerical value is highly model dependent. According to the chosen corrections, several determinations of $G_A$  can be found, considering the same experiment and cross-section data. 

The matrix element for  pion electroproduction on the nucleon is fully described by six amplitudes, functions of five independent kinematical variables. Assuming that the interaction occurs through the exchange of a virtual photon, the cross-section in the center of mass frame of the final $\pi N$ system, ${d\sigma}/{d\Omega_\pi}$,
can be decomposed into a transverse $\sigma_T$, longitudinal  $\sigma_L$,  and two interference parts, $\sigma_{LT}$, $\sigma_{TT}$, related to out-of plane kinematics: 

\begin{equation}
\displaystyle\frac{d\sigma}{d\Omega_\pi}= \displaystyle\frac{d\sigma_T}{d\Omega_\pi}+\epsilon_L \displaystyle\frac{d\sigma_L}{d\Omega_\pi}+ \sqrt{\epsilon_L(1+\epsilon_T)}\, \displaystyle\frac{d\sigma_{LT}}{d\Omega_\pi}\cos\phi+\epsilon_T \displaystyle\frac{d\sigma_{TT}}{d\Omega_\pi}\cos2\phi,
\label{SLT}
\end{equation}
where $\epsilon_{L(T)}$ is the longitudinal(transverse) polarisation of the virtual photon, $\phi$ is the angle between the $eN$-scattering plane and the $\pi N$ plane. For in-plane kinematics,  one can disentangle the longitudinal and transverse components by cross-section measurements at fixed $Q^2$ through a Rosenbluth separation. At low energies, the connection with the theory is done through multipole expansion, in terms of two S-wave multipole amplitudes, called $E_{0+}$ and $L_{0+}$, related respectively to  the transverse and longitudinal couplings of the virtual photon to the nucleon spin.

Both for pion electroproduction data and neutrino scattering data the dipole approximation is assumed a priori and the axial meson mass is determined from a fit  of the data. The corresponding value obtained from (anti)neutrino scattering, $M_A=(1.026\pm 0.021)$ GeV, is lower than the value found from electroproduction experiments  $M_A=(1.069\pm 0.018) $ GeV, leading to an axial radius difference of about $5\%$. 
In Ref.~\cite{Bernard:2001rs}  this apparent discrepancy was solved  in the framework of baryon chiral perturbation theory (CHPT), pointing out an additional model-independent contribution to a low energy theorem  
that relates an electric dipole amplitude to the axial radius. Such contribution can not be obtained by current algebra methods.
Isospin symmetry (as well as SU(3) ) is not a perfect symmetry of the Standard Model. There are two sources of violation: the mass difference between up and down quarks and electromagnetic corrections. Electromagnetic interactions are mostly responsible for the mass difference among the pion states. At leading order in CHPT, form factors are $q^2$ independent and are the same for proton and neutron, whereas at higher orders,  loops with mesons and nucleons of different masses contribute. The contribution from the exchange of virtual photon will also be different. The  corrections to isospin breaking  are expected to be small within SU(2), and somewhat larger within SU(3). In order to pin down isospin breaking effects, one needs to treat simultaneously the electromagnetic and strong contributions. In this respect, CHPT constitutes a very powerful tool, as electric charge is assigned dimension one in the power counting based on the observation that $e^2/(4\pi)\approx m^2_\pi/(4\pi F_\pi)^2\approx 1/100$  (with $m_\pi$ and $F_\pi$ the pion mass and decay constant respectively).

In the absence of new precise measurements of neutrino cross-sections on hydrogen and deuterium, the only way to quantify the error related to the knowledge of $G_A$ is to compare the predictions of available theoretical models for the axial nucleon structure in the kinematical conditions of ongoing neutrino experiments. The dipole parametrization of form factors is predicted by perturbative QCD, when the transferred momentum is transmitted to all three quarks leaving the nucleon in its ground state. In a non relativistic approach, the dipole distribution is the Fourier transform of an exponential charge distribution.  Other models and parametrizations are available (axial-Vector dominance~\cite{Amaro:2015lga}, neural-network Bayesian analyses~\cite{Alvarez-Ruso:2018rdx}) as well as recent results from lattice QCD~\cite{Alexandrou:2017hac}. In this letter the axial form factor is calculated in framework of Vector Meson Dominance (VDM). We consider the parametrization of the axial form factor from Ref.~\cite{Adamuscin:2007fk} and apply it to the study of (anti)neutrino-nucleus cross-sections. This parametrization is inspired by the two-components model of Iachello, Jackson, Land\'e model (IJL)~\cite{Iachello:1972nu} for nucleon electromagnetic form factors, further extended to the time-like region~\cite{Bijker:2004yu}.  
\subsection{The neutrino-nucleus cross-section}
For a given energy $E_\nu$ of the incident neutrino, the double differential cross-section for inclusive CC neutrino-nucleus scattering, $(\nu_l,l)$, can be written in the Rosenbluth form:
\begin{equation}
    \frac{d^2\sigma}{dp_l d\cos\theta_l} = \sigma_0 \left(v_{CC} R_{CC}+v_{CL} R_{CL}+v_{LL} R_{LL}+v_T R_T\pm v_{T'} R_{T'}\right) ,
    \label{eq:cs}
\end{equation}
where $p_l$ and $\theta_l$ are the outgoing lepton momentum and scattering angle, respectively, $\sigma_0$ is the Mott-like cross-section for weak interactions, $v_K$ are factors depending on the lepton kinematics~\cite{Amaro:2004bs} and $R_K(q,\omega)$ are the five response functions, embodying the nuclear physics content of the problem and depending on the transferred momentum and energy $q$ and $\omega$. The $\pm$ signs correspond to $\nu$ and $\bar\nu$ scattering, respectively.
The $CC$, $CL$, $LL$ and $T$ response functions can be decomposed into vector-vector $(VV)$ and $(AA)$ parts
\begin{equation}
    R_K = R_K^{VV}+R_K^{AA} ,\ \ \ \ \ \ \ \ K=CC,CL,LL,T,
\end{equation}
arising from the contraction of the $V$ and $A$ leptonic and hadronic current, whereas the response $R_{T'}=R_{T'}^{VA}$ comes from the interference of the leptonic vector and hadronic axial currents.
According to the the values of $q$ and $\omega$, different reactions contribute to the nuclear responses, which depend upon both the single-nucleon form factors entering the elementary process and the model employed to describe the initial and final nuclear state.  
In the quasielastic (QE) region, centered at $\omega=\frac{|Q^2|}{2m_N}$, the dominant process is the elastic scattering of the probe with a bound moving nucleon, corresponding to the current appearing in Eq.~\eqref{eq:mat}. Therefore the $AA$ and $VA$ quasielastic responses are affected by the axial form factor uncertainties. The effect of different parametrizations of the vector form factors on the cross-section has been explored in Refs.~\cite{Megias:2017PhD} and found to be negligible at the relevant kinematics for current neutrino experiments~\cite{Megias:2013aa}.

In comparing theoretical predictions with neutrino data the cross-section \eqref{eq:cs} must be folded with the experimental flux, which varies in mean energy and broadness depending on the specific experiment.  As a consequence, unlike the case of electron scattering where the beam energy is precisely known, for (anti)neutrino scattering it is impossible to disentangle the genuine quasielastic reaction - {\it i.e.,} the excitation of one-particle-one-hole (1p1h) states - from other processes leading to the same final state. When only the outgoing lepton is detected and no pions are present in the final state, the so-called CC$0\pi$ cross-section receives contributions not only from QE scattering, but also from processes induced by two-body meson-exchange currents (MEC), that can excite both 1p1h and 2p2h states. These must be accounted for in the comparison with experimental data and can be more or less sizeable depending on the kinematics~\cite{Megias:2016lke,Megias:2016fjk,Megias:2017PhD}. The MEC also depend on various form factors, the most important ones being related to the weak $N\to\Delta$ transition. In this work we will stick to the form factors used in Ref.~\cite{Simo:2016ikv} for the 2p2h responses and focus on the sensitivity of the 1p1h response to the nucleon axial form factor $G_A$.

  \section{New evaluation of the axial form factor in the two-component model}
  \label{sec:model}
We perform for the first time a joint analysis of neutrino-deuterium scattering data and pion electroproduction data in order to evaluate the sensitivity on the axial form factor of the nucleon, in the framework of the two-component model~\cite{Iachello:1972nu}. Such model was used in Ref.~\cite{Adamuscin:2007fk} to analyze pion electroproduction data. In Ref~\cite{Bodek:2007ym} a comprehensive analysis of neutrino-deuterium data was performed with the BBA07 parametrization. We rely on the latter for the analysis of neutrino-scattering data reinterpreting the tabulated values of the form factor in the two-component model. Further constraints on the axial form factor parametrization are extracted from a re-analysis of the electromagnetic form factors in electron scattering data, updating the analysis reported in Ref.~\cite{Bijker:2004yu}.
\subsection{Features of two-component model for the axial form factor of the nucleon}
\label{sec:IJL}
The picture of the nucleon, where the three quarks are concentrated in a hard core of radius $r \approxeq 0.34$ fm surrounded by a meson cloud, was suggested by  Iachello, Jackson, Land\'e (IJL model)~\cite{Iachello:1972nu}  since 1973. In particular, this model predicted the decrease of the electric to magnetic form factor ratio, much earlier than precise data, based on the recoil proton polarization Akhiezer- Rekalo method~\cite{Akhiezer:1968ek,Akhiezer:1974em}, were collected~\cite{Puckett:2017flj}. This approach was successful in describing the four nucleon electromagnetic form factors (electric and magnetic, for proton and for neutron)~\cite{Bijker:2004yu,Wan:2005ds,Iachello:2004aq}, the strange form factors of the proton~\cite{Bijker:2005pe} and was applied to the deuteron as well~\cite{TomasiGustafsson:2005ni}.  Advantages of this model are that it contains a limited number of parameters and can be applied both in the space- and time-like regions. The extension to axial form factors has been done in \cite{Adamuscin:2007fk}.

Following Ref.~\cite{Iachello:1972nu}, the axial nucleon FF can be parametrized as:
\begin{eqnarray}
G_A(Q^2) &=& G_A(0) \, g(Q^2) \left[ 1-\alpha +\alpha \frac{m_A^2}{m_A^2+Q^2} \right] ~, 
\nonumber\\
g(Q^2) &=& \left (1+\gamma Q^2\right )^{-2} ~,
\label{eq:eq1}
\end{eqnarray}
where $Q^2>0$ in the space-like region and $\alpha$ is a fitting parameter which corresponds to the coupling of the photon with an axial meson. 
One can fix $m_A= 1.230$ GeV, corresponding to the mass of the axial meson $a_1(1260)$ with  $I^G(J^{PC})=1^-(1^{++})$. The form factor $g(Q^2)$ describes the coupling to the intrinsic structure (three valence quarks) of the nucleon.

In non-relativistic approximation (and in a relativistic framework but in the Breit reference frame) form factors are Fourier transforms of the charge and magnetic densities. The Fourier transform of the dipole \eqref{eq:dipole} form is an exponential
\be
FT(G_A^D(Q^2))=G_A(0)\exp(-r/M_A) ~.
\label{eq:FT}
\ee
The largest is $M_A$ the softer is the density that expands to larger distances.
Similarly to the charge radius, the nucleon axial radius ($r_A$) is defined as:
\be
<r_A^2>=-6\frac{1}{G_A(0)}\frac{dG_A(Q^2)}{dQ^2}|_{Q^2=0} ~,
\label{eq:radius}
\ee
that gives 
\be
\frac{G_{A}(Q^2)}{G_{A}(0)}=1-\frac{1}{6}Q^2 <r_{A}^2>+O(Q^4).
\label{eq:radiusSeries}
\ee

Therefore from the slope of the axial form factor at 
$Q^2\to 0$ one can deduce a value of  the axial radius $\sqrt{<r_A^2>}=0.60$ fm for Ref.~\cite{Iachello:1972nu} and a slightly larger value of $0.62$ fm for Ref.~\cite{Bijker:2004yu}, slightly smaller than the values obtained from 
the dipole parametrization : $\sqrt{<r_A^2>}=0.64$ fm  $M_A=1.069$ GeV corresponding to  charged pion electroproduction and $0.67$ fm for $M_A=1.026$ GeV as found in neutrino scattering.  Note that from the two-component model, one can disentangle the contribution of the quark core and of the meson cloud to the axial radius, finding that the meson cloud dominates by a factor of ten. Updated values of the proton radius are extracted from the joint fit of electron scattering and neutrino scattering data in the next section.

\subsection{Fit of the electromagnetic form factor to electron scattering data}
\label{sec:EMfit}
In the two component model, the intrinsic form factor related with the three-quark structure of the nucleon has the following parametrization, common to the axial and electromagnetic interaction
\begin{equation}
 g(Q^2)=\frac{1}{(1+\gamma Q^2)^2}~.
 \label{eq:gQ2}
\end{equation}
Notably, this form is consistent with partonic QCD, even if the model was introduced before the development of partonic QCD. The $\gamma$ parameter can be extracted from a fit to the electromagnetic form factors, as reported in Ref.~\cite{Bijker:2004yu}. Such analysis has been repeated here, including additional data made available since then and the correction for the logarithmic dependence of perturbative QCD, which was suggested but not included in the fit in Ref.~\cite{Bijker:2004yu}:
\begin{equation}
  \label{eq:EMFF_QCDcorr}
Q^2 \rightarrow Q^2 \frac{\ln[(\Lambda^2+Q^2)/\Lambda_{QCD}^2]}{\ln(\Lambda^2/\Lambda_{QCD}^2)},
\end{equation}
with $\Lambda=2.27$~GeV and $\Lambda_{QCD}=0.29$~GeV~\cite{Gari:1986rj}. Such correction may give non-negligible effects at relatively high $Q^2$ ($>10$\% above 1~GeV$^2$). 

Following Ref.~\cite{Bijker:2004yu}, the electromagnetic Sachs form factors are expressed as
\begin{eqnarray}
  G_{M_p}&=&(F_1^S+F_1^V)+(F_2^S+F_2^V),\\
  G_{E_p}&=&(F_1^S+F_1^V)-\frac{Q^2}{4M_p^2}(F_2^S+F_2^V),\\
  G_{M_n}&=&(F_1^S-F_1^V)+(F_2^S-F_2^V),\\
  G_{E_n}&=&(F_1^S-F_1^V)-\frac{Q^2}{4M_n^2}(F_2^S-F_2^V),
\end{eqnarray}
as a function of the Dirac (Pauli) isoscalar or isovector, $F^S_1(Q^2)$ ($F^S_2(Q^2)$) or $F^V_1(Q^2)$ ($F^V_2(Q^2)$) form factors which are parametrized as
\begin{eqnarray}
  F_1^S(Q^2)&=&\frac{1}{2} g(Q^2)\left[ 1- \beta_\omega - \beta_\phi + \beta_\omega \frac{m^2_\omega}{m^2_\omega+Q^2}+ \beta_\phi \frac{m^2_\phi}{m^2_\phi+Q^2}\right],\\
  F_1^V(Q^2)&=&\frac{1}{2} g(Q^2)\left[ 1- \beta_\rho + \beta_\rho \frac{m_\rho^2+8\,\Gamma_\rho m_\pi/\pi}{m_\rho^2+Q^2+(4m^2_\pi+Q^2)\Gamma_\rho\alpha(Q^2)/m_\pi}\right],\\
  F_2^S(Q^2)&=&\frac{1}{2} g(Q^2)\left[ (\mu_\pi+\mu_n-1-\alpha_\phi)\frac{m^2_\omega}{m^2_\omega+Q^2}+ \alpha_\phi\frac{m^2_\phi}{m^2_\phi+Q^2}\right],\\
  F_2^V(Q^2)&=&\frac{1}{2} g(Q^2)\left[\frac{(\mu_\pi-\mu_n - 1 -\alpha_\rho)}{1+\gamma Q^2} +\alpha_\rho \frac{m_\rho^2+8\,\Gamma_\rho m_\pi/\pi}{m_\rho^2+Q^2+(4m^2_\pi+Q^2)\Gamma_\rho\alpha(Q^2)/m_\pi}\right],\\
\end{eqnarray}
with
\begin{equation}
  \alpha(Q^2)=\frac{2}{\pi}\left[ \frac{4m_\pi^2+Q^2}{Q^2}\right]^{1/2} \ln\left(\frac{\sqrt{4m_\pi^2+Q^2}+\sqrt{Q^2}}{2m_\pi}\right) ~,
  \end{equation}
$\mu_p=2.793$, $\mu_n=-1.913$, $m_\pi=0.1396$~GeV, $m_{\rho}=0.776$~GeV, $m_{\omega}=0.783$~GeV, $m_{\phi}=1.019$~GeV and $\beta_\rho$, $\beta_\omega$, $\beta_\phi$, $\alpha_\rho$, $\alpha_\phi$, $\gamma$ free parameters in the fit to electron-scattering data.

\begin{figure}
 \begin{center}
  \includegraphics[width=8cm]{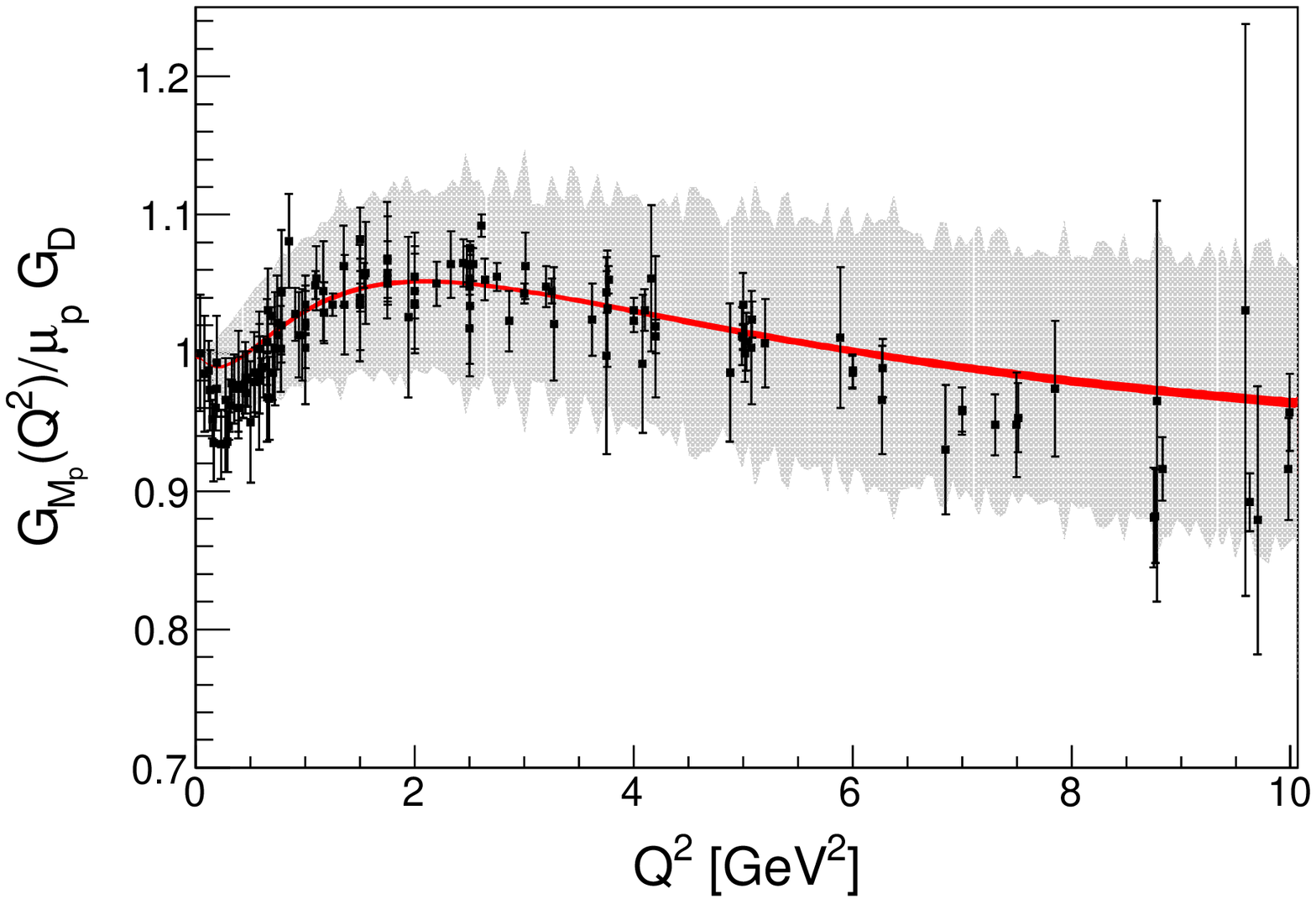}
    \includegraphics[width=8cm]{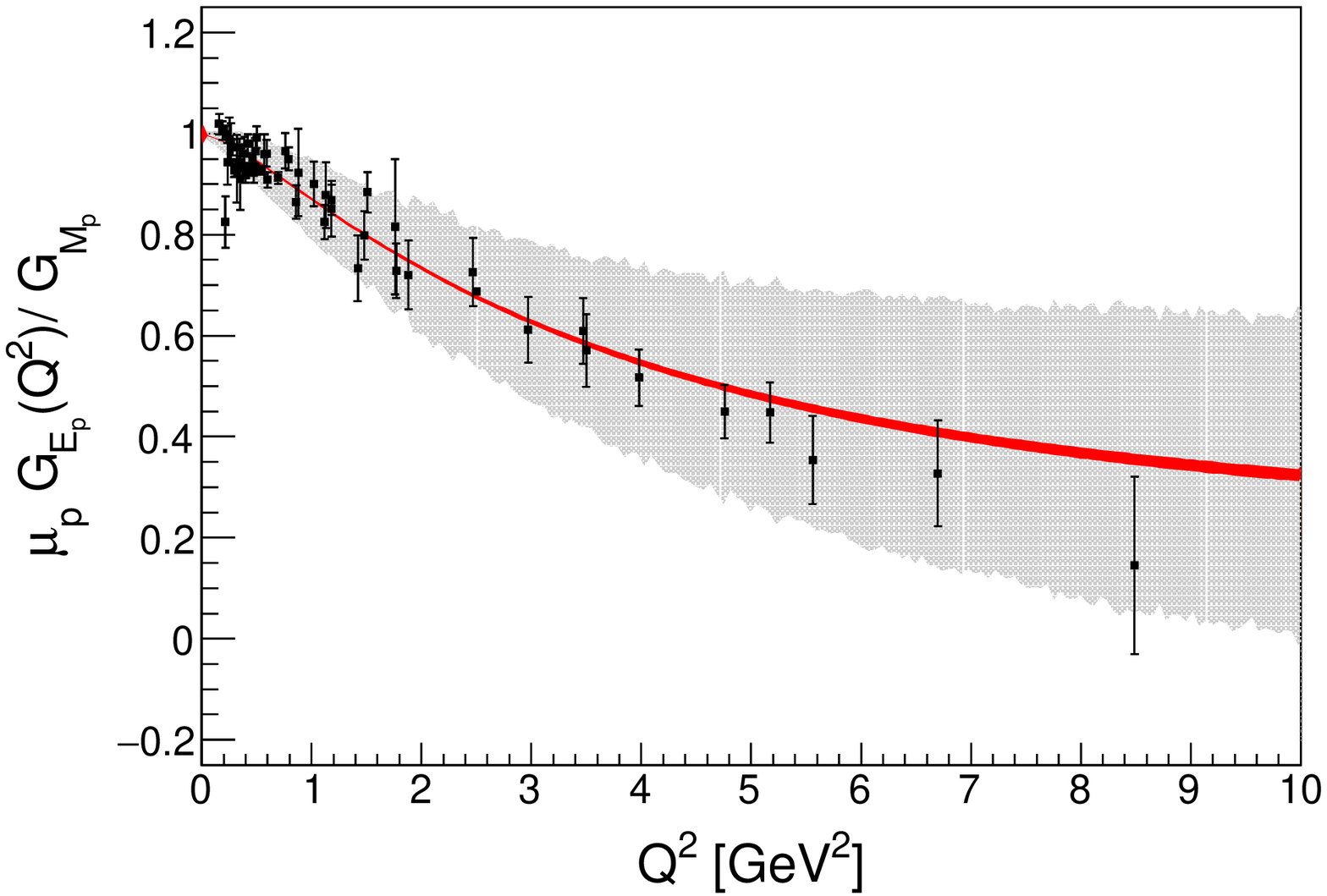}\\
     \includegraphics[width=8cm]{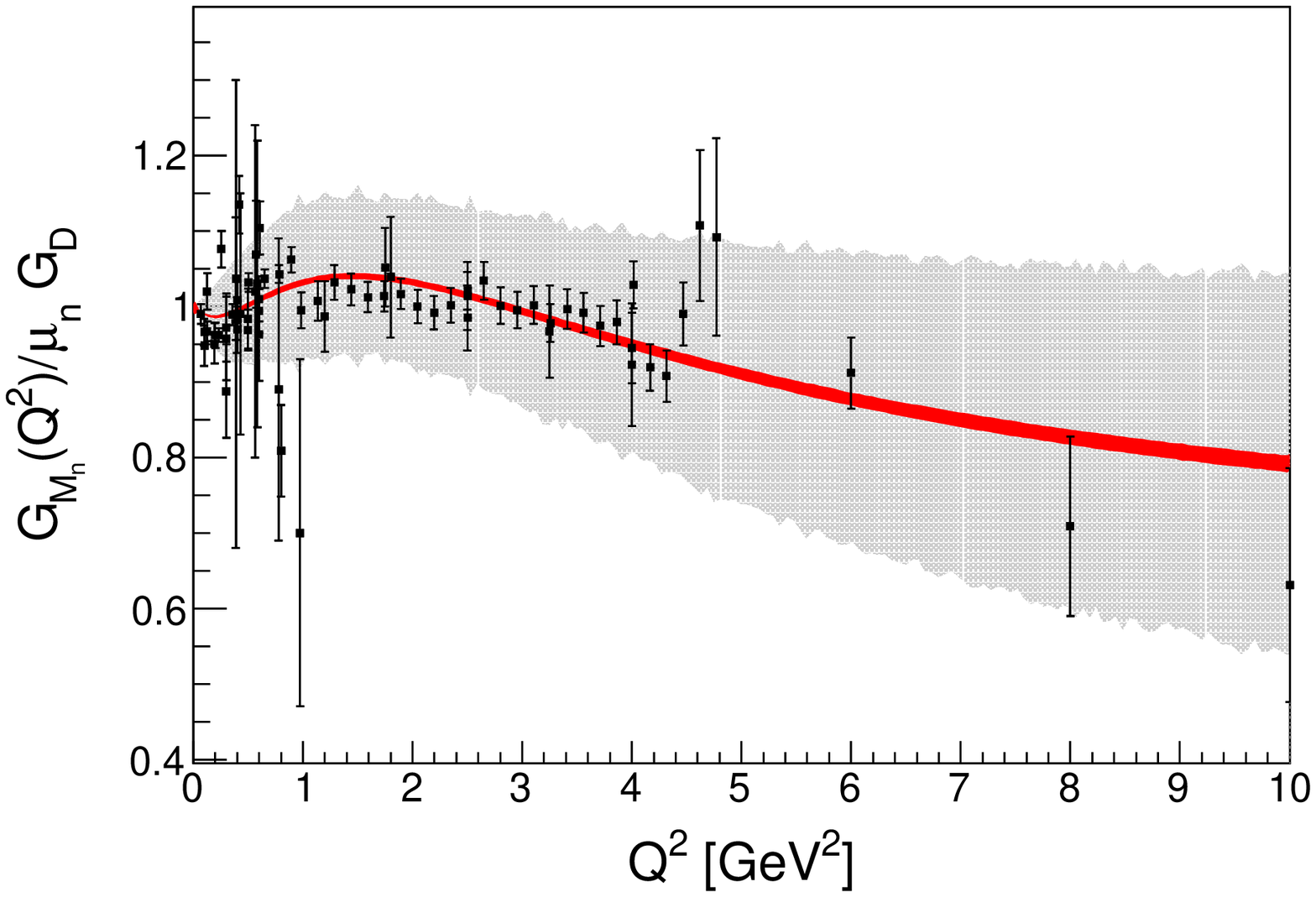}
    \includegraphics[width=8cm]{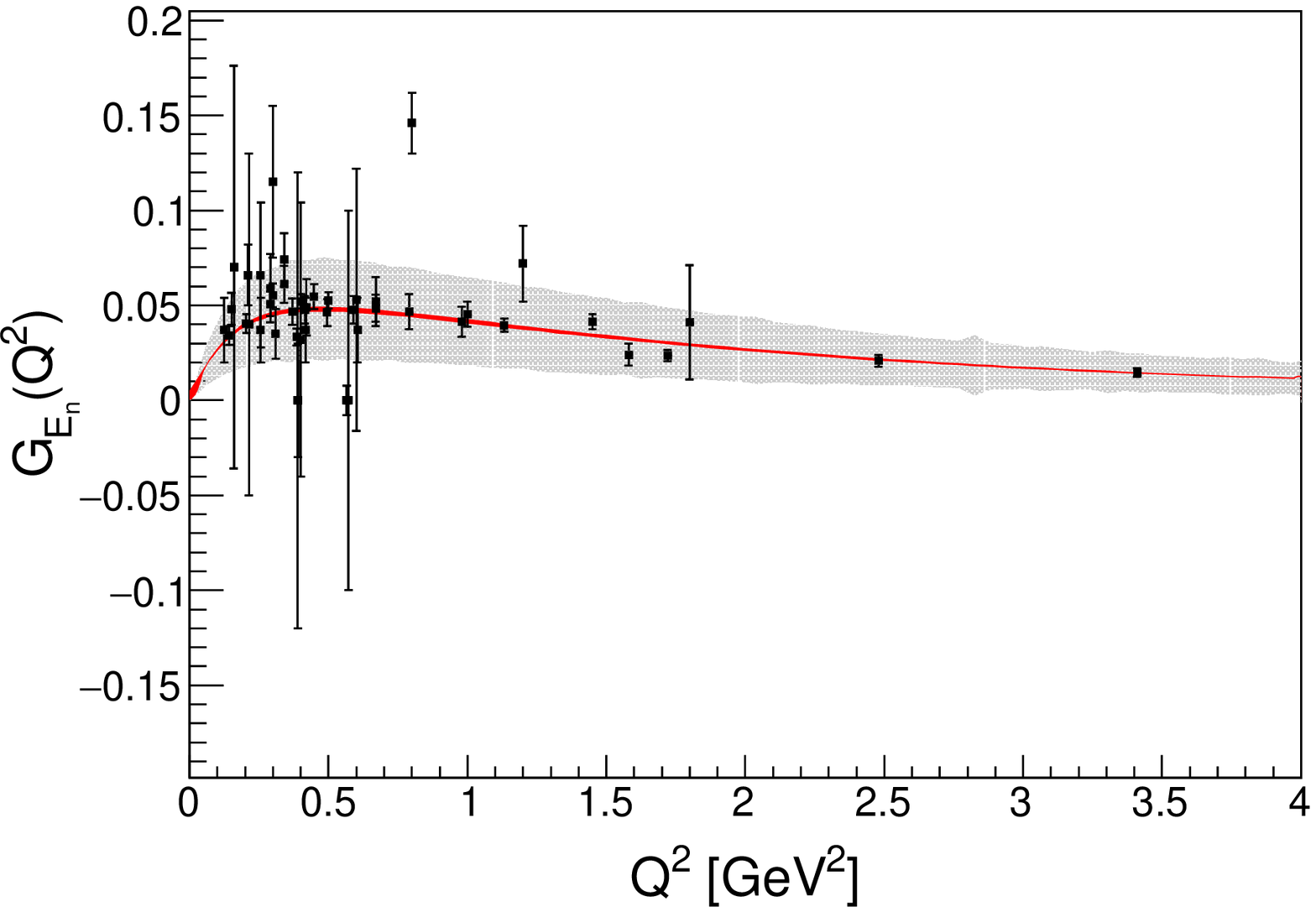}
\end{center}
\caption{Data and fit to the electromagnetic nucleon form factor in the two-component model. The red (grey) band shows the statistical uncertainty before (after) the error inflation procedure}
 \label{fig:EMFFfit}
\end{figure}

We have analyzed the data from Ref.~\cite{Akerlof:1964zz,Anderson:2006jp,Anklin:1998ae,Bartel:1973rf,Bermuth:2003qh,Bruins:1995ns,Becker:1999tw,Eden:1994ji,Geis:2008aa,Glazier:2004ny,Golak:2000nt,Hanson:1973vf,Herberg:1999ud,Kirk:1972xm,Litt197040,Kubon:2001rj,Lachniet:2008qf,Meyerhoff:1994ev,Passchier:1999cj,Plaster:2005cx,Riordan:2010id,Ron:2011rd,Rohe:1999sh,Rock:1982gf,Ostrick:1999xa,Stein:1966ke,Schlimme:2013eoz,JonesWoodward:1991ih,Sulkosky:2017prr,Warren:2003ma,Zhu:2001md,Xu:2002xc,Albrecht:1965ki,Andivahis:1994rq,Christy:2004rc,PhysRevLett.20.292,Crawford:2006rz,PhysRev.142.922,PhysRevD.4.45,Jones:1999rz,Gayou:2001qt,Jones:2006kf,MacLachlan:2006vw,Puckett:2011xg,Qattan:2004ht,PhysRevD.48.29,Walker:1993vj,Zhan:2011ji,Berger:1971kr,Puckett:2017flj,Lung:1992bu}. A reasoned selection of available data has been performed, notably removing old measurements when new and more precise ones are available for the same $Q^2$ region; only data below 10~GeV$^{2}$ have been considered. We did not include the very precise data from Ref.~\cite{Bernauer:2010wm} since they would constrain strongly the fit with a very large statistics at very low $Q^2$. Given that this region is not the most relevant for the neutrino-scattering experiments considered here, we leave the inclusion of those data and the corresponding discussion for a further work.
Those data are anyway well inside the final uncertainty of the present fit (including the inflation error procedure described below). 
The results of the fit are presented in Fig.~\ref{fig:EMFFfit} and in Tab.~\ref{tab:EMFFfit}. 

The fit results and the $\chi^2$ evaluation may be affected by correlations between the data points. Large correlations are possible for data points obtained by the same experiment due to experimental systematic uncertainties. Unfortunately such correlations are not published. Correlations between different data sets are also plausible due to theoretical uncertainties in the extraction of the form factors from the measured cross-sections. In order to have conservative enough uncertainties, notably for the $\gamma$ parameter which is used for the axial form factor evaluation in Sec.~\ref{sec:fitaxial}, the statistical treatment of error inflation as in Ref.~\cite{PinzonGuerra:2018rju} is applied. This treatment is based on the discussion in Ref.~\cite{Pumplin:2000vx} and consists in inflating the errors of the post-fit form factor parameters to ensure that the pulls between the fit and the data have RMS$=1$. The pulls are defined as
\begin{equation}
\frac{G^{fit}-G^{measured}}{\Delta G(f\times \Delta p)},
\end{equation}
where $G$ are the different form factors shown in Fig.~\ref{fig:EMFFfit} and $\Delta G$ is the error on such form factors computed propagating numerically the uncertainties ($\Delta p$) on the parameters of Tab.~\ref{tab:EMFFfit} inflated by a factor $f$, conserving the fit correlations. This procedure provides over-coverage of the data and thus ensures conservative uncertainties on the form factor parameters to be used in the following.

\begin{table}[htbp]
\centering
   \caption{\label{tab:EMFFfit} Results of the fit of the electromagnetic nucleon form factor to electron scattering data in the two-component model and comparison to the previous analysis of Ref.~\cite{Bijker:2004yu}. Uncertainties and fit $\chi^2$ are not available for the latter. For the present analysis the uncertainties before and after error inflation are reported. NDOF here refers to number of degrees of freedom.}
   \begin{tabular}{|l|ll|l|}
\hline \textbf{Parameter} &   & with error inflation & previous analysis~\cite{Bijker:2004yu}\\
\hline
$\beta_\rho$ & $0.475 \pm 0.007$ &  $ \pm 0.14$ & $0.512$ \\
$\beta_\omega$ & $1.19\pm 0.03$ & $\pm 0.59$ &  $1.129$ \\
$\beta_\phi$ & $-0.37 \pm 0.03$ & $\pm 0.66$ &  $-0.263$ \\
$\alpha_\rho$ & $2.65 \pm 0.03$ & $\pm 0.59$ & $2.675$ \\
$\alpha_\phi$ & $-0.27 \pm 0.03$ & $ \pm 0.60$ &  $-0.200$ \\
$\gamma$ (GeV$^{-2}$)& $0.529\pm0.004$ &  $\pm0.093$ & $0.515$\\
\hline
$\chi^2/$NDOF (NDOF) & \multicolumn{2}{c|}{2.01 (307)} & - \\
\hline
   \end{tabular}
\end{table}

\subsection{Fit of axial form factor from pion electro-production and neutrino-scattering data}
\label{sec:fitaxial}
The joint fit of the electron- and neutrino-scattering data is performed taking into account the pion loop corrections from Ref.~\cite{Bernard:2001rs}. Therefore two separated functions are defined for the nucleon form factor in pion electroproduction and neutrino-scattering data
\begin{eqnarray}
  \label{eq:GApi}
  \widetilde{G_A^\pi}(Q^2)&=&\frac{g_A}{(1+\widetilde{\gamma_\pi} Q^2)^2}\left(1-\widetilde{\alpha_\pi}+\widetilde{\alpha_\pi}\frac{m^2_A}{m^2_A+Q^2}\right)\\
  \label{eq:GAnu}
  G_A^\nu(Q^2)&=&\frac{g_A}{(1+\gamma_\nu Q^2)^2}\left(1-\alpha_\nu+\alpha_\nu\frac{m^2_A}{m^2_A+Q^2}\right)
\end{eqnarray}
with $m_A=1.23$~GeV, $g_A=G_A(0)=1.2695$~\cite{Tanabashi:2018oca} and $\alpha^{\pi,\nu}$, $\gamma^{\pi,\nu}$ free parameters of the fit. It should be noted that $\widetilde{G_A^\pi}(Q^2)$ is an effective form factor parametrization, not corrected for Ref.~\cite{Bernard:2001rs}.
The neutrino form factor and the pion electroproduction effective form factor are related by joint constraints in the likelihood minimization which include the correction on the nucleon radius of Ref.~\cite{Bernard:2001rs}:
\begin{equation}
  \chi^2=\sum^{\nu data}\left[\frac{G_A^{\nu}(Q^2)-x}{\delta x}\right]^2 + \sum^{\pi data}\left[\frac{\widetilde{G_A^{\pi}}(Q^2)-x}{\delta x}\right]^2 + \left(\frac{\widetilde{\gamma_\pi}-0.529}{0.093}\right)^2 + \left(\frac{\gamma_\nu-0.529}{0.093}\right)^2 +\left(\frac{r^2_\nu-\widetilde{r^2_\pi}-0.0456}{0.0050}\right)^2~,
\end{equation}
where $x$ are measurements of the form factors in pion electroproduction and neutrino-scattering data and $\widetilde{r^{\pi}},r^{\nu}$ is the nucleon radius from Eq.~\eqref{eq:radius} evaluated using the form factors of Eq.~\eqref{eq:GApi},~\eqref{eq:GAnu}, respectively. The second and third term of the likelihood are penalty terms to include the constraints on $\gamma$ from the fit to the electromagnetic form factors of Sec.~\eqref{sec:EMfit} and the correction to pion electroproduction values due to loop corrections from Ref.~\cite{Bernard:2001rs}, with the corresponding uncertainty.

We fit the pion electroproduction data as reported in~\cite{Adamuscin:2007fk}, from various sources~\cite{Nambu:1970jm,Benfatto:1973dq,Furlan:1970cw,Dombey:1973ve,Joos:1976ng}, and the neutrino-scattering data as reported in~\cite{Bodek:2007ym}, from various sources~\cite{Allasia:1990uy,Baker:1981su,Son:1983xh,Kitagaki:1990vs,Miller:1982qi}. The pion electroproduction data can be interpreted in different theory frameworks to extract the form factor. We consider here separately several cases: the Soft Pion approximation~\cite{Nambu:1970jm}, the Partially Conserved Axial Current approximation (PCAC)~\cite{Benfatto:1973dq}, the Furlan approximation (enhanced soft pion production)~\cite{Furlan:1970cw} and the Dombey and Read approximation~\cite{Dombey:1973ve}. Data corresponding to $\Delta$ excitation~\cite{Joos:1976ng} are considered separately. The spread between the different approximations and sets of data is sizable and should be considered as an intrinsic systematic uncertainty in the extraction of the form factor. 

\begin{figure}
 \begin{center}
 \includegraphics[width=8cm]{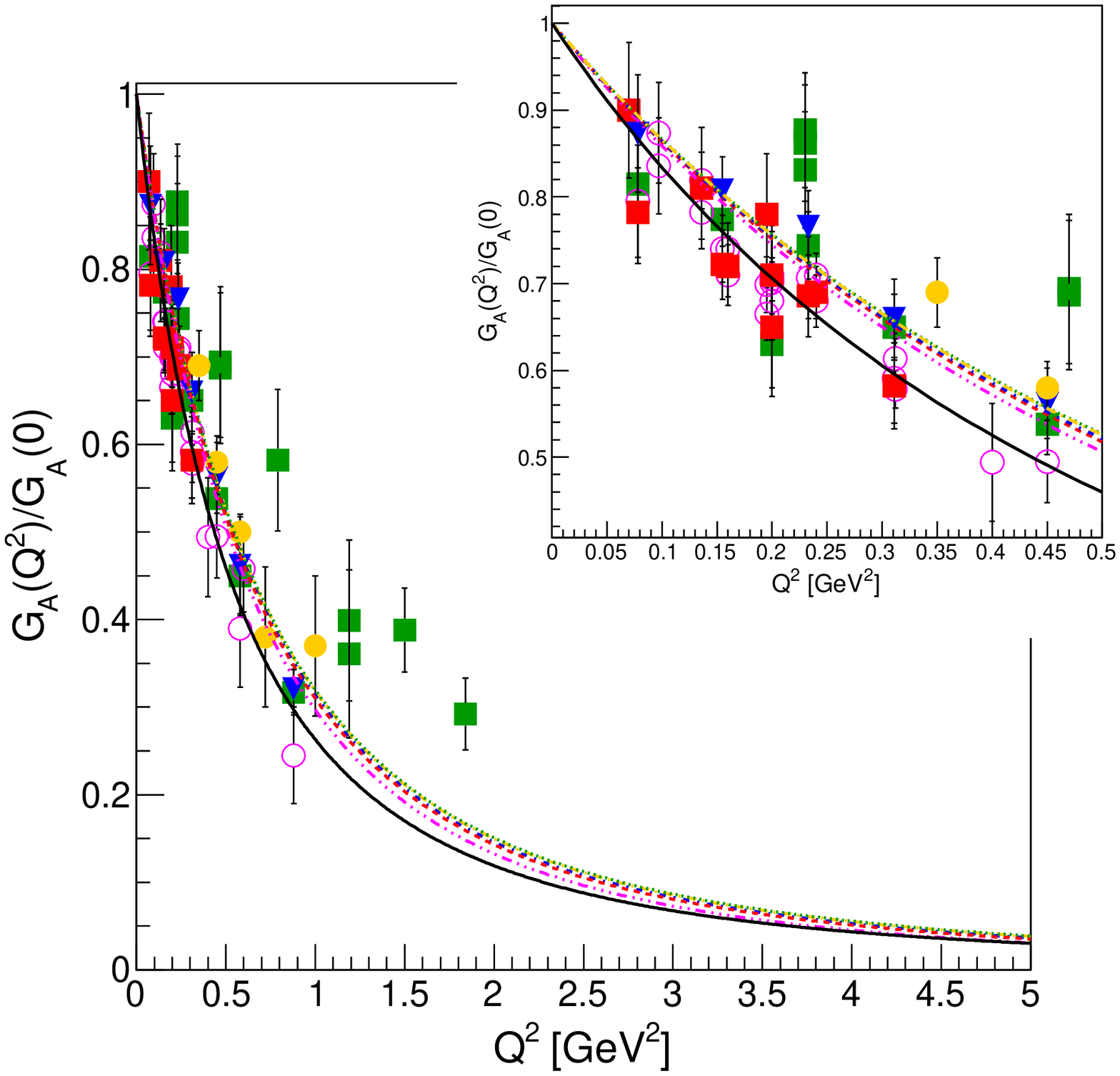}
    \includegraphics[width=5cm]{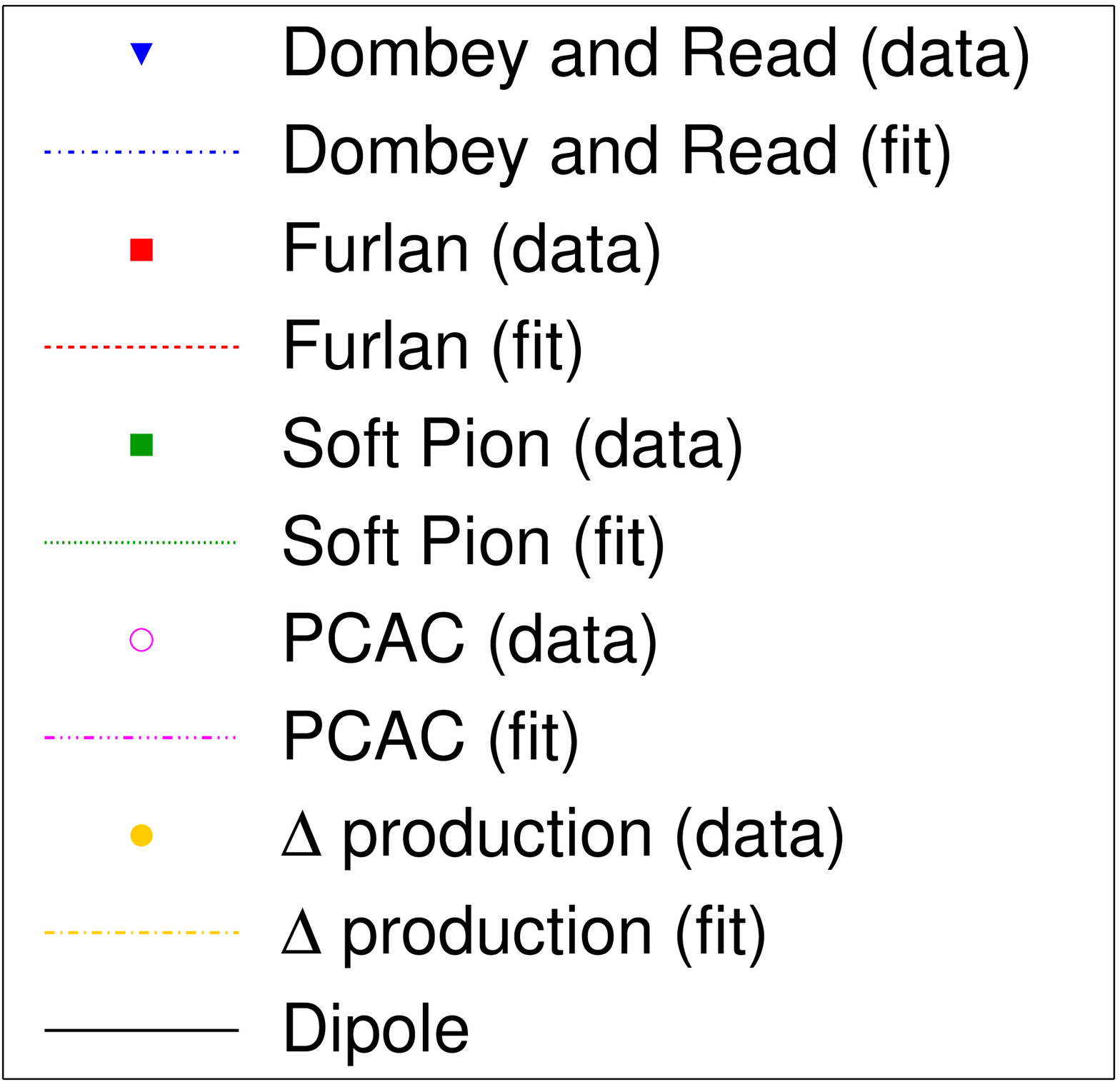}
    \includegraphics[width=8cm]{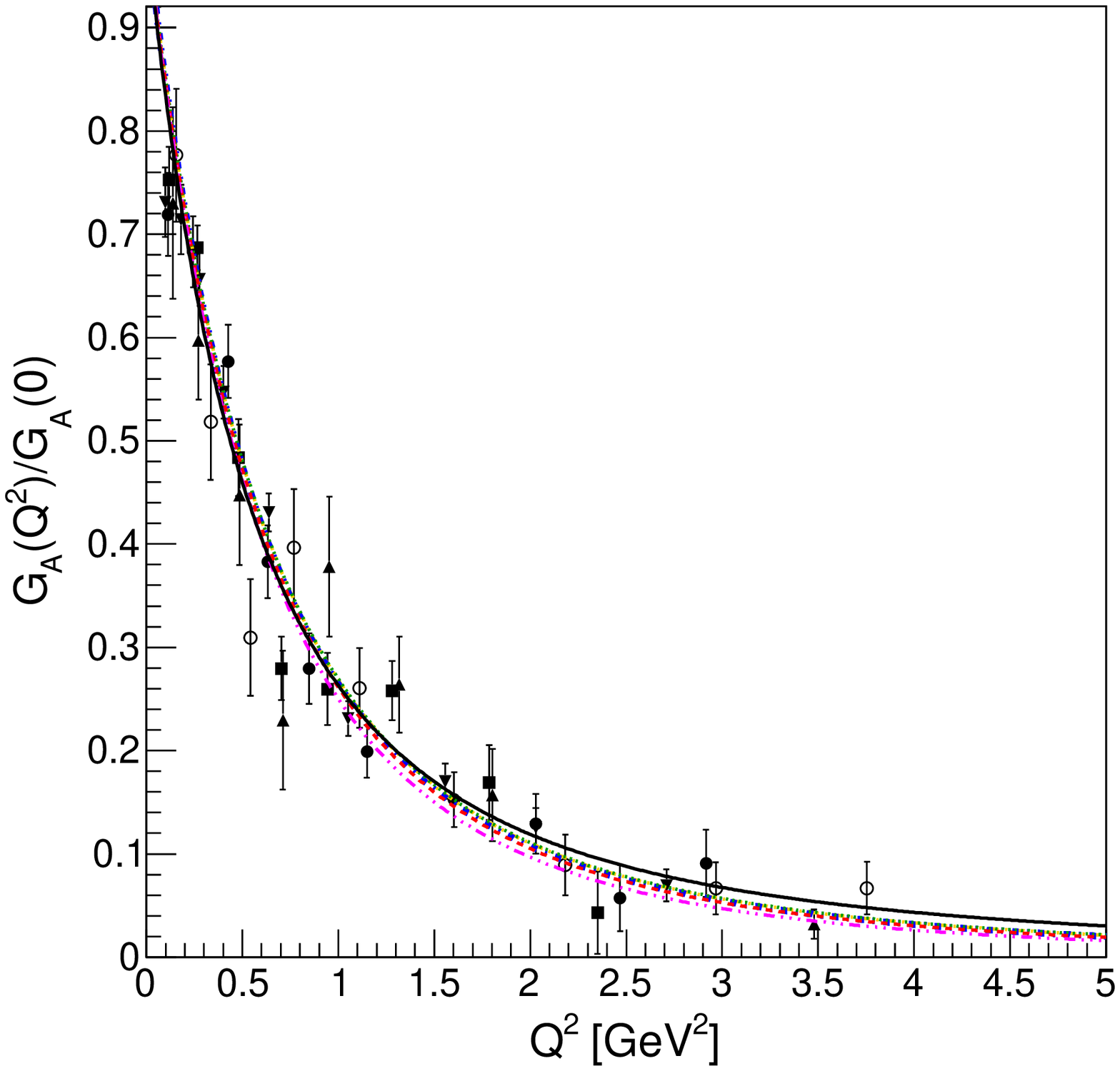}
    \includegraphics[width=5cm]{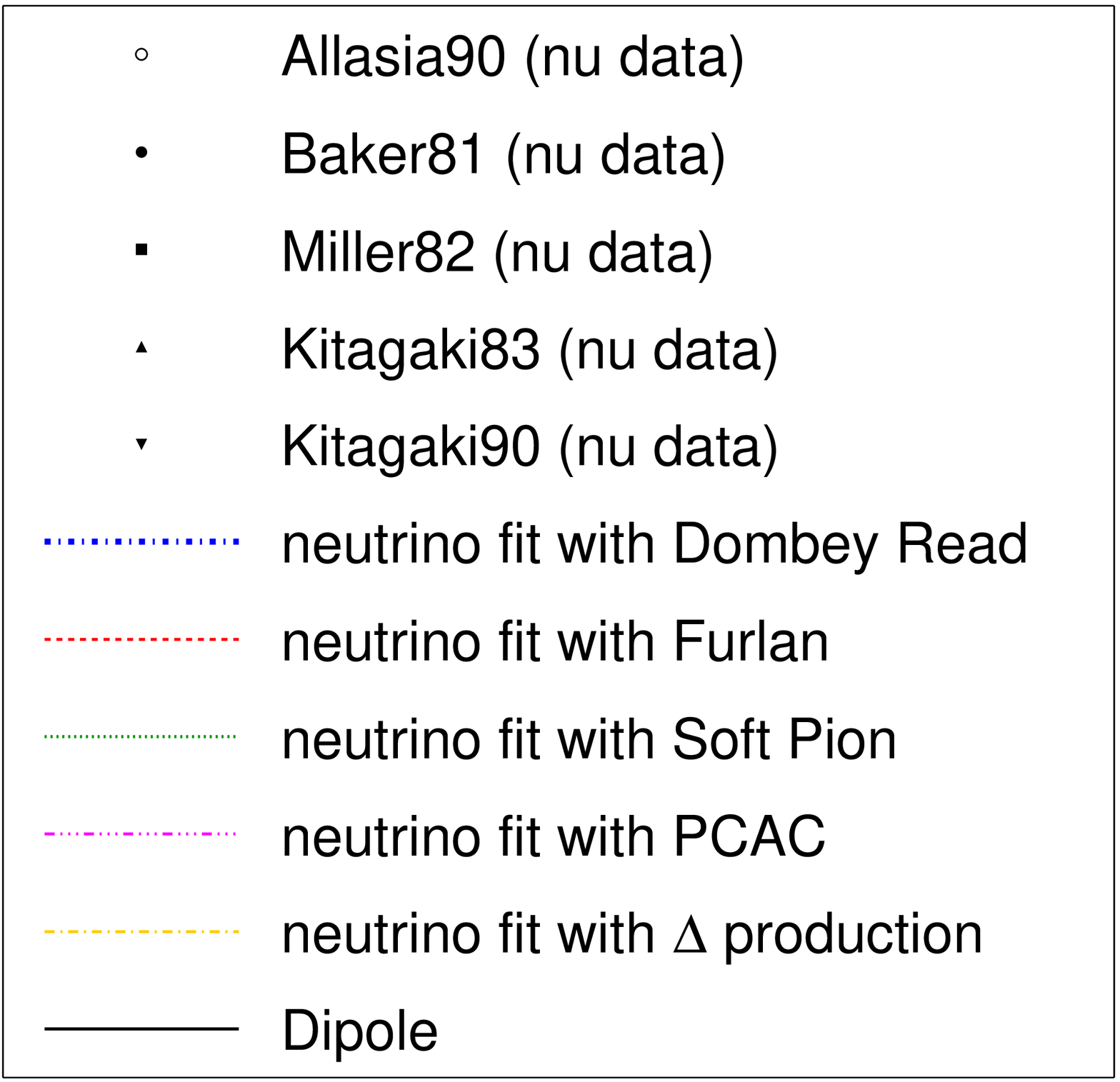}
\end{center}
\caption{Axial form factor in the two-components model evaluated with joint fits to the pion electroproduction data (top) and neutrino scattering data (bottom) for different interpretations of the pion data. The dipole form factor is shown for comparison.
 \label{fig:FFfitNuPion}}
\end{figure}

\begin{figure}
 \begin{center}
  \includegraphics[width=8cm]{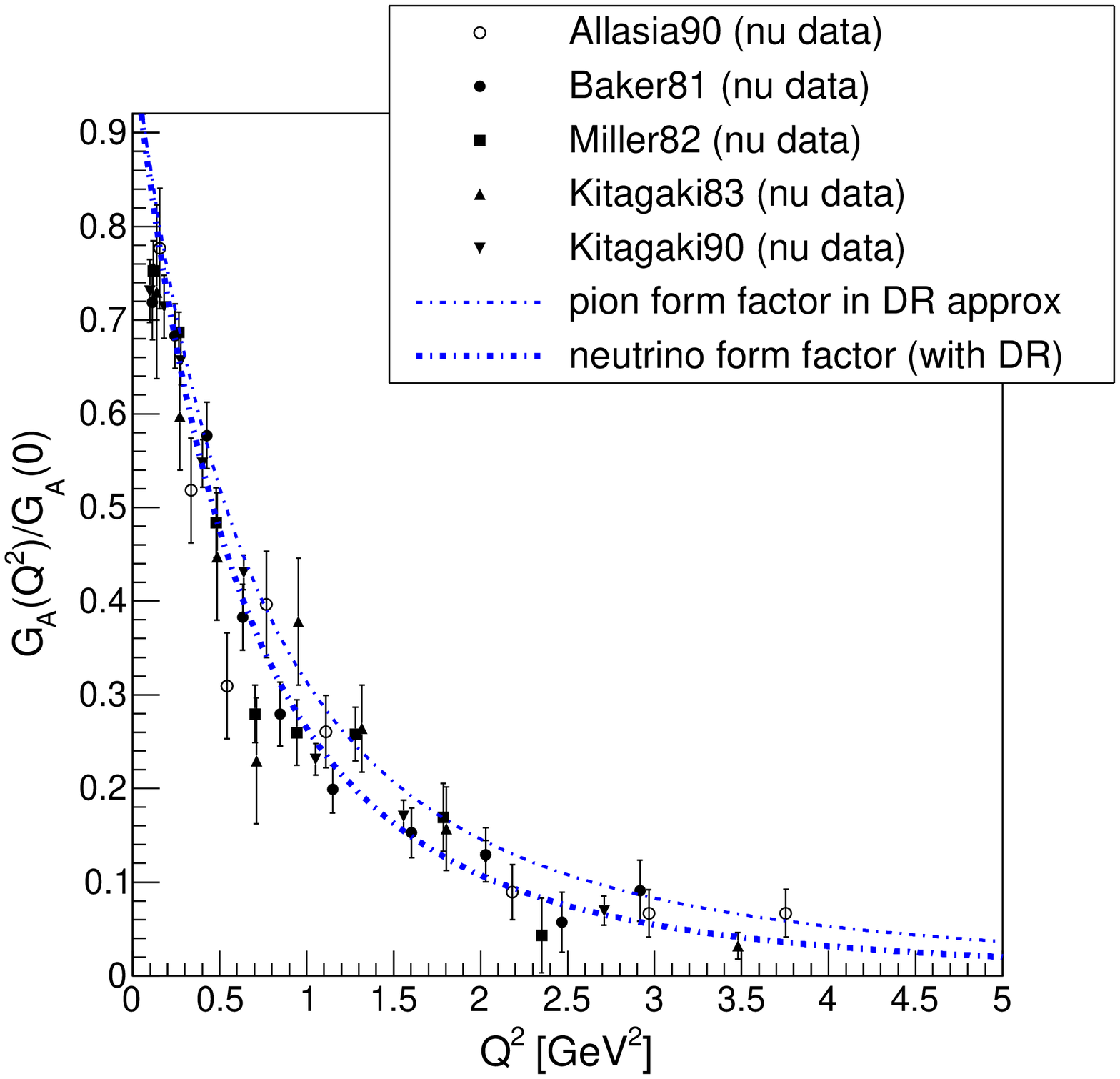}
    \includegraphics[width=8cm]{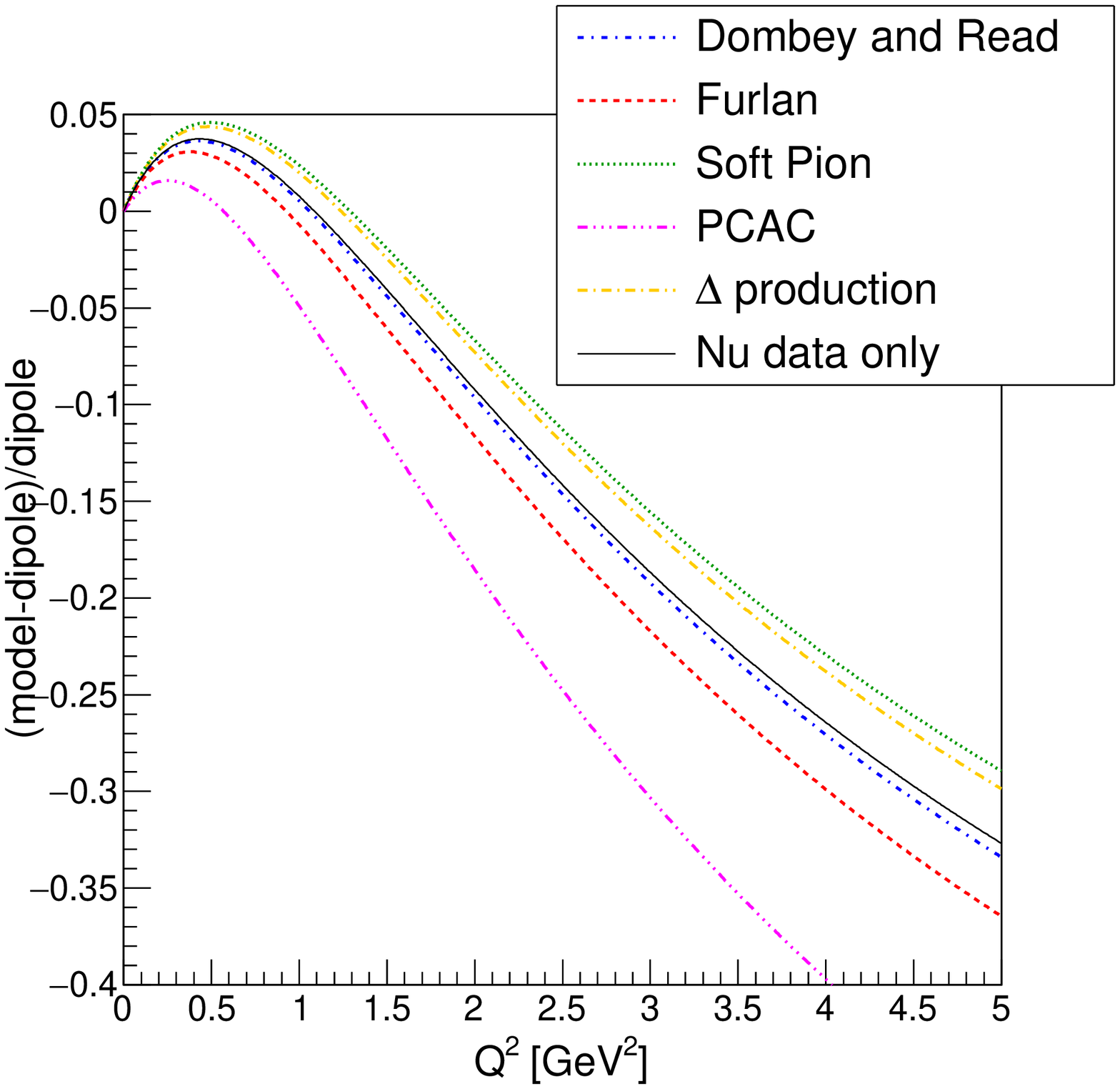}
\end{center}
\caption{Left: comparison of axial form factor for neutrino scattering data and pion electroproduction data (in the Dombey-Read approximation) obtained by a joint fit. Right: comparison of dipole axial form factor to axial form factors in the two-component model evaluated with fits to neutrino scattering data and to different interpretations of pion electroproduction data. 
 \label{fig:FFfitNuPion2}}
\end{figure}

The results of the fits are shown in Fig.~\ref{fig:FFfitNuPion} and Tab.~\ref{tab:FFfitNuPion}. The statistical uncertainties are a factor $\approx$~2 smaller than the spread from different theoretical interpretations of the pion electroproduction data, the PCAC and Soft Pion approximations corresponding to the two opposite extreme cases. In Fig.~\ref{fig:FFfitNuPion2} (left), the difference between the form factor in neutrino-scattering and in pion electroproduction is shown: according to the pion loop corrections, the latter measures a slightly smaller nucleus radius, before correction, thus they exhibit a less steep $Q^2$ dependence. The increasing of the fit $\chi^2$ in case of Soft Pion, PCAC and Furlan models indicates a tension between pion electroproduction data, in such interpretations, and the neutrino data. It should be noted, though, that no information is available on the correlation of uncertainties between the different data inside the same data set, nor between different data sets.

As can be seen in Fig.~\ref{fig:FFfitNuPion2} (right), all the fits show clear differences with respect to the dipole form factor (here evaluated with axial mass $M_A=1.026$~GeV): 
in neutrino-scattering,  the two-component model gives a larger form factor than the dipole below about 1~GeV$^2$ (up to 5\% difference), while above 1~GeV$^2$ the two-component model predicts a smaller form factor than the dipole model. Around 3~GeV$^2$ the difference is of the order of 15-30\%. In the same figure, the impact of including pion electroproduction data can be appreciated by comparing the fit using only neutrino-scattering data. As already shown in Tab~\ref{tab:FFfitNuPion}, the Dombey-Read interpretation of pion electroproduction data is the most in agreement with neutrino data.

\begin{table}[htbp]
\centering
   \caption{\label{tab:FFfitNuPion} Results of fit to pion electroproduction data and neutrino scattering data for different extractions of the pion data. The fit using only neutrino scattering data is also shown for comparison. The value of the proton radius extracted from Eq.~\eqref{eq:radius}, and including the corrections from Ref.~\cite{Bernard:2001rs} for pion electroproduction data ($r^2_\pi=\widetilde{r^2_\pi}-0.0456$), is reported.}
   \begin{tabular}{|l|l|l|l|l|l|l|l|}
\hline \textbf{Dataset} &  \textbf{$\widetilde{\alpha_\pi}$}&  \textbf{$\widetilde{\gamma_\pi}$} (GeV$^{-2}$)& \textbf{$\alpha_\nu$}&  \textbf{$\gamma_\nu$} (GeV$^{-2}$)& \textbf{$\chi^2/NDOF (NDOF)$} & \textbf{$r_\pi$} (fm)& \textbf{$r_\nu$} (fm)\\
\hline
Dombey-Read $+\nu$ & $0.67\pm0.05$ & $0.53\pm0.01$ & $0.95\pm0.04$ &  $0.53\pm0.01$  & 1.79 (46) & $0.630 \pm 0.007$ & $0.630 \pm 0.007$ \\
Furlan $+\nu$ & $0.69\pm0.05$ & $0.53\pm0.01$ & $0.96\pm0.04$ &  $0.0.53\pm0.01$  & 1.97 (50) & $0.633 \pm 0.006$ & $0.633 \pm 0.006$ \\
Soft Pion $+\nu$ & $0.63\pm0.05$ & $0.53\pm0.01$ & $0.93\pm0.04$ &  $0.53\pm0.01$  & 2.50 (57) & $0.626\pm 0.004$ & $0.626 \pm 0.005$ \\
PCAC $+\nu$ & $0.78\pm0.05$ & $0.53\pm0.01$ & $1.01\pm0.04$ &  $0.54\pm0.01$  & 2.10 (64) & $0.642 \pm 0.003$ & $0.642 \pm 0.004$ \\
$\Delta$ excitation $+\nu$ & $0.64\pm0.05$ & $0.53\pm0.01$ & $0.93\pm0.04$ &  $0.53\pm0.01$  & 1.93 (44) & $0.627 \pm 0.007$ & $0.627 \pm 0.007$ \\
$\nu$ data only & N/A & N/A & $0.94\pm0.04$ &  $0.53\pm0.01$  & 1.94 (41) & N/A & $0.630 \pm 0.004$ \\
\hline
   \end{tabular}
\end{table}

\section{Impact of form factor uncertainties on neutrino-nucleus interaction cross-sections}
\label{sec:neutrinoNucleus}
We focus here on CCQE and 2p2h cross-sections evaluated with the SuSAv2 model. We compare the cross-section with different axial form factors: the dipole form factor and the form factor in the two-components model evaluated with neutrino scattering and pion electroproduction data. For the latter we consider the two most extreme variations given by the Soft Pion and PCAC approximations in the interpretation of the pion electroproduction data. While the comparison of these form factor parametrizations allows the evaluation of the form factors uncertainties in a specific model of the nucleon, most of the arguments developed below, highlighting the phase space regions which are most affected by such uncertainties, has general relevance and can be applied to any form factor model.

\subsection{Features of the SuSAv2 model of the nucleus}
\label{sec:SuSAv2}

The SuSAv2 model, where "SuSA" stands for Super-Scaling Approach and "v2" for its improved version, is based on the idea that in order to test and constrain nuclear models to be used in the analyses of neutrino experiments, it is necessary to use the information provided by other experiments, in particular electron-nucleus scattering data. 

The model, first introduced in~\cite{Amaro:2004bs}, exploits the scaling and superscaling properties exhibited by electron scattering data in order to predict neutrino-nucleus observables. In its more recent version, SuSAv2~\cite{Gonzalez-Jimenez:2014eqa}, the model also takes into account the behaviour of the responses provided by the Relativistic Mean Field (RMF): in particular, the natural enhancement of the transverse electromagnetic response provided by RMF, a genuine dynamical relativistic effect, is incorporated in the SuSAv2 approach. 
However, while the RMF approach works properly at low to intermediate values of the momentum transfer $q$, where the effects linked to the treatment of the final-state interactions (FSI) are significant, it fails at higher $q$ due to the strong energy-independent scalar and vector RMF potentials, whose effects should instead become less and less important with increasing momentum transfer. In this regime the relativistic plane-wave impulse approximation (RPWIA) is indeed more appropriate. Therefore, the SuSAv2 model incorporates both approaches, RMF and RPWIA, and combines them through a $q$-dependent "blending" function that allows a smooth transition from low/intermediate (validity of RMF) to high (RPWIA-based region) $q$-values.

The SuSAv2 predictions for inclusive $(e,e')$ scattering on $^{12}$C have been presented in~\cite{Megias:2016lke}, where they are shown to provide a remarkably good description of the data for very different kinematical situations. In order to perform such comparison the SuSAv2 model has been extended from the the quasielastic (QE) domain to the inelastic region by employing phenomenological fits to the single-nucleon inelastic electromagnetic structure functions. Furthermore, ingredients beyond the impulse approximation, namely two-particle-two-hole (2p2h) excitations, have been added to the model. These contributions, corresponding to the coupling of the probe to a pair of interacting nucleons and associated to two-body meson exchange currents (MEC), are known to play a very significant role in the "dip" region between the QE and $\Delta$ peaks. In the SuSAv2 approach 2p2h excitations are treated within the Relativistic Fermi Gas (RFG) model, which allows for an exact and fully relativistic calculation, as required for the extended kinematics involved in neutrino reactions. The sum of CCQE and 2p2h cross-sections are called CCQE-like in the following.
Comparisons of the model predictions to charged-current neutrino scattering observables are shown in Ref.~\cite{Megias:2016fjk} and a good agreement with all available neutrino cross-sections is obtained.
\subsection{Impact of the axial form factor on the CCQE-like 
cross-section}
\label{sec:results}
\begin{figure}\vspace{0.08cm}
	\begin{center}\vspace{0.80cm}
		\includegraphics[scale=0.29, angle=270]{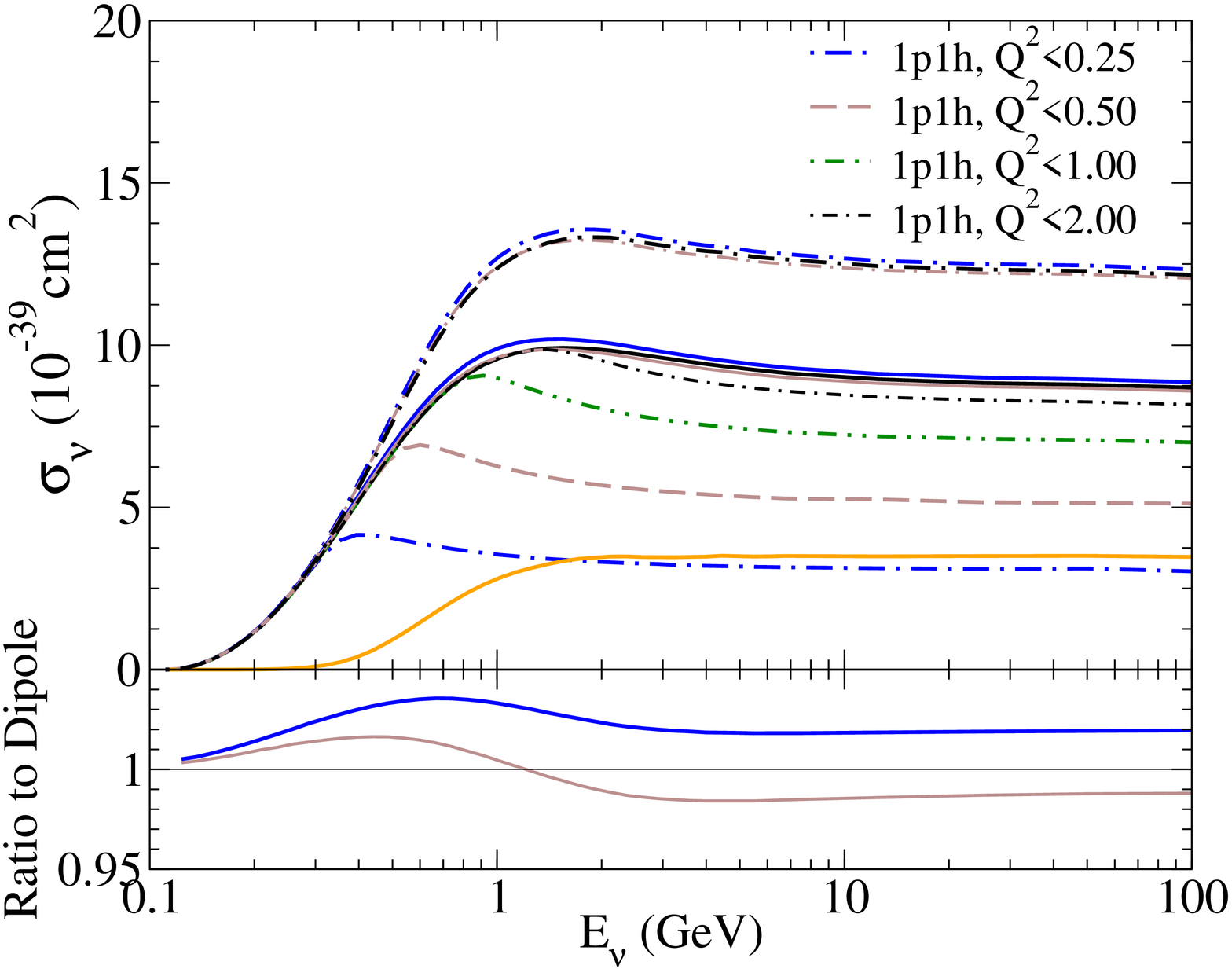}\hspace*{-0.15cm}
		\includegraphics[scale=0.29, angle=270]{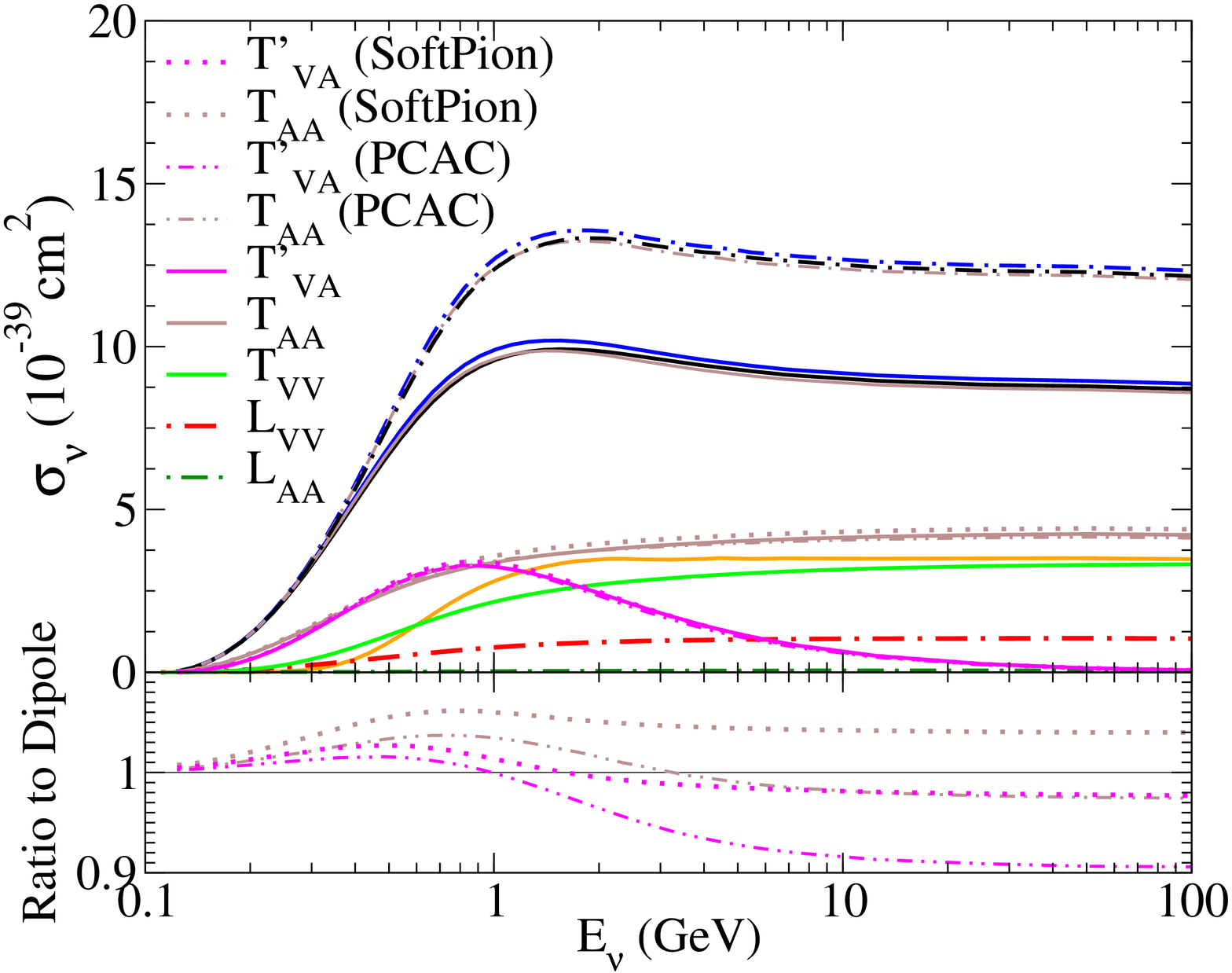}\\
		\includegraphics[scale=0.29, angle=270]{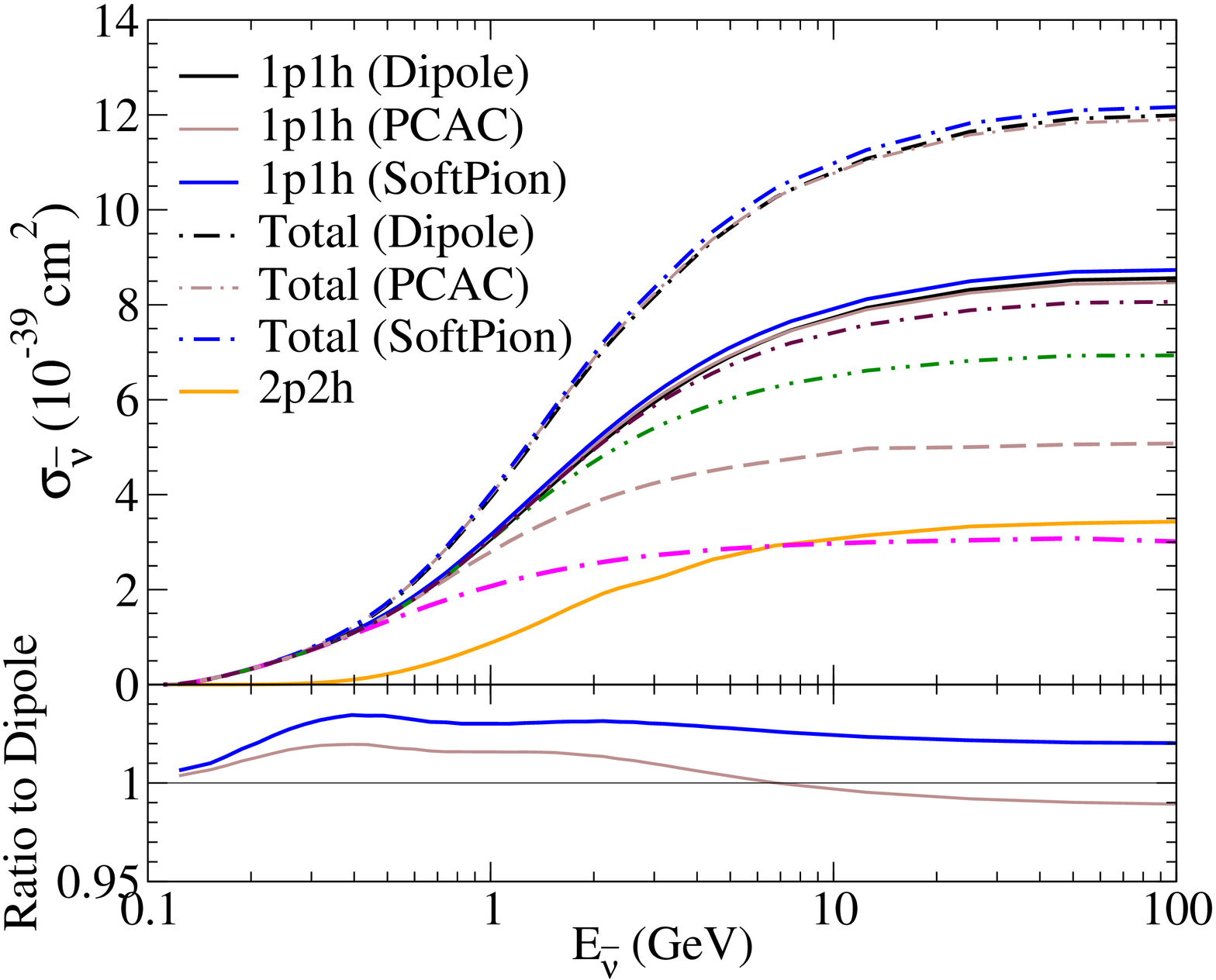}\hspace*{-0.15cm}\includegraphics[scale=0.29, angle=270]{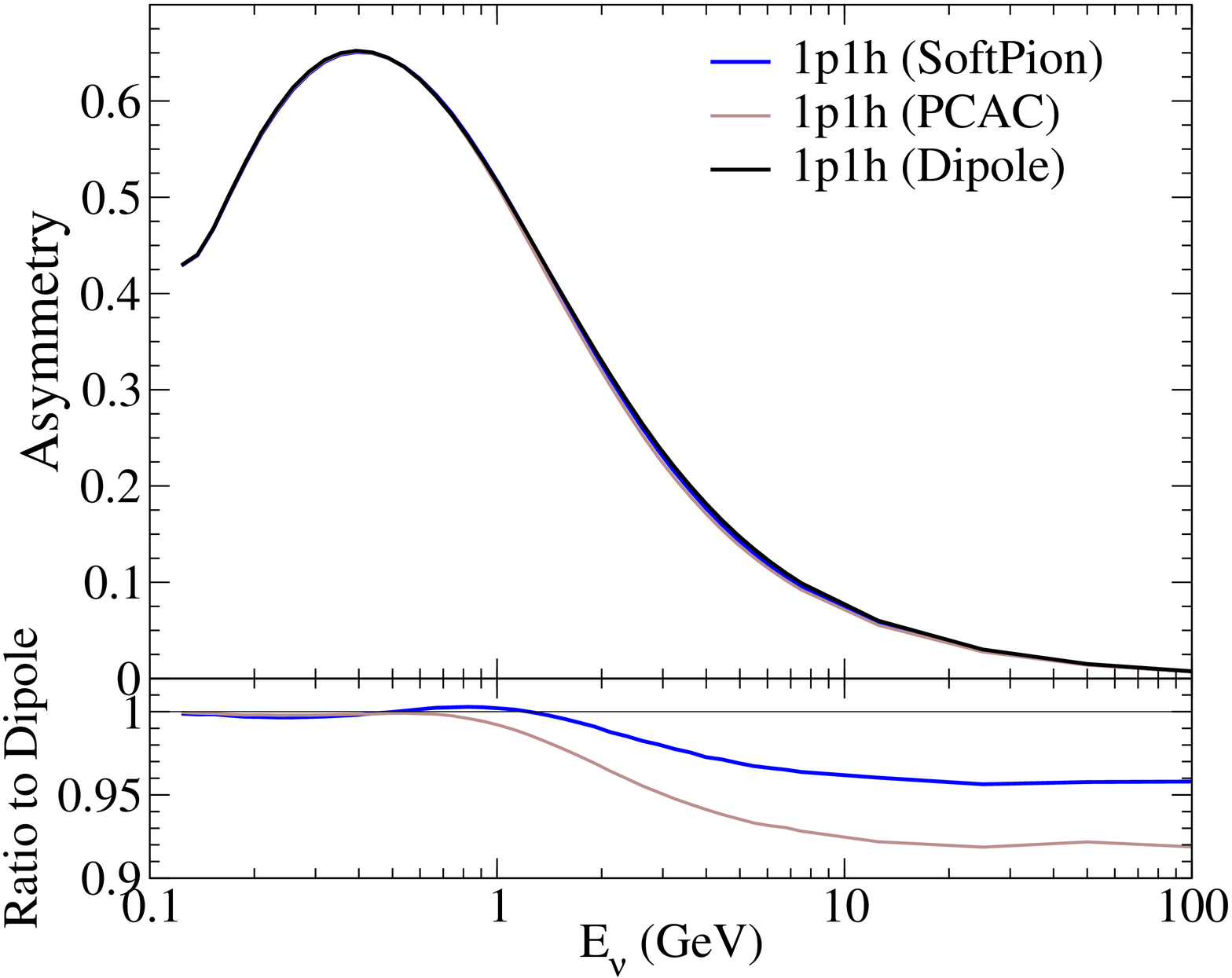}\\
		\begin{center}
			\vspace{-1cm}
		\end{center}
	\end{center}
	\caption{Total CCQE and 2p2h neutrino and antineutrino cross-section on $^{12}$C within the SuSAv2 model for different prescriptions of the axial form factor. The contributions from different $Q^2$ ranges for neutrino (top left) and for antineutrino (bottom left) are shown for the case of the dipole axial form factor. Similar results are found with other axial form factors parameterizations. The neutrino-antineutrino difference is also displayed (bottom right). The ratio between the 1p1h cross-section for alternative form factors over the dipole case is shown in the bottom of each figure. The different axial-vector contributions for neutrino is also shown (top right).  
	\label{fig:total}}
\end{figure}

Figure~\ref{fig:total} illustrates the total CCQE muon-neutrino and antineutrino cross-sections on $^{12}$C and their difference, evaluated in the SuSAv2 model as a function of the (anti)neutrino energy. The cross-sections for different axial form factors are compared and the contribution of different ranges of $Q^2$ is shown. For completeness, we also display the total result when considering 2p2h contributions, where the form factors are the ones used in Ref.~\cite{RuizSimo:2016ikw} and specified in Ref.~\cite{Hernandez:2007qq}. The SoftPion case is always above the dipole, as expected from Fig.~\ref{fig:FFfitNuPion2}, since the region of $Q^2<1.2$~GeV$^2$ dominates the cross-section for all values of $E_\nu$. The PCAC case is above the dipole for $E_\nu<$1~GeV, where the contribution of $Q^2<0.5$~GeV$^2$ dominates the cross-section, and below the dipole for larger $E_\nu$ where larger $Q^2$ values dominate. The differences in the cross-section using the different form factors can reach 4\%. 
As previously observed, the axial form factor only affects the axial-axial and vector-axial contribution to the cross-section. Due to the cancellation between the $T_{AA}$ and the $T'_{VA}$ contributions in antineutrino cross-section, the effect of the different axial form factors is different with respect to neutrino. Notably, above 3~GeV, a region more relevant for MINERvA, the alternative form factors stay below the dipole up to larger energies.
This is due to the fact that in the antineutrino cross-section lower $Q^2$ values are dominant with respect to neutrino cross-section. In practice such differences result in effects on neutrino-antineutrino cross-section difference as large as +4\% at 1~GeV and -8\% (-12\%) at 3~GeV (above 10~GeV).

\begin{figure}\vspace{0.08cm}
	\begin{center}\vspace{-0.10cm}
        	\includegraphics[scale=0.35, angle=0]{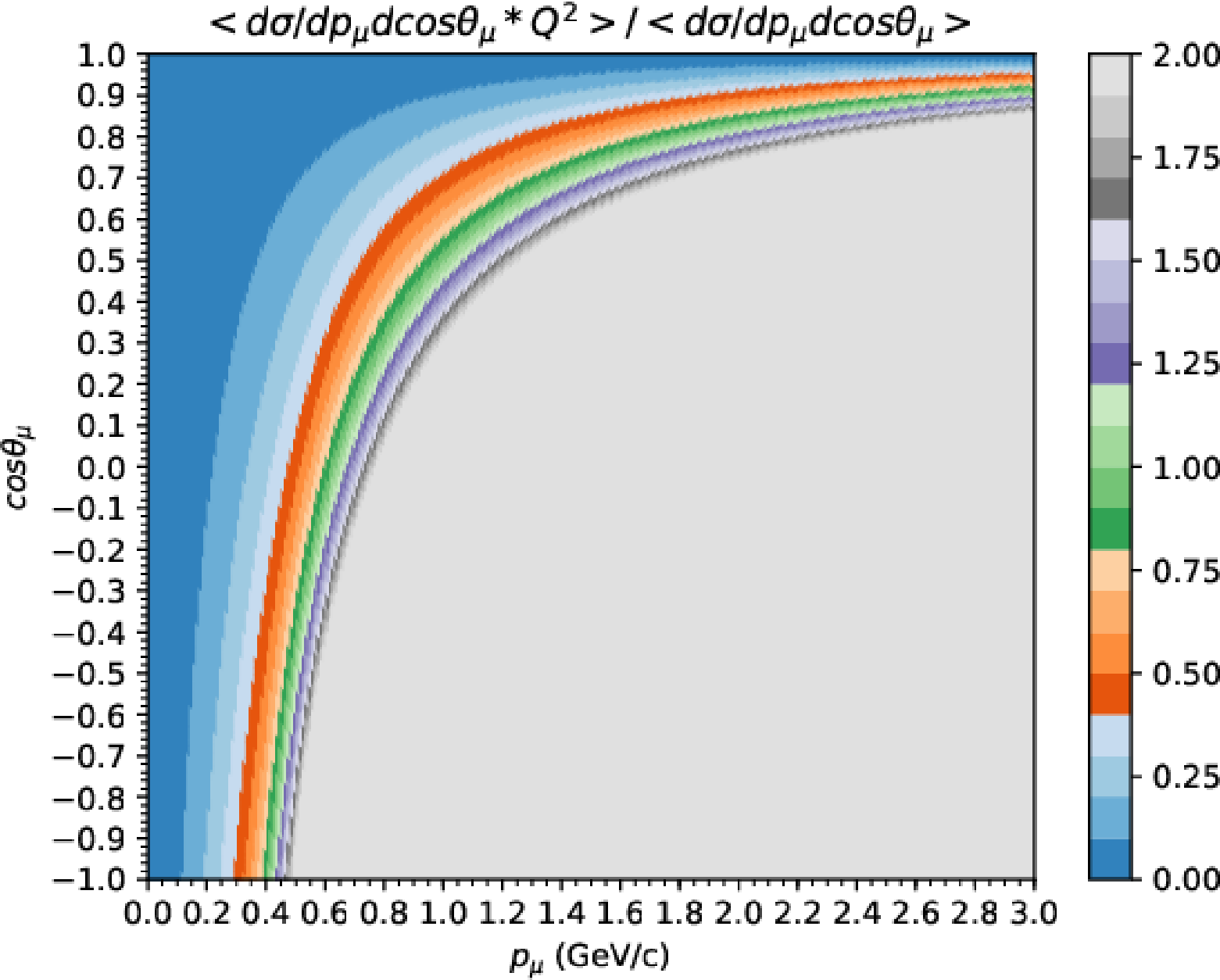}\hspace{0.195cm}
		\includegraphics[scale=0.35, angle=0]{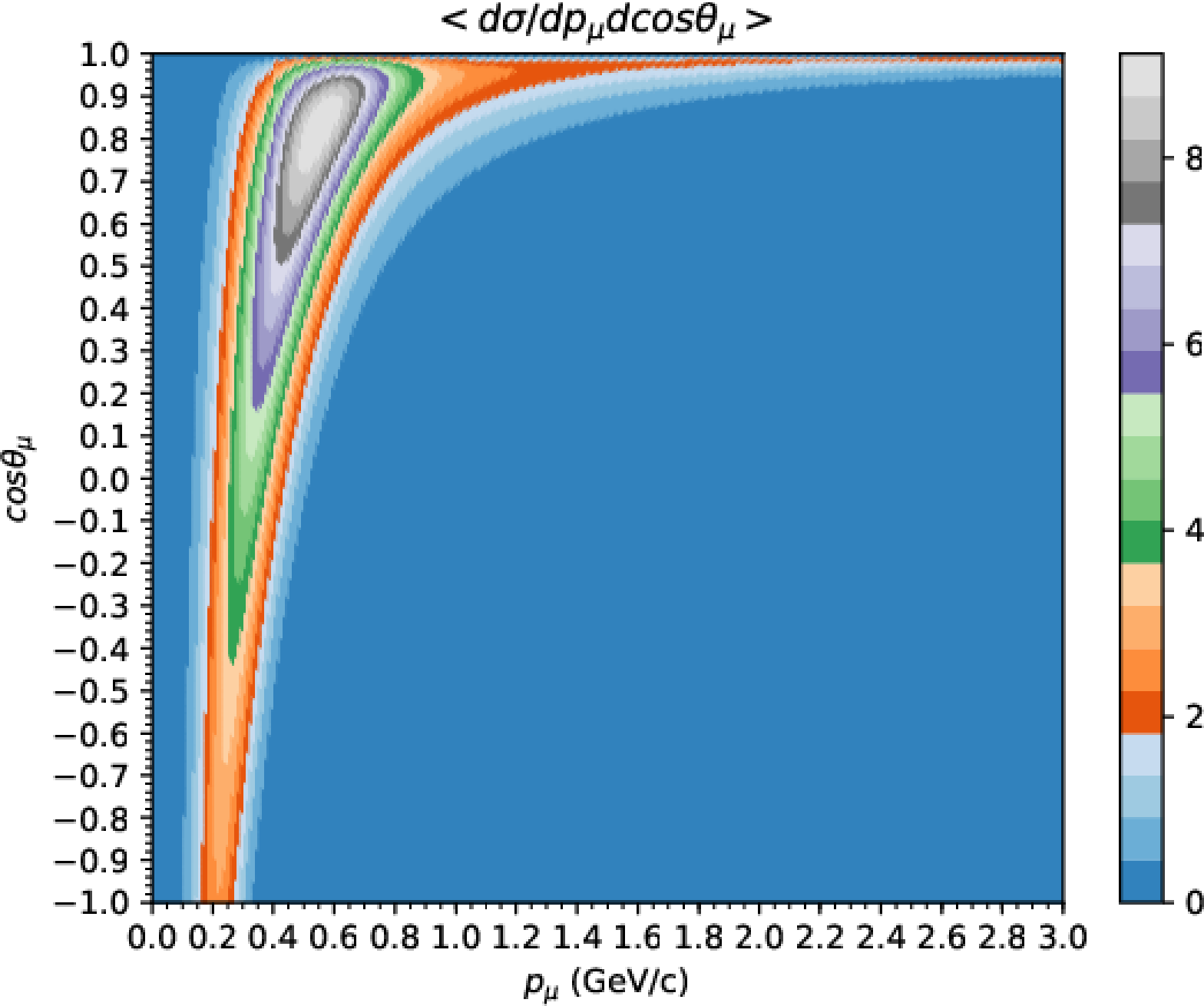}\\\vspace*{0.35cm}		
        \includegraphics[scale=0.35, angle=0]{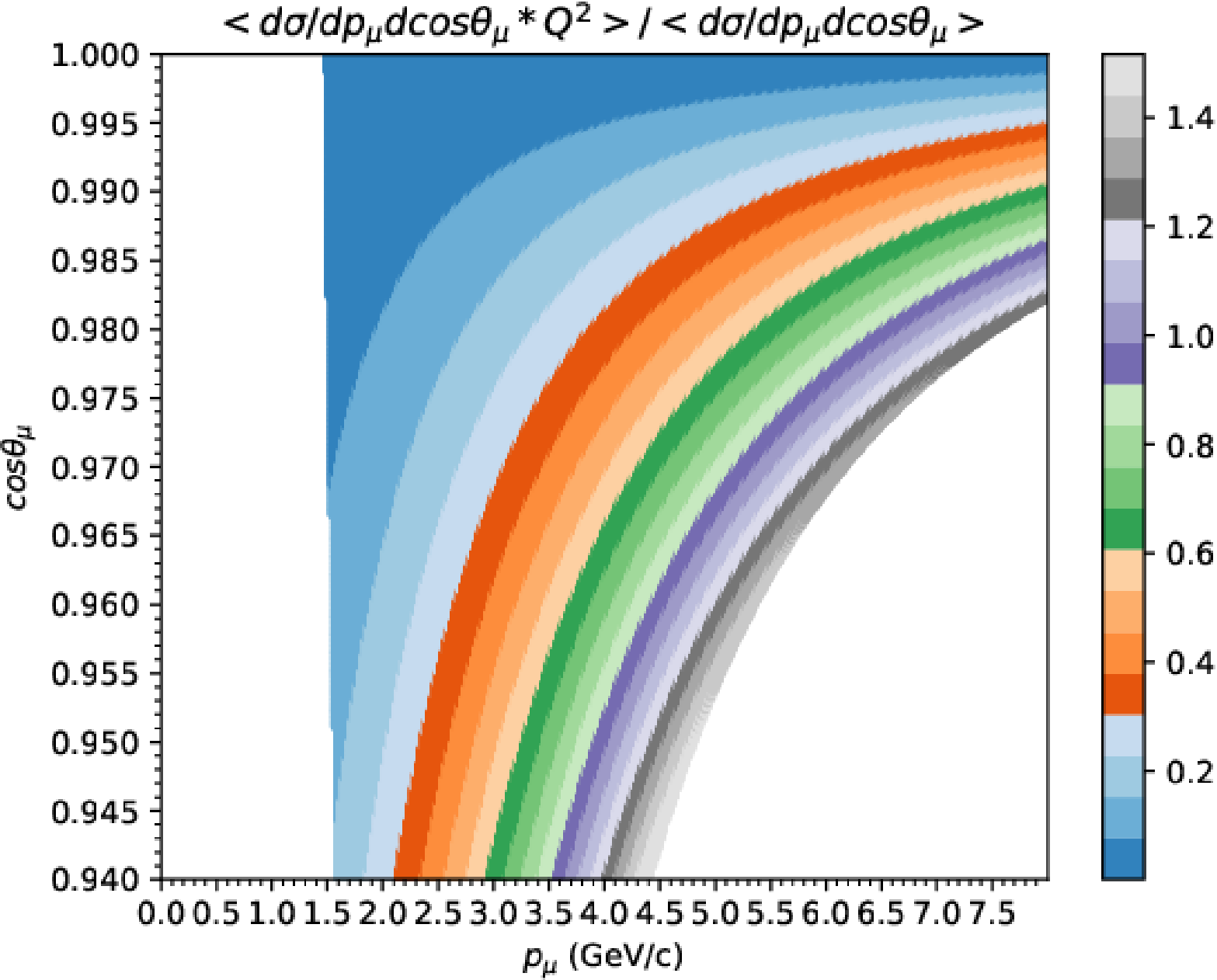}\hspace{0.195cm}
		\includegraphics[scale=0.35
		, angle=0]{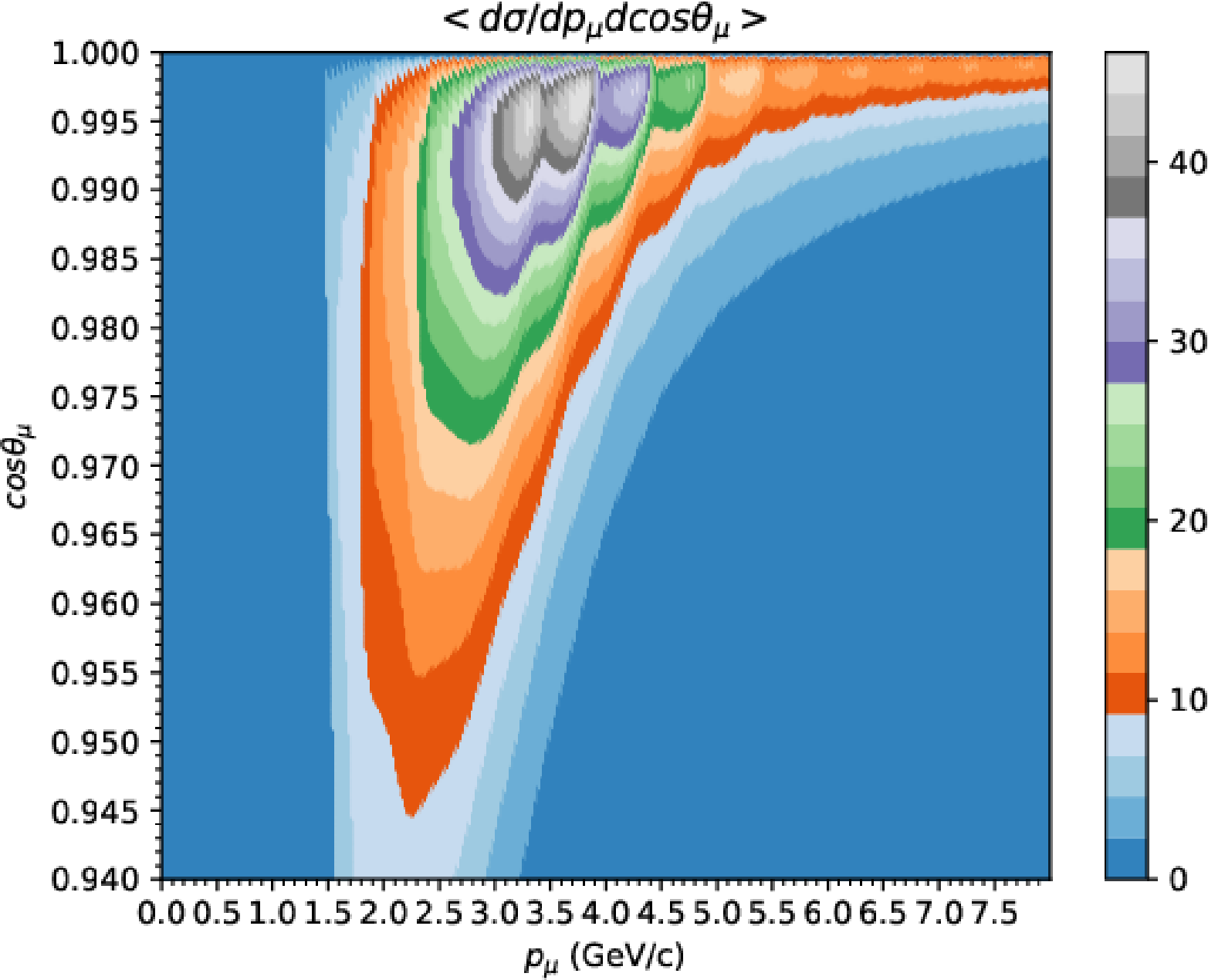}\\\vspace*{0.35cm}
        \includegraphics[scale=0.35, angle=0]{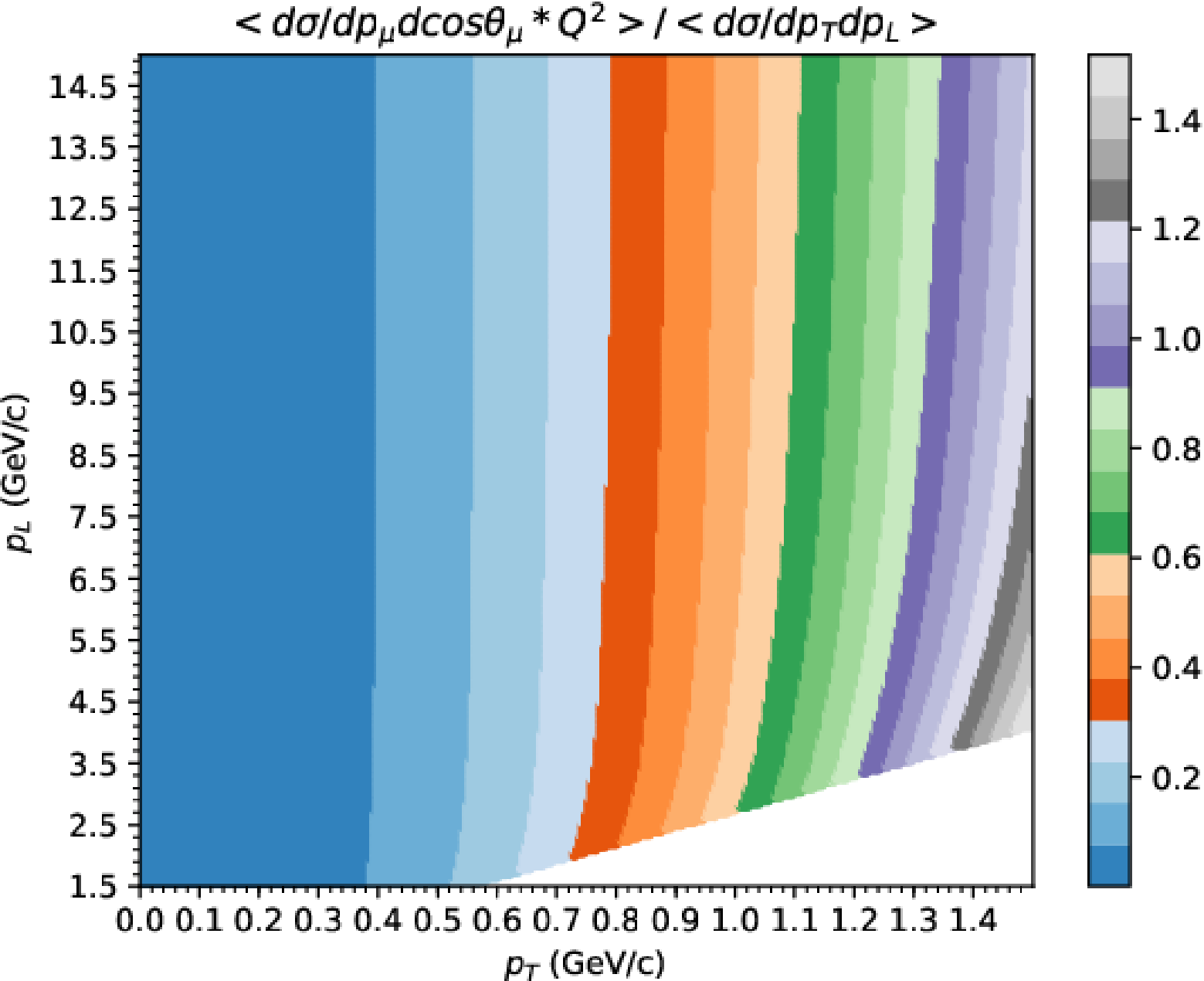}\hspace{0.195cm}
		\includegraphics[scale=0.35, angle=0]{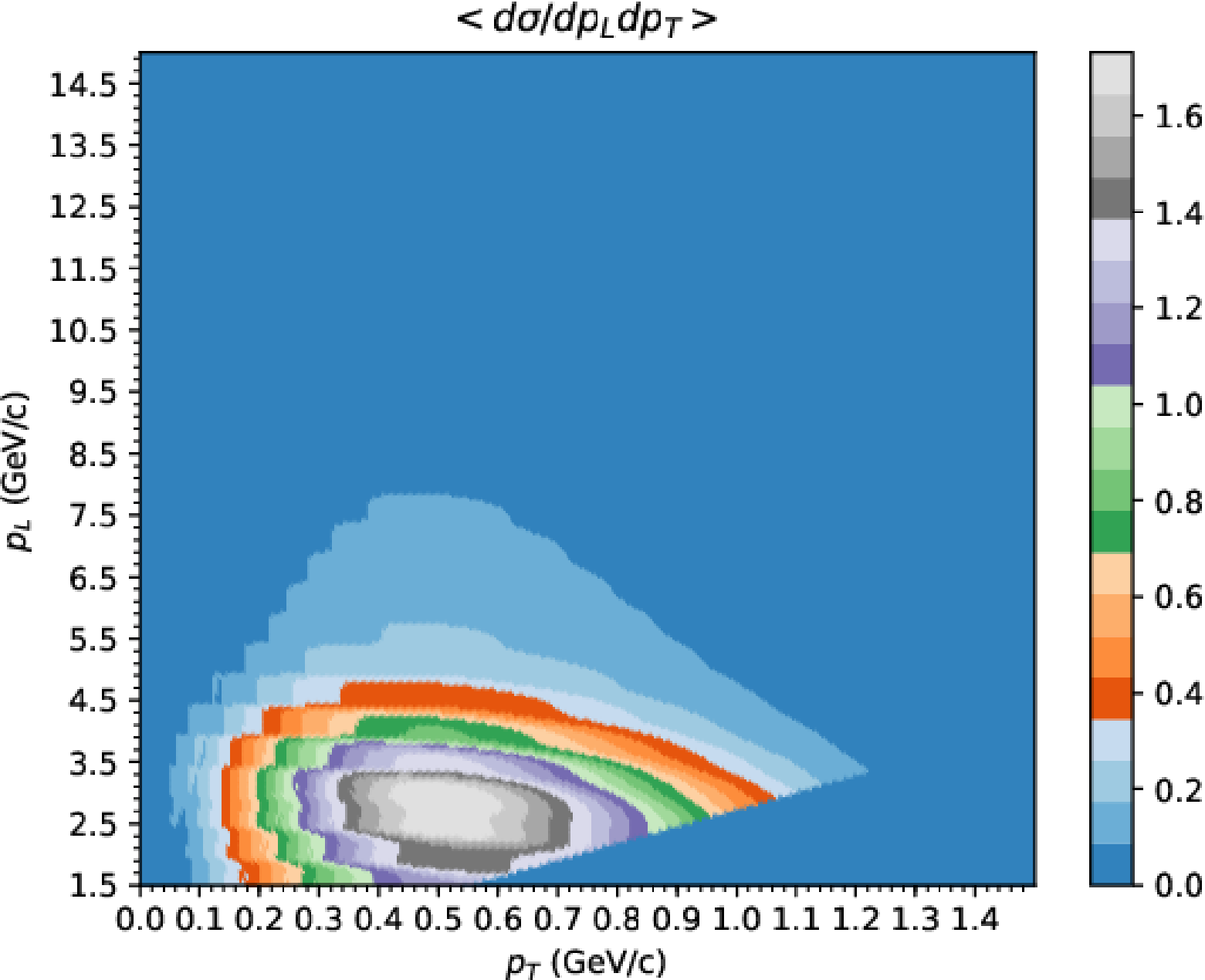}\vspace*{-0.25cm}
	\end{center}
	\caption{Left panels: flux-averaged $Q^2$ distribution as a function of muon kinematics in units of GeV$^2$ per nucleon. Right panels: double differential flux-averaged cross-section as a function of muon kinematics in units of 10$^{-39}$ cm$^2$/GeV per nucleon for the upper and mid panels and 10$^{-39}$ cm$^2$/GeV$^2$ per nucleon for the lower panel. Top panels: T2K kinematics. Second and third row: MINERvA kinematics with limited acceptance in $\cos\theta_\mu$, $p_L$ and $p_T$. 
	}\label{fig:weighted}
\end{figure}

The impact of the axial form factor on the neutrino-nucleus cross-section is clearly driven by the $Q^2$ dependence of the cross-section. A more detailed assess of such impact can be performed by studying the $Q^2$ ranges as a function of the muon kinematics.
The relevant $Q^2$ values for T2K~\cite{Abe:2011ks} and MINERvA~\cite{Aliaga:2013uqz} are presented in Fig.~\ref{fig:weighted} as function of the muon kinematics and compared to the flux-averaged double differential cross-section (CCQE-like $\nu_\mu$-$^{12}$C). The figure shows the flux-averaged $Q^2$ value defined as
\begin{equation}
    \frac{\int_{E_{min}(p_\mu,\cos\theta_\mu)}^{E_{max}(p_\mu,\cos\theta_\mu)} Q^2(E_\nu,p_\mu,\cos\theta_\mu) \frac{d\sigma}{dE_\nu} f(E_\nu) dE_\nu}{\int \frac{d\sigma}{dE_\nu}f(E_\nu) dE_\nu},
\end{equation}
where $f$ is the neutrino flux. In T2K, the $Q^2$ at the maximum of the cross-section ($\cos\theta_\mu$:0.7-0.9, $p_\mu$: 0.4-0.7~GeV) lies around 0.1-0.2 GeV$^2$ with tail to larger values up to 0.8~GeV$^2$ in the backward angle region. In MINERvA the $Q^2$ at the maximum of the cross-section ($\cos\theta_\mu >$0.99, $p_\mu$: 3-4~GeV) lies around 0.2-0.3 GeV$^2$ with tail to larger values up to 0.8~GeV$^2$ in the region of $\cos\theta_\mu \approx$0.95. The differences in kinematics between T2K and MINERvA are simply due to the different energy of the neutrino flux. The MINERvA results include cuts on the detector acceptance as used in Ref.~\cite{PhysRevD.97.052002} and they are shown also as a function transverse and longitudinal muon momentum ($p_T, p_L$). These latter variables have a more direct mapping into $Q^2$: a given bin of $p_T$ corresponds to a limited range of $Q^2$.

The SuSAv2 model, with different axial form factors, are compared to the T2K and MINERvA data in the next Subsections. It is important to remind that the so-called CC$0\pi$ data, where no pions are observed in the final state, also contain possible pion-absorption effects, which are small at T2K kinematics ($<15\%$)~\cite{Dolan:2019bxf} and larger for MINERvA ($\gtrsim20\%$)~\cite{Megias:2018ujz}, as well as the contribution of two-particle-two-hole (2p2h) excitations, corresponding to the coupling of the probe to a two-body meson-exchange current, {\it i.e.} to a pair of correlated nucleons. 
The contribution of 2p2h excitations within the SuSAv2 model has been widely explored in past work~\cite{Amaro:2010sd,Amaro:2011aa,Megias:2014qva,Simo:2016ikv,RuizSimo:2016ikw,Megias:2016fjk,Megias:2017cuh}
and the results have been successfully compared with all existing electron- and (anti)neutrino scattering data on carbon and oxygen. The results are in qualitative agreement with those of other microscopic calculations~\cite{Martini:2009uj,Martini:2010ex,Nieves:2011pp,Nieves:2011yp}, although at a quantitative level some differences emerge. 
Although 2p2h are not the focus of the present study,  these reaction mechanisms are also added in the following plots to show how  they modify the pure QE cross-section at T2K kinematics: their contribution is peaked at lower $p_\mu$ than the QE one and tends to increase the cross-section by about 15\%, yielding better agreement with the data. Further  details on the model and additional comparison with data can be found in the above mentioned references.

\subsubsection{Effects of the axial form factor at T2K kinematics}
\label{sec:T2K}

\begin{figure}\vspace{0.08cm}
	\begin{center}\vspace{0.80cm}
		\includegraphics[scale=0.23, angle=270]{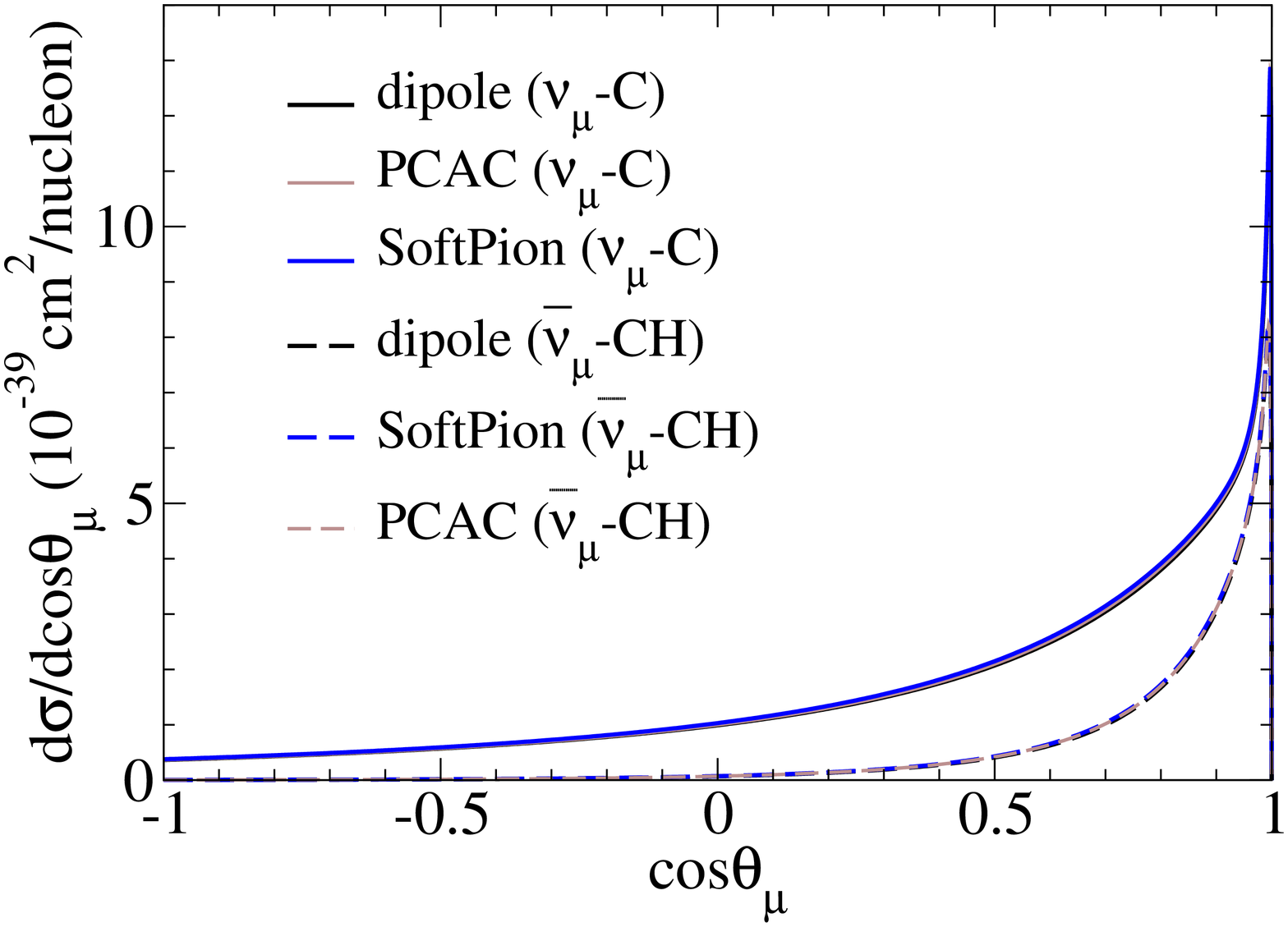}\hspace*{-0.15cm}%
		\includegraphics[scale=0.23, angle=270]{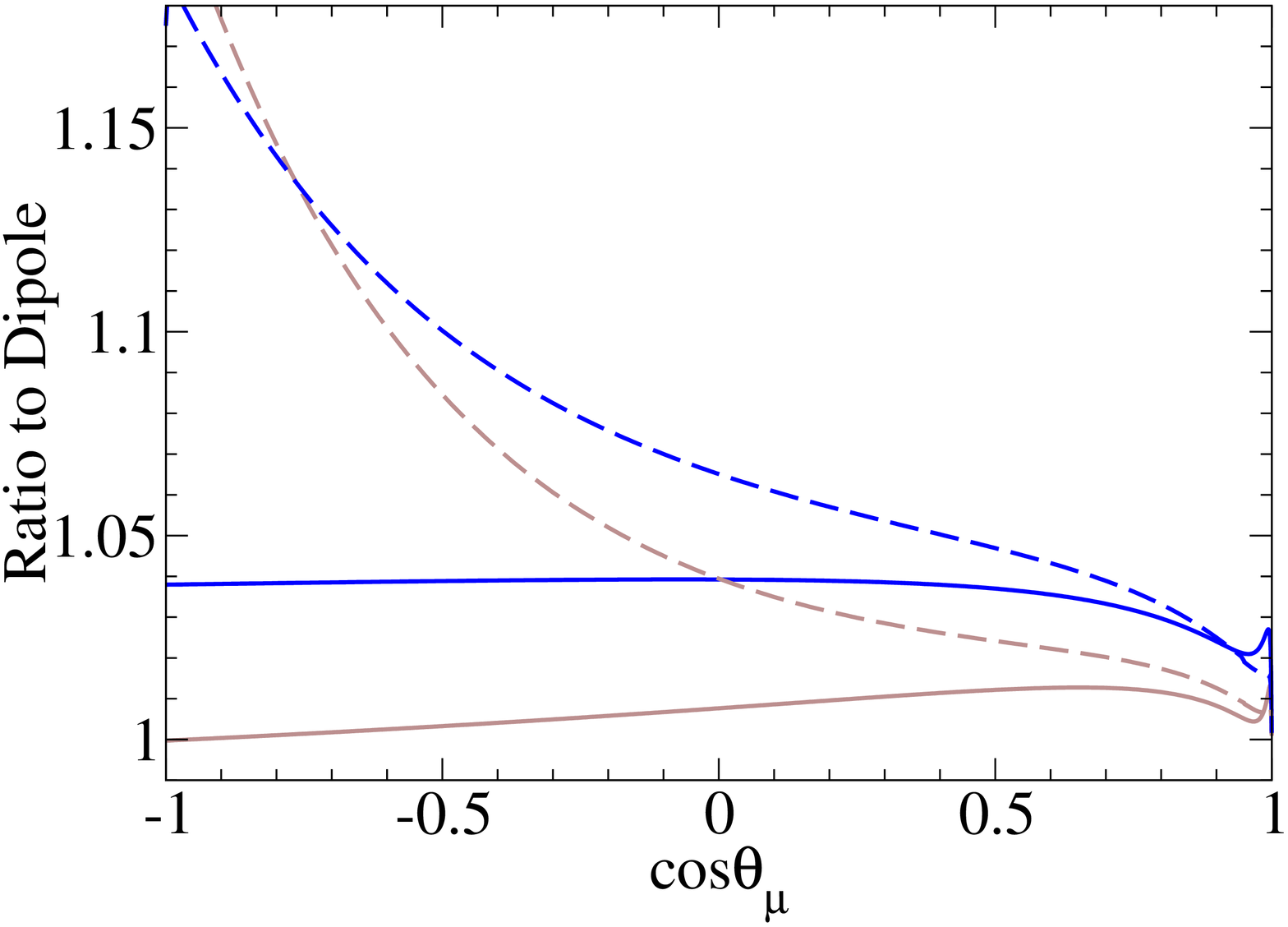}\\\vspace*{-0.285cm}%
				\includegraphics[scale=0.23, angle=270]{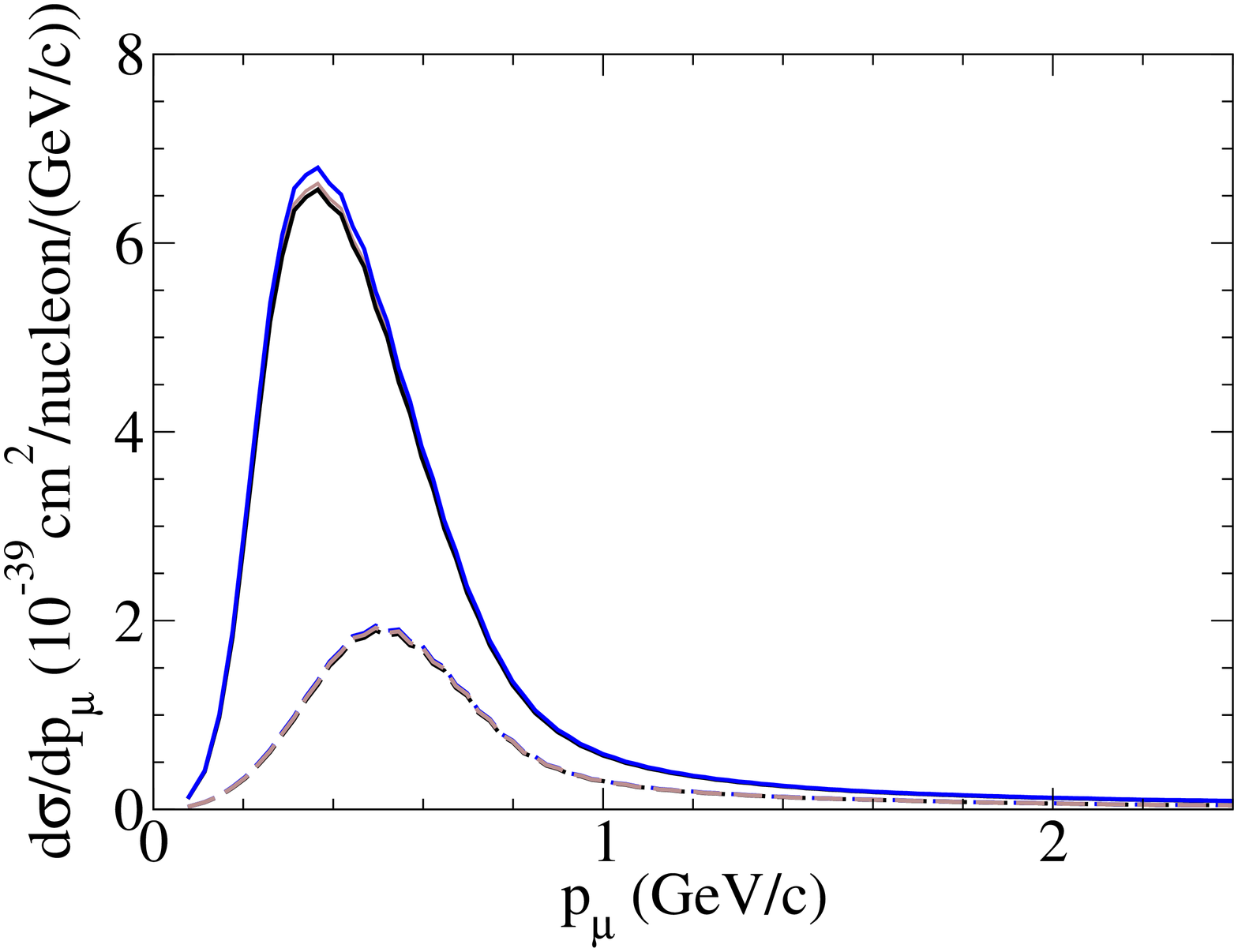}\hspace*{-0.15cm}%
		\includegraphics[scale=0.23, angle=270]{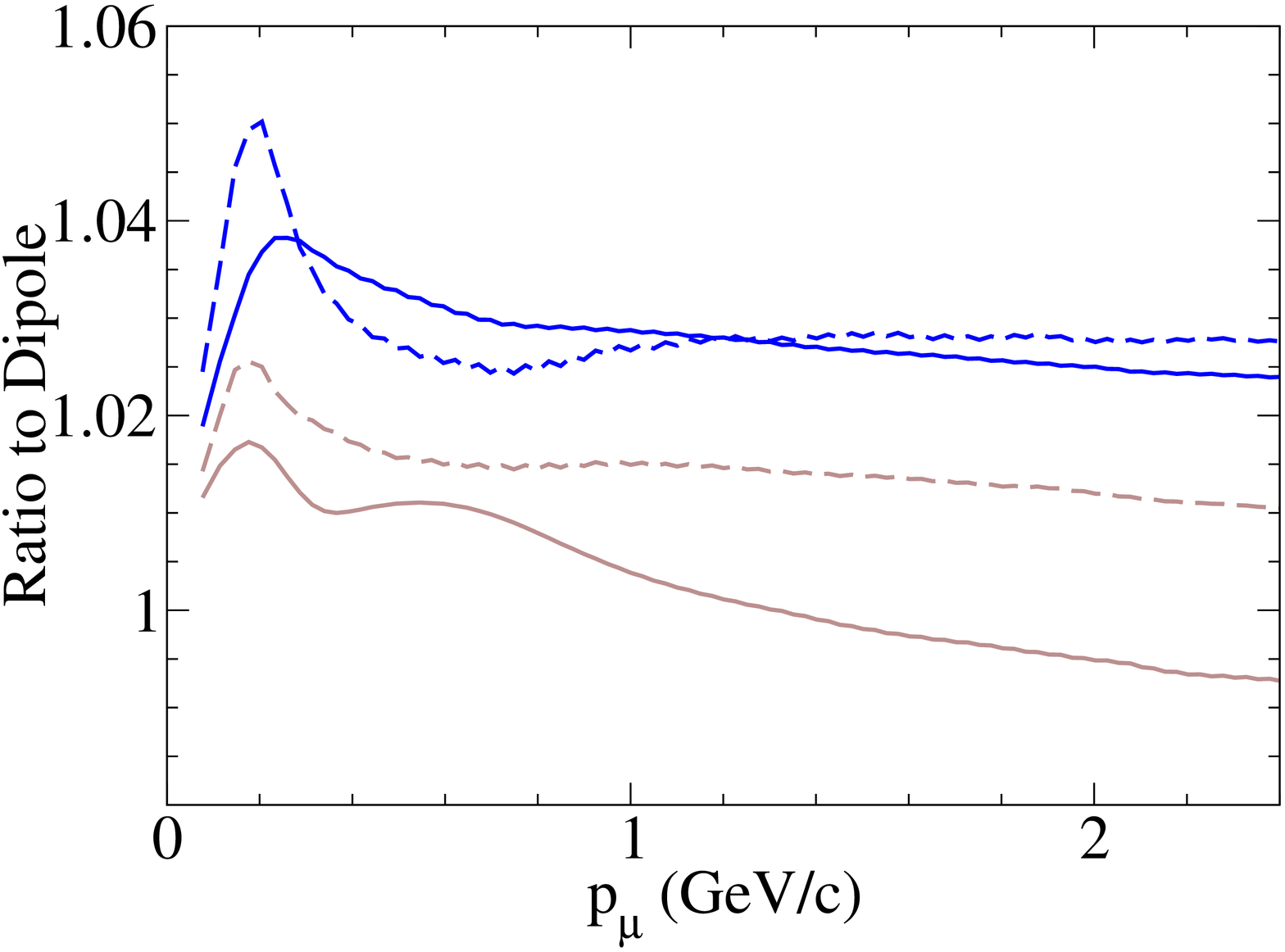}\\	
	\end{center}
	\caption{Single differential cross-section (left) for neutrino and antineutrino as a function of muon angle (first row) and muon momentum (second row) for different nucleon form factors at T2K kinematics. 
	The ratio of the different form factors with respect to dipole is also shown (right). 
	\label{fig:T2K_singlediff}}
\end{figure}

In Fig.~\ref{fig:T2K_singlediff} the impact of the different parametrizations of the axial form factor on the single-differential cross-section as a function of muon angle or momentum is shown. The effects can be quite large and different between neutrino and antineutrino, notably in the backward region of very low cross-section. 

\begin{figure}\vspace{0.08cm}
	\begin{center}\vspace{0.80cm}
			\hspace*{-0.95cm}\includegraphics[scale=0.192, angle=270]{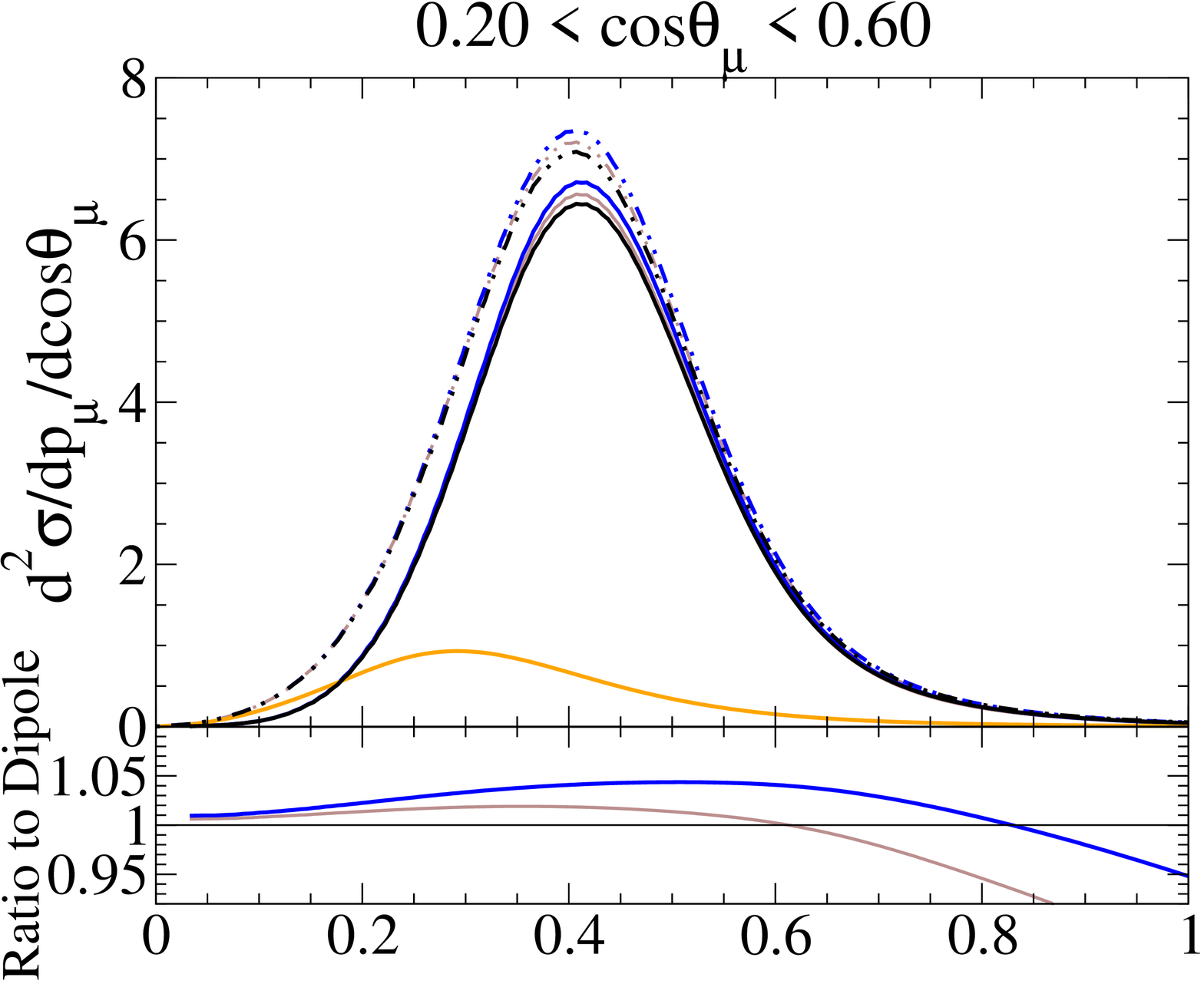}\hspace*{-0.295cm}%
		\includegraphics[scale=0.192, angle=270]{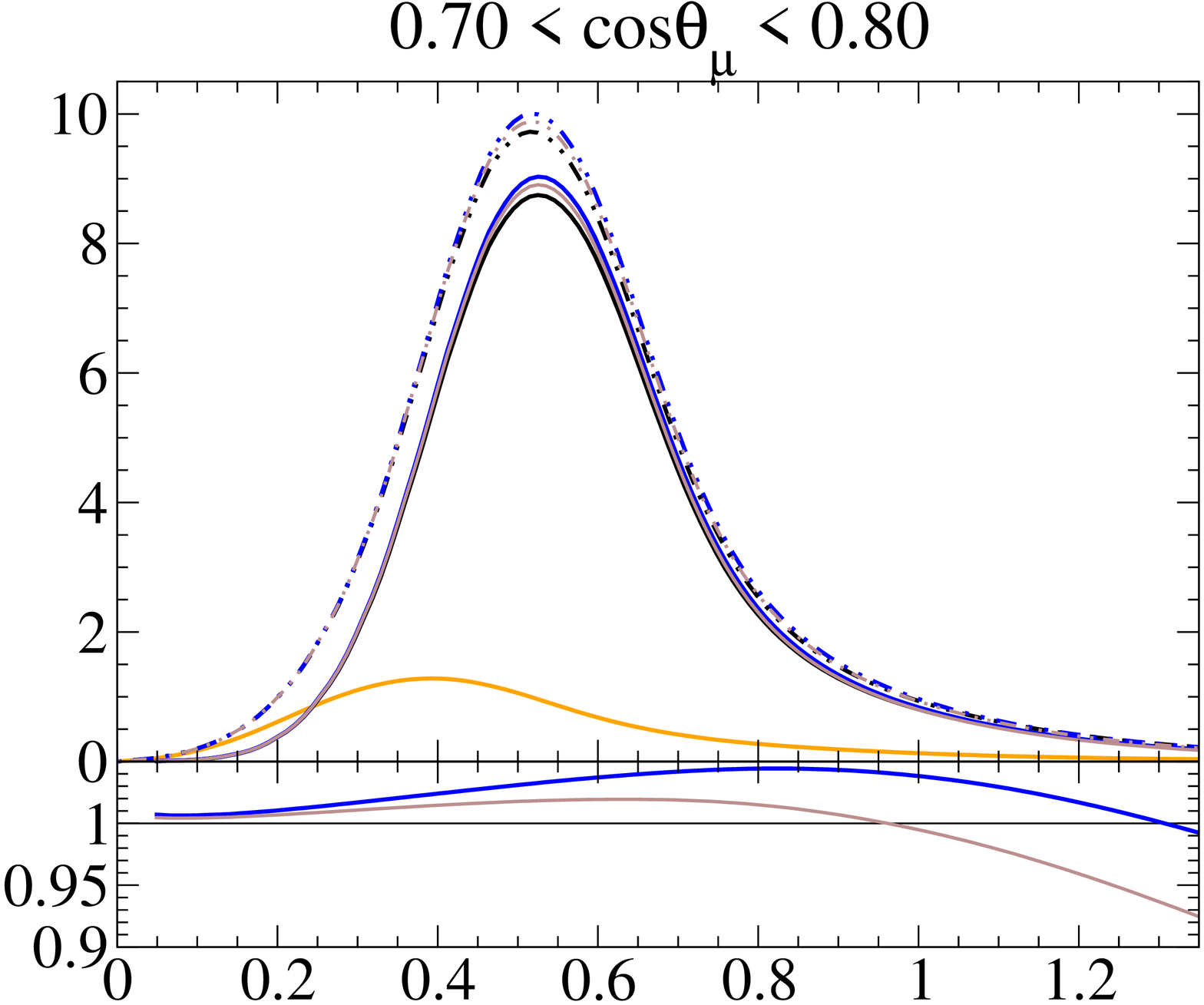}\hspace*{-0.495cm}%
		\includegraphics[scale=0.192, angle=270]{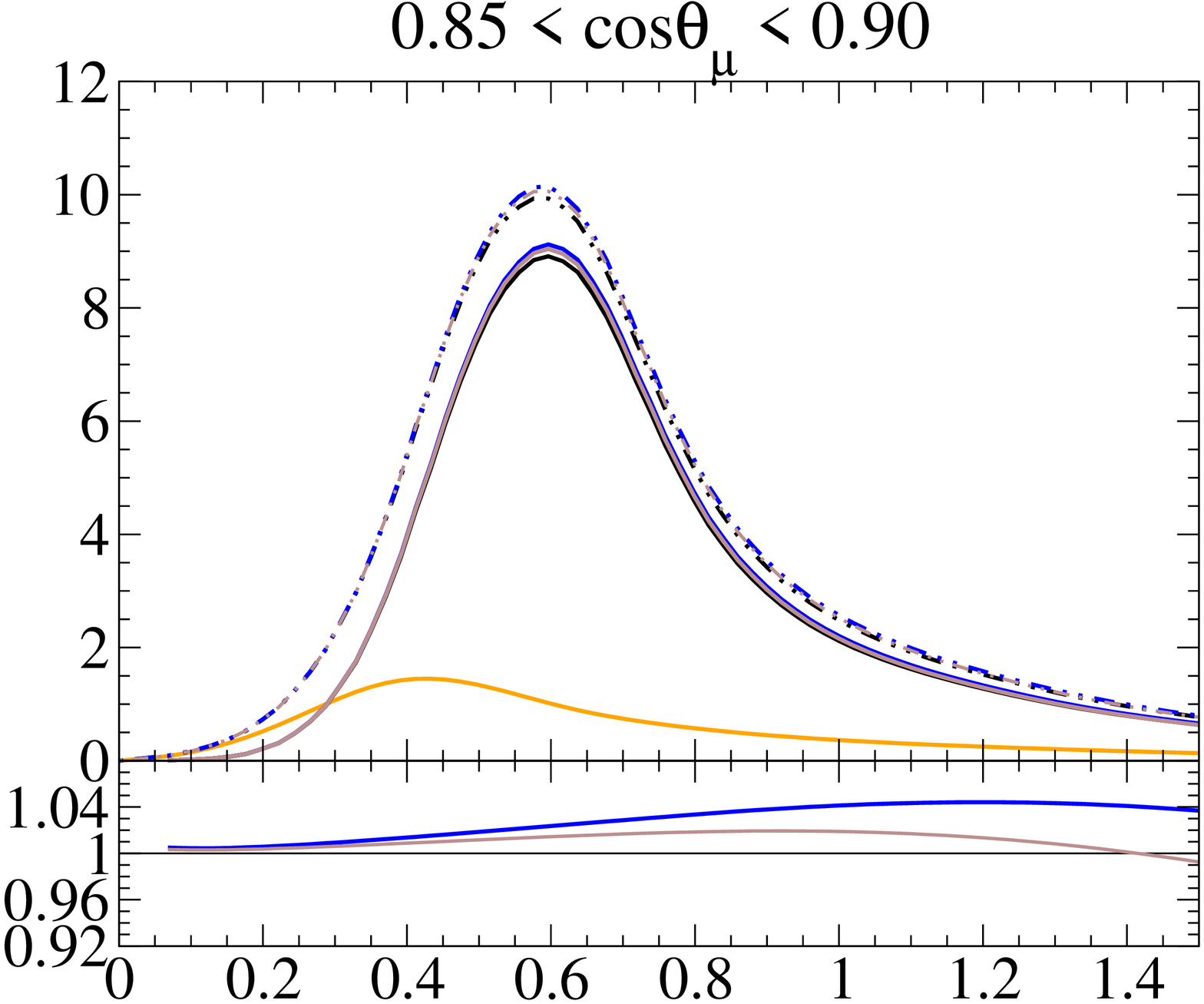}\hspace*{-0.495cm}%
		\includegraphics[scale=0.192, angle=270]{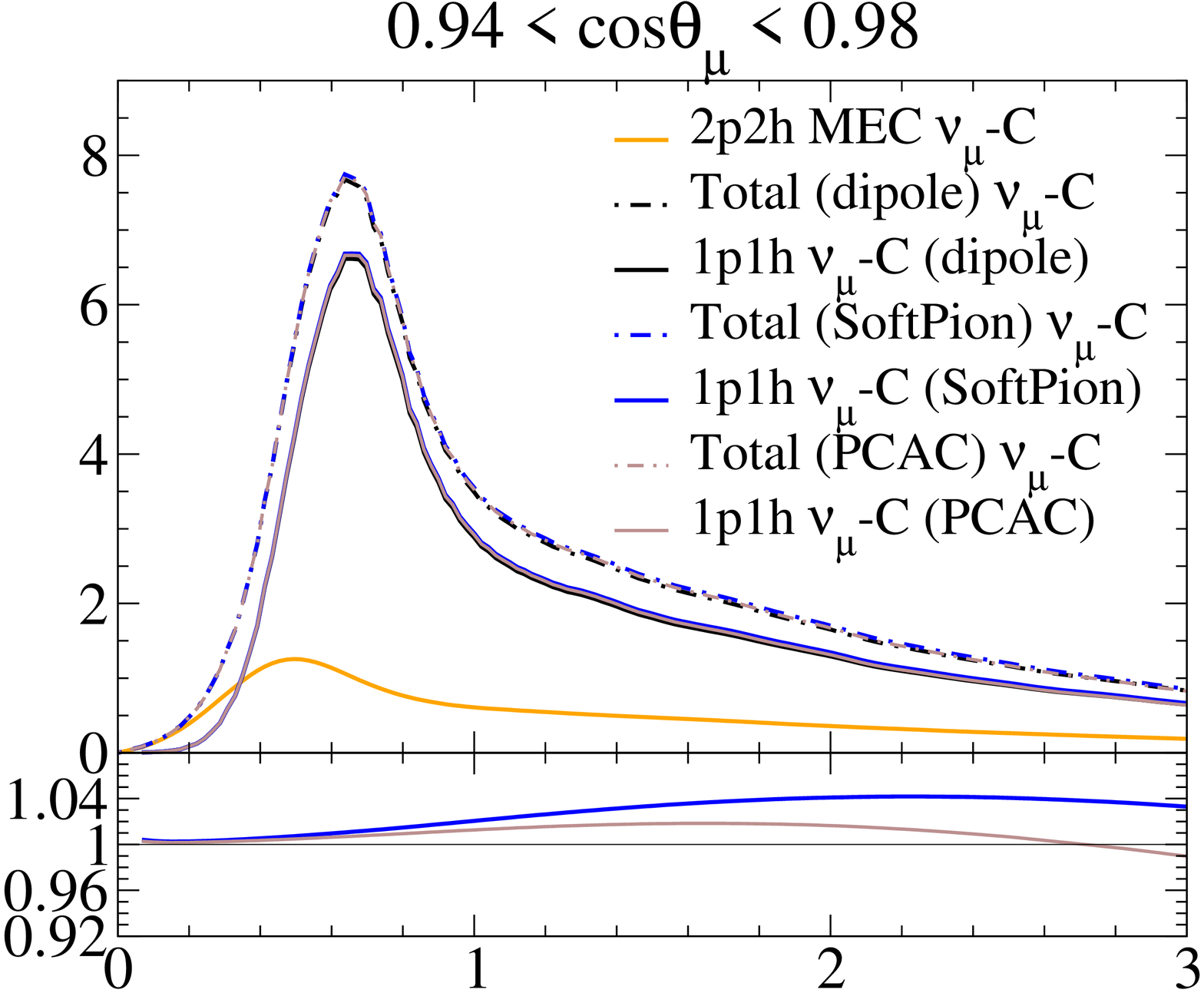}
		\\
		\hspace*{-0.95cm}\includegraphics[scale=0.192, angle=270]{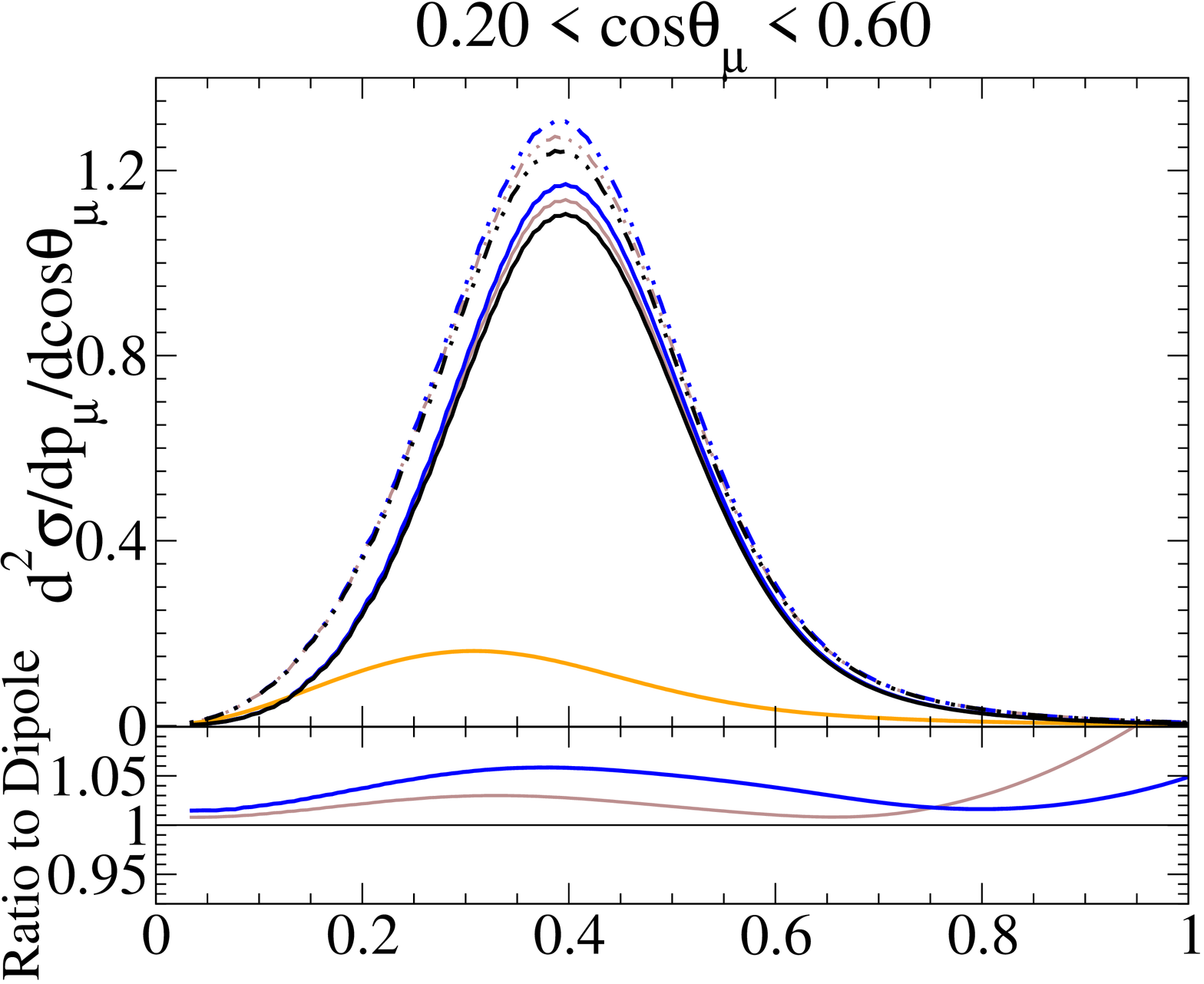}\hspace*{-0.295cm}%
		\includegraphics[scale=0.192, angle=270]{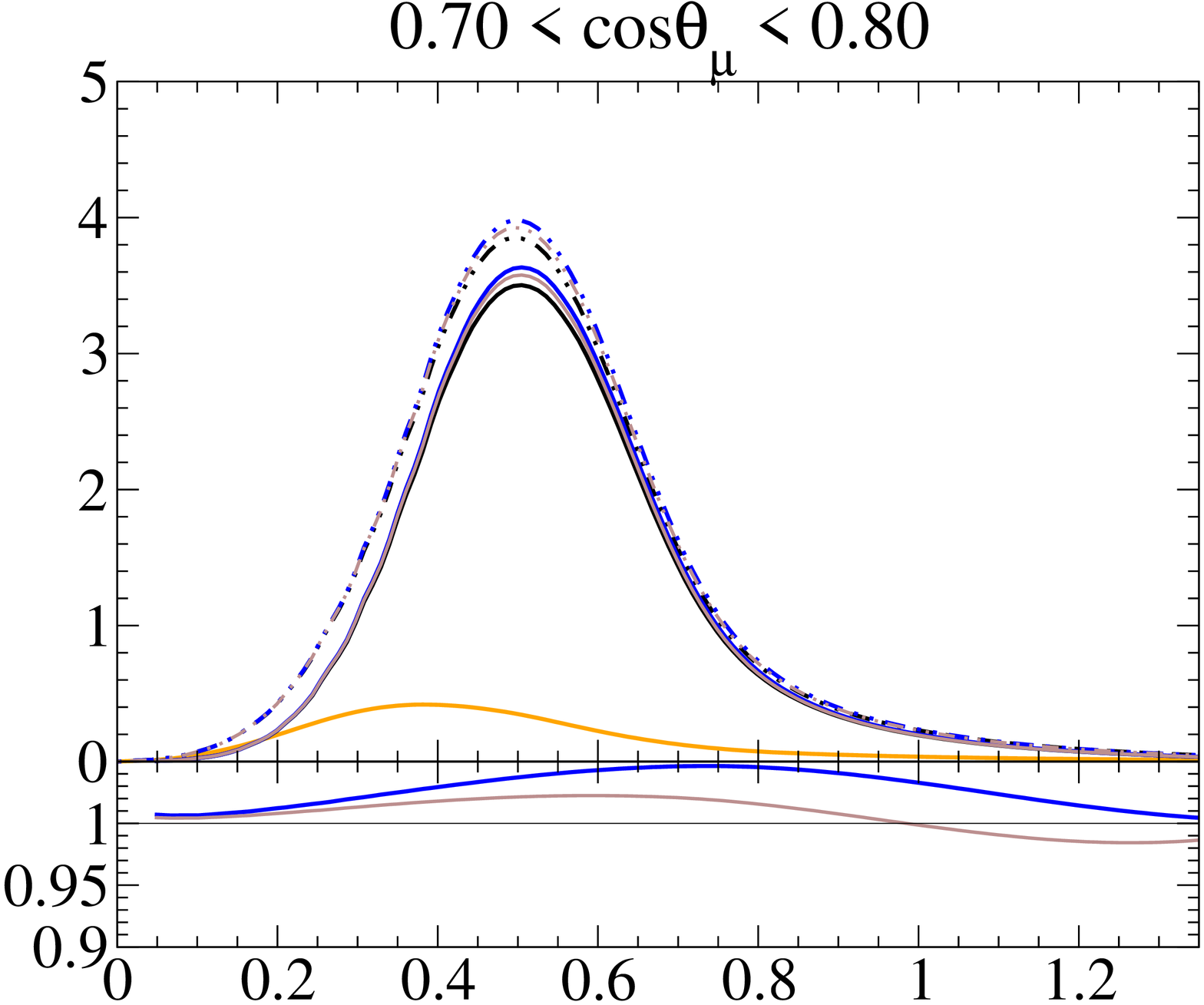}\hspace*{-0.495cm}%
		\includegraphics[scale=0.192, angle=270]{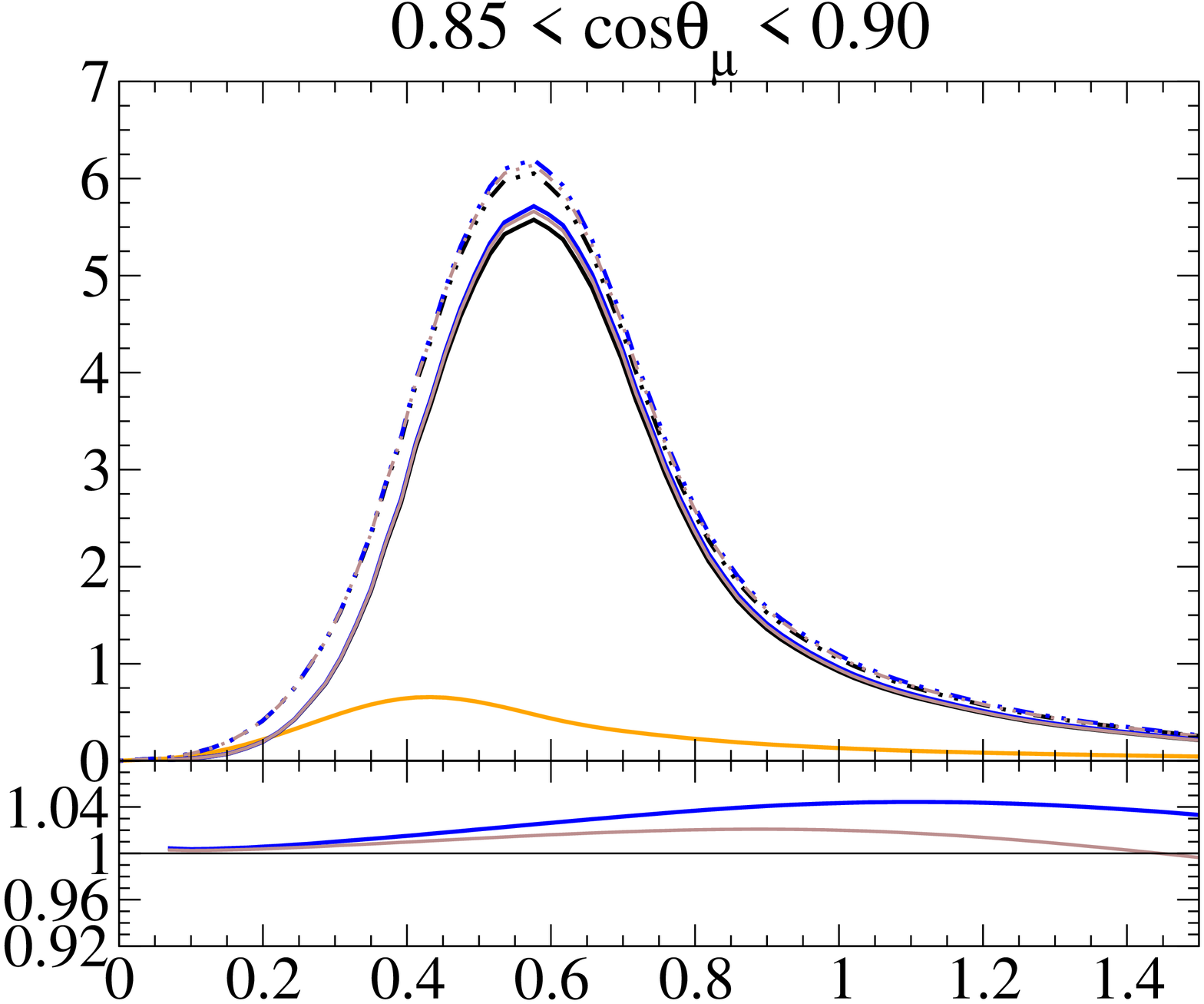}\hspace*{-0.495cm}%
		\includegraphics[scale=0.192, angle=270]{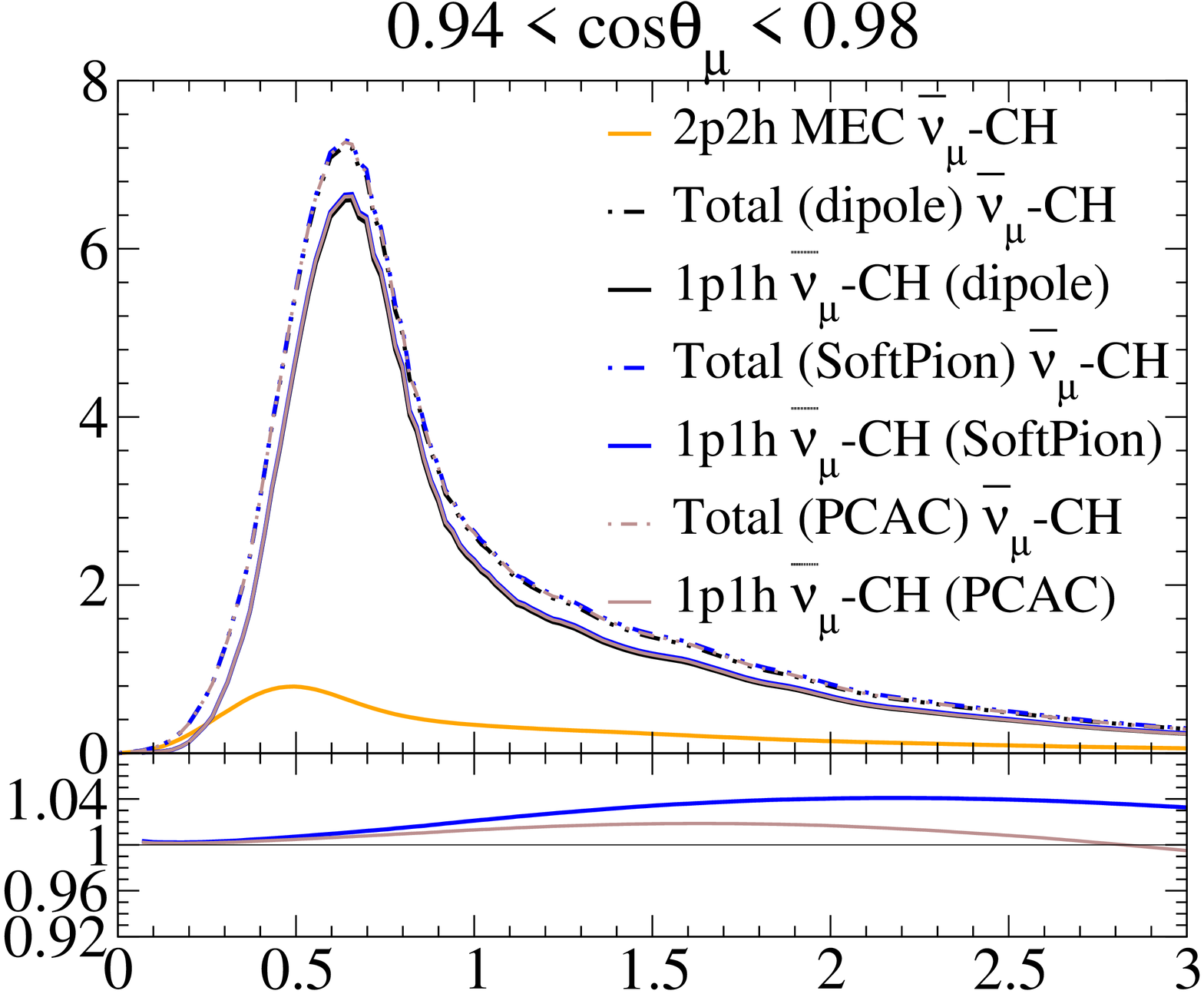}		\\
		\hspace*{-0.95cm}\includegraphics[scale=0.192, angle=270]{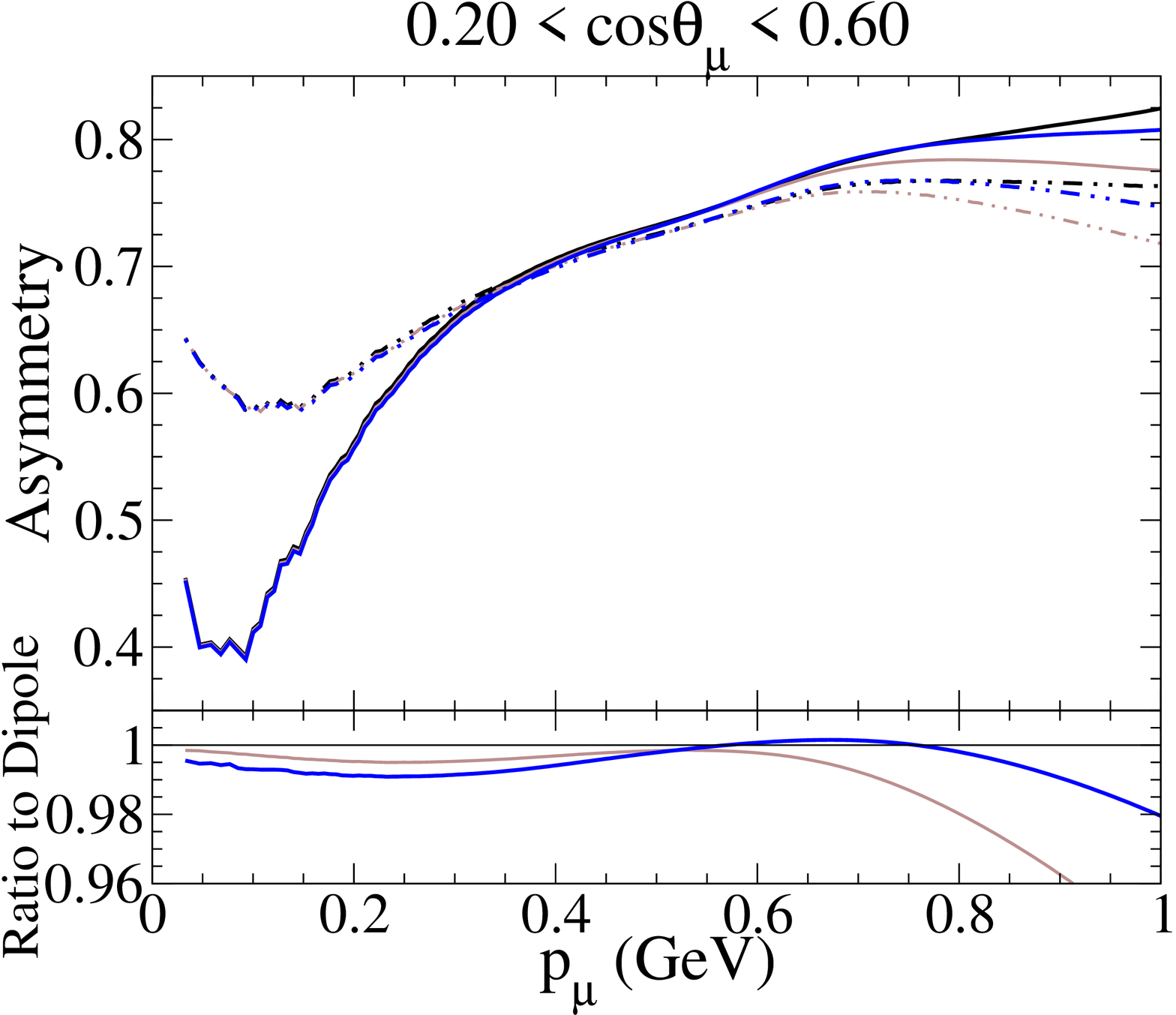}\hspace*{-0.295cm}%
		\includegraphics[scale=0.192, angle=270]{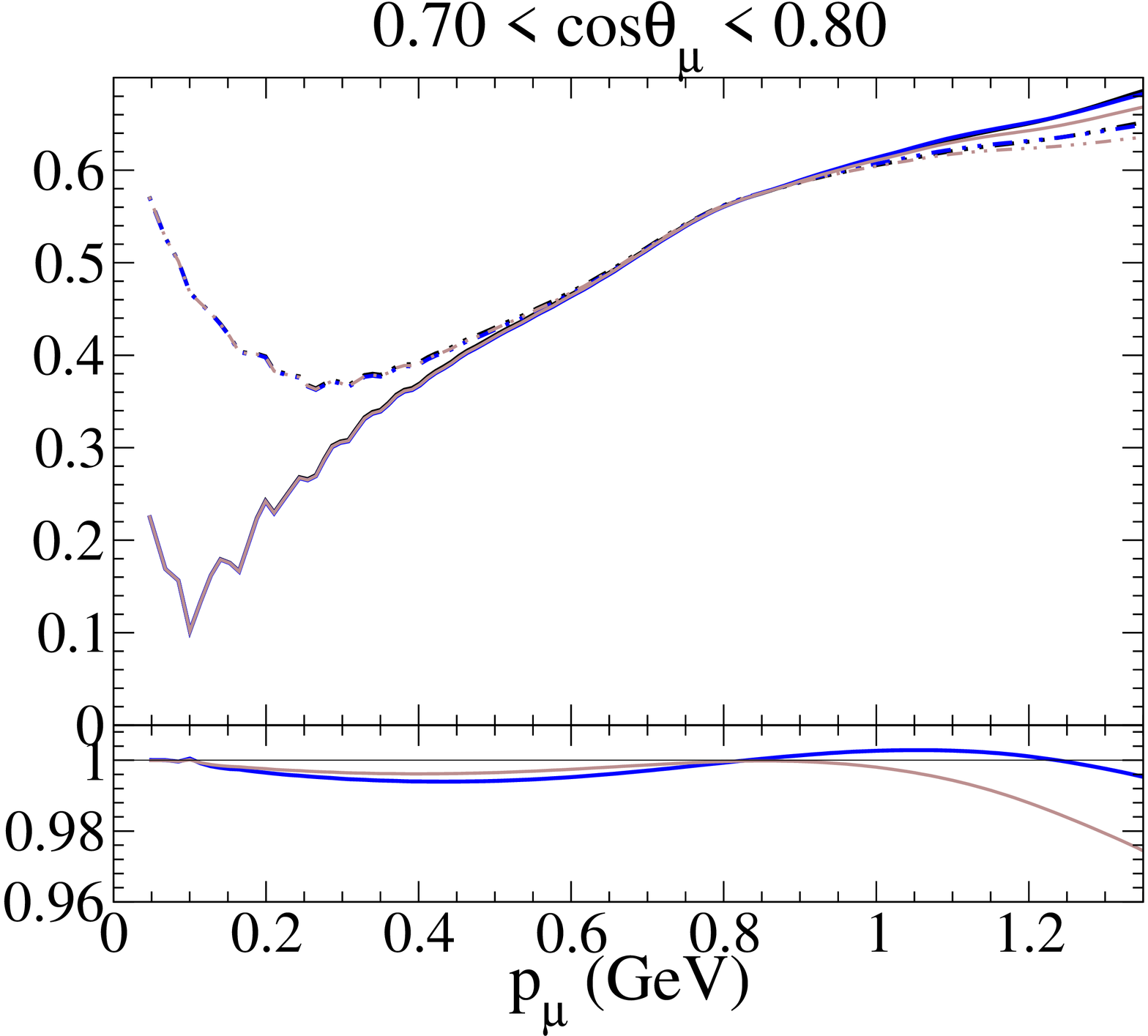}\hspace*{-0.495cm}%
		\includegraphics[scale=0.192, angle=270]{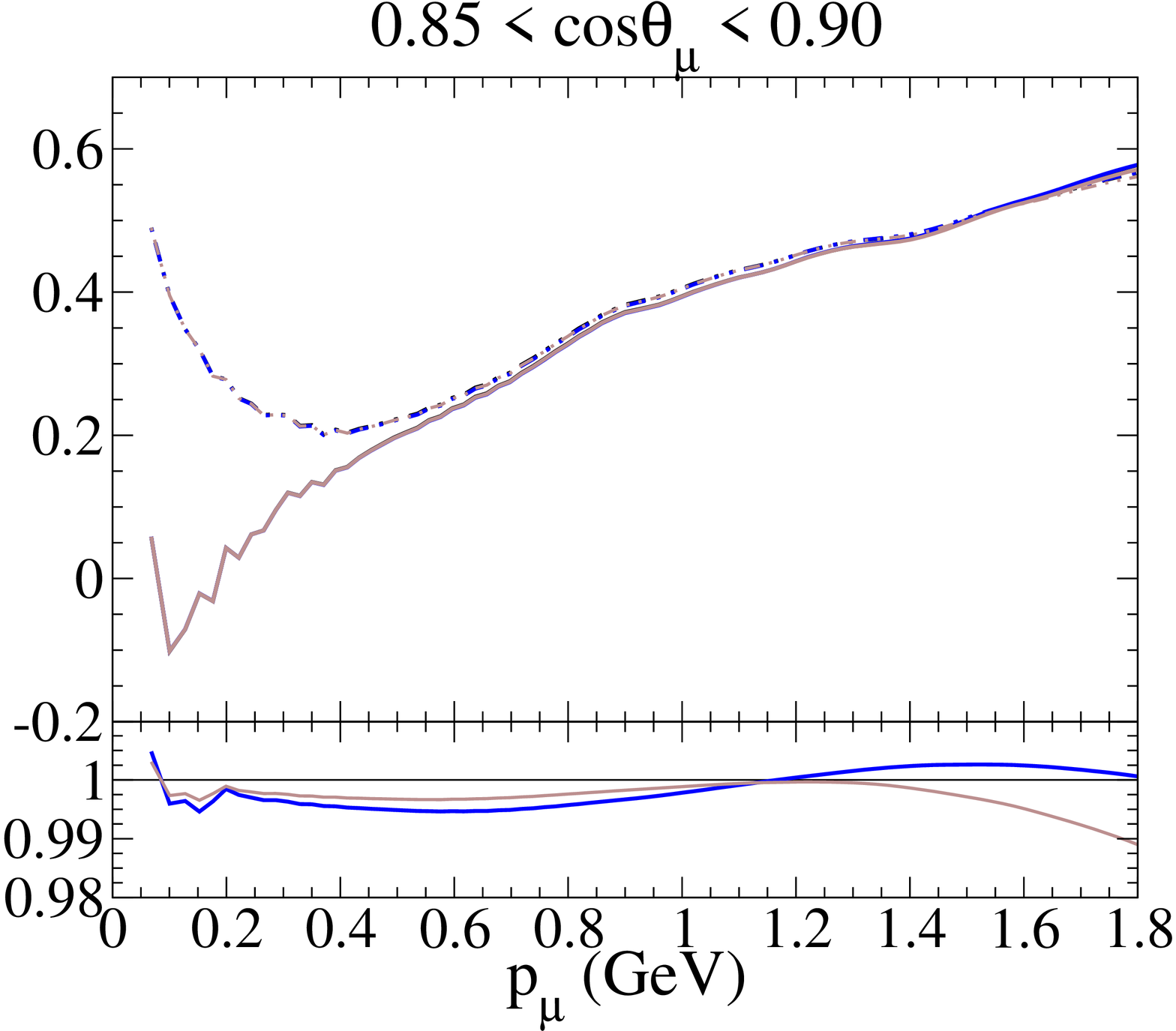}\hspace*{-0.495cm}%
		\includegraphics[scale=0.192, angle=270]{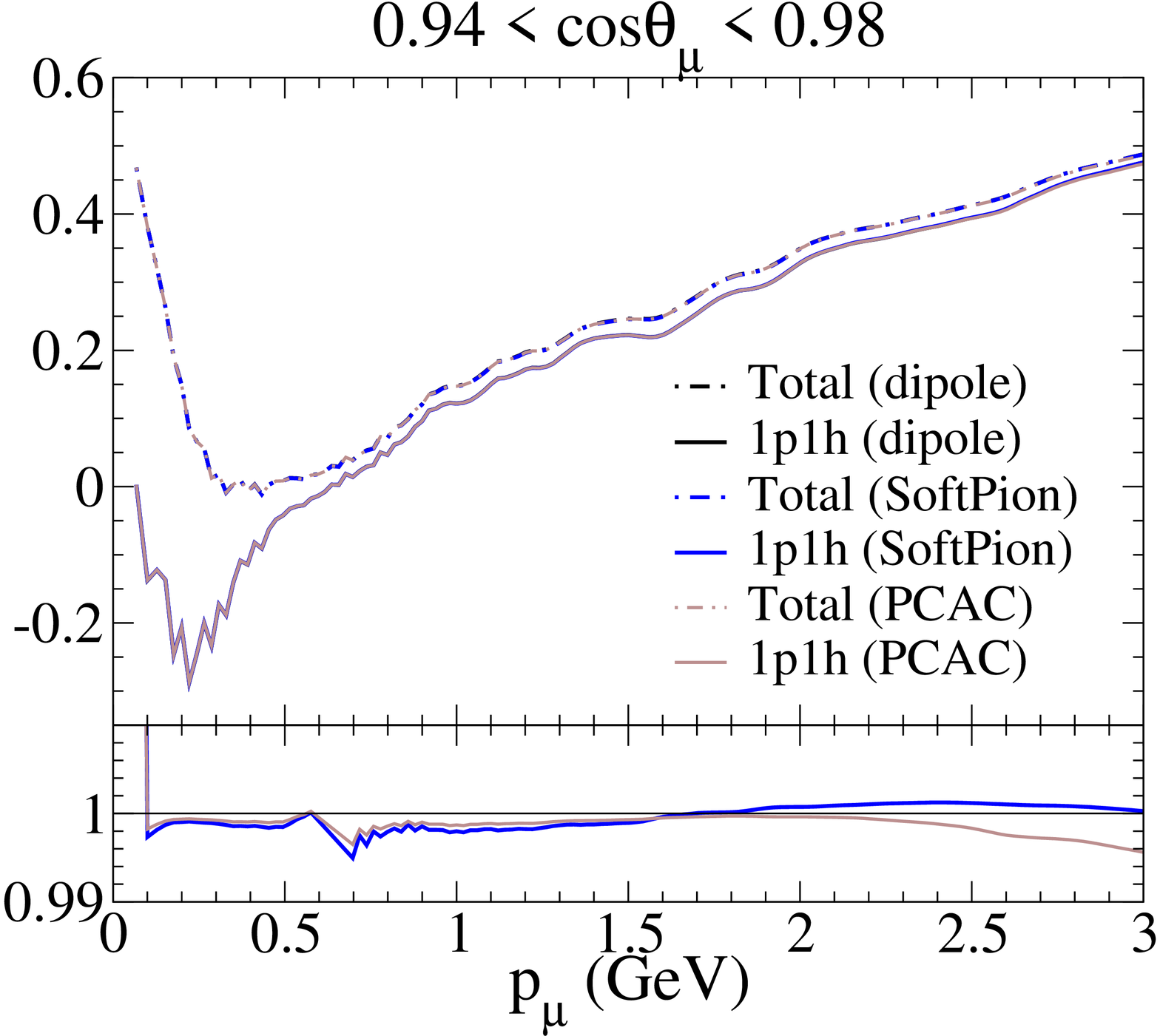}
	\end{center}
	\caption{The T2K flux-integrated  CCQE and 2p2h double-differential cross-section for neutrino (first row), antineutrino (second row) and the asymmetry \eqref{eq:asym} 
(third row) for scattering on $^{12}$C, within the SuSAv2-MEC model, using different form factors. Results are displayed for different bins of the muon scattering angle as functions of the muon momentum. Double differential cross sections are shown in units of 10$^{-39}$ cm$^2$/GeV per nucleon. 
	\label{fig:T2K_d2s}}
\end{figure}

\begin{figure}\vspace{0.08cm}
	\begin{center}\vspace{0.80cm}
	    	\includegraphics[scale=0.225, angle=270]{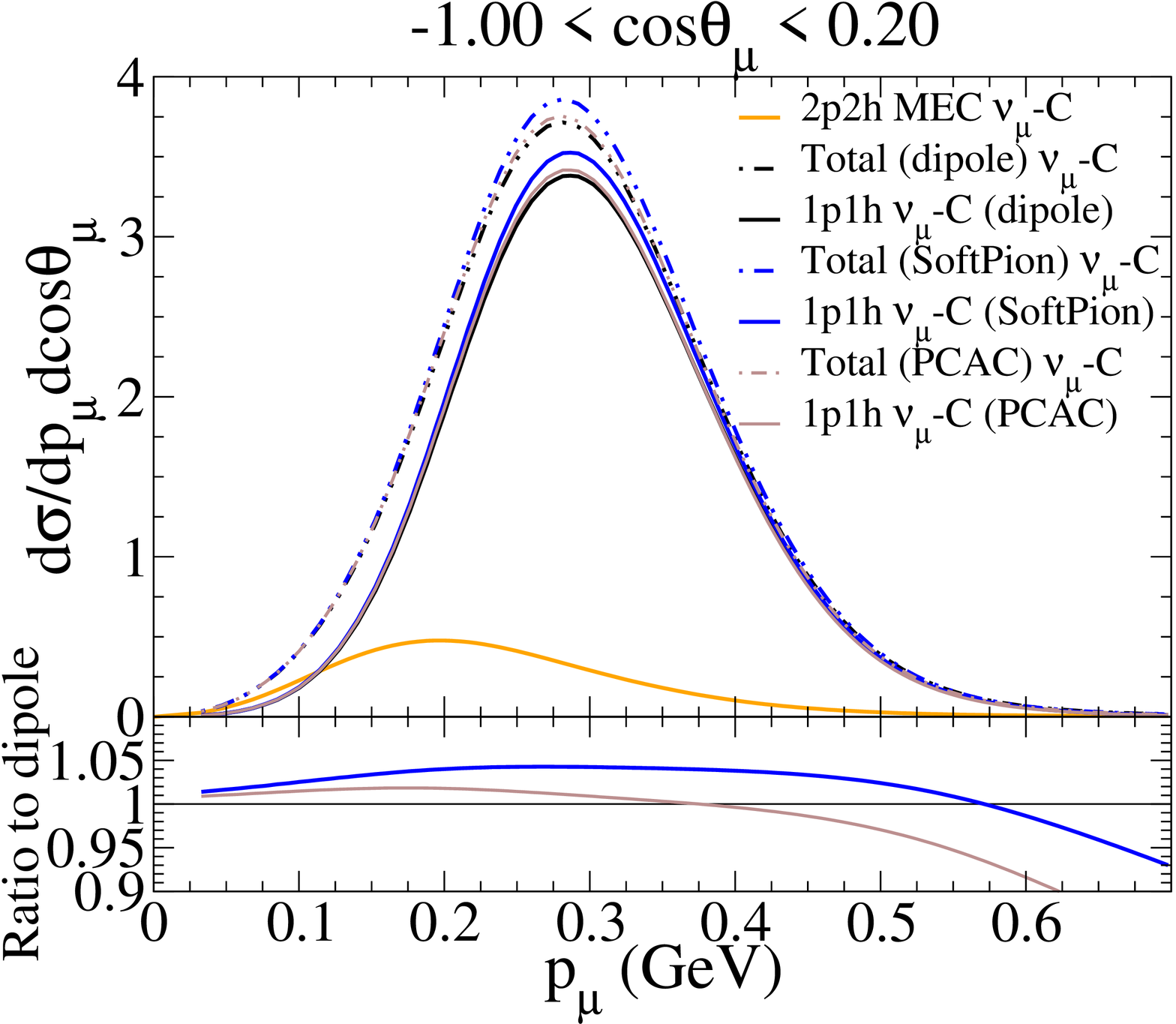}\hspace*{-0.25cm}
				\includegraphics[scale=0.225, angle=270]{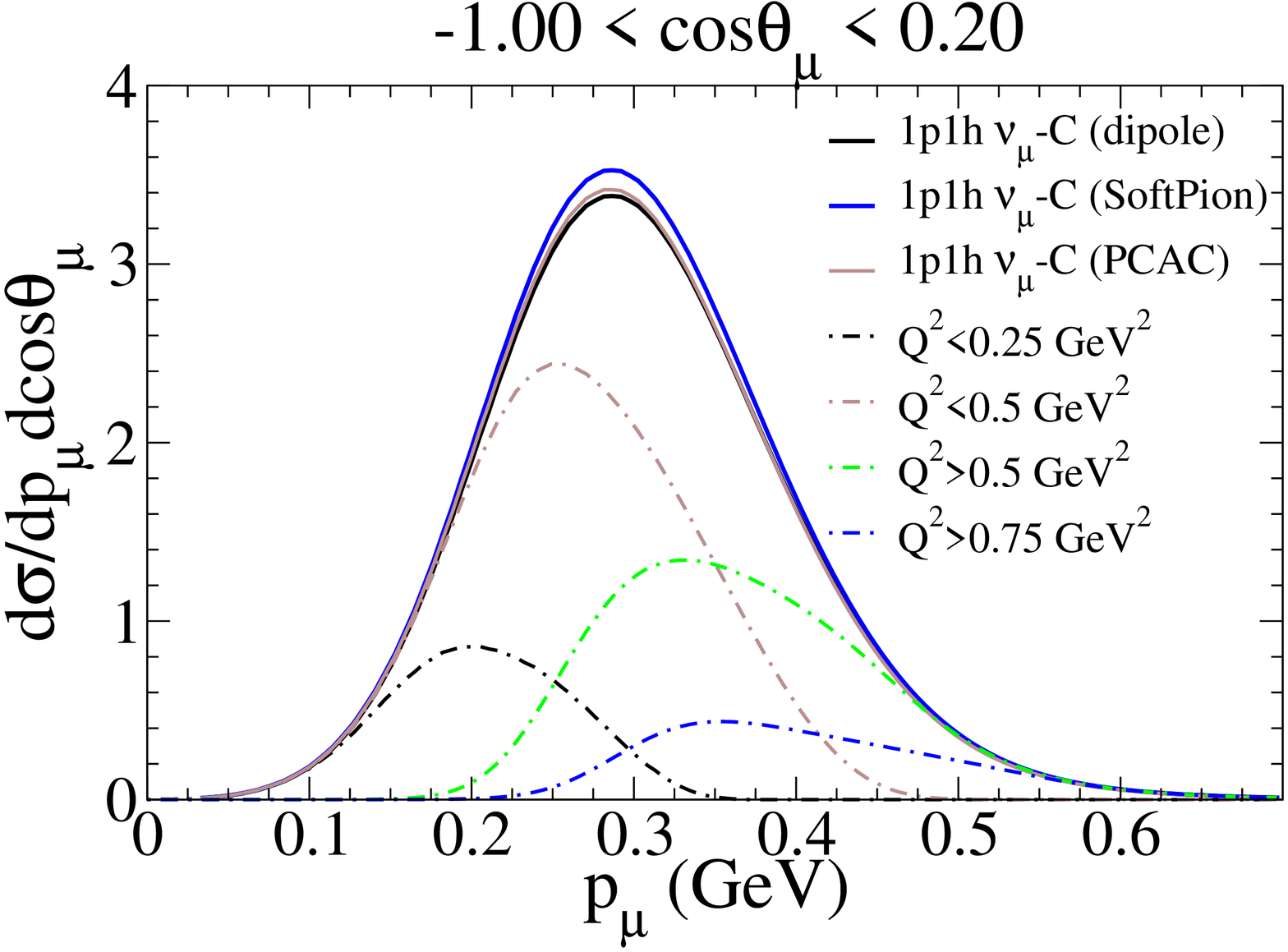}\hspace*{-0.25cm}
				\includegraphics[scale=0.225, angle=270]{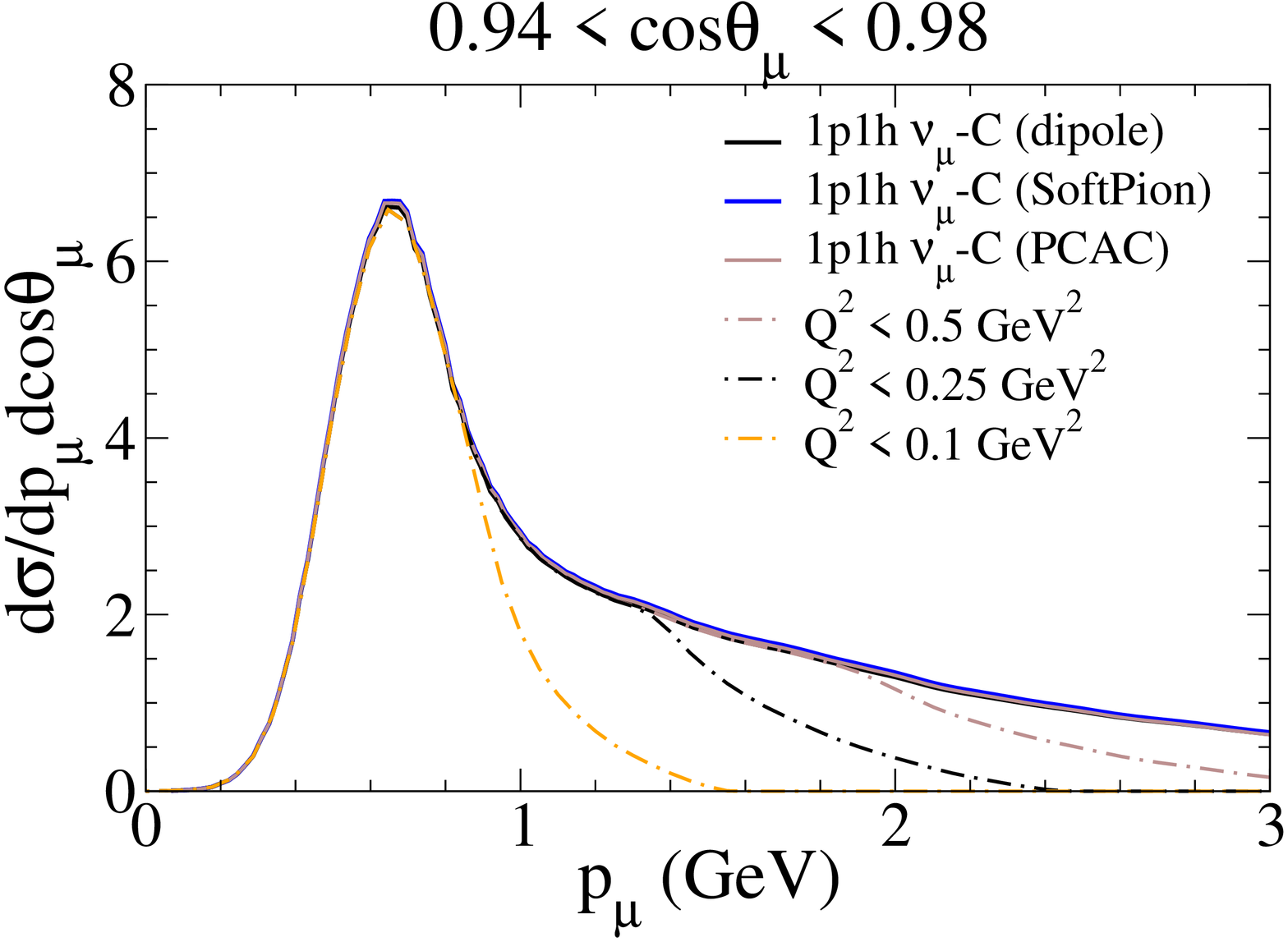}
	\end{center}
	\caption{T2K flux-integrated double-differential neutrino cross-section (as in Fig.~\ref{fig:T2K_d2s}) for backward angle (left and center) and for forward angle (right) using different form factors. Different $Q^2$ contributions to the cross-section is shown (center and right).Double differential cross sections are shown in units of 10$^{-39}$ cm$^2$/GeV per nucleon.
	\label{fig:T2K_nucuts}}
\end{figure}

\begin{figure}\vspace{0.08cm}
	\begin{center}
		\includegraphics[scale=0.225, angle=270]{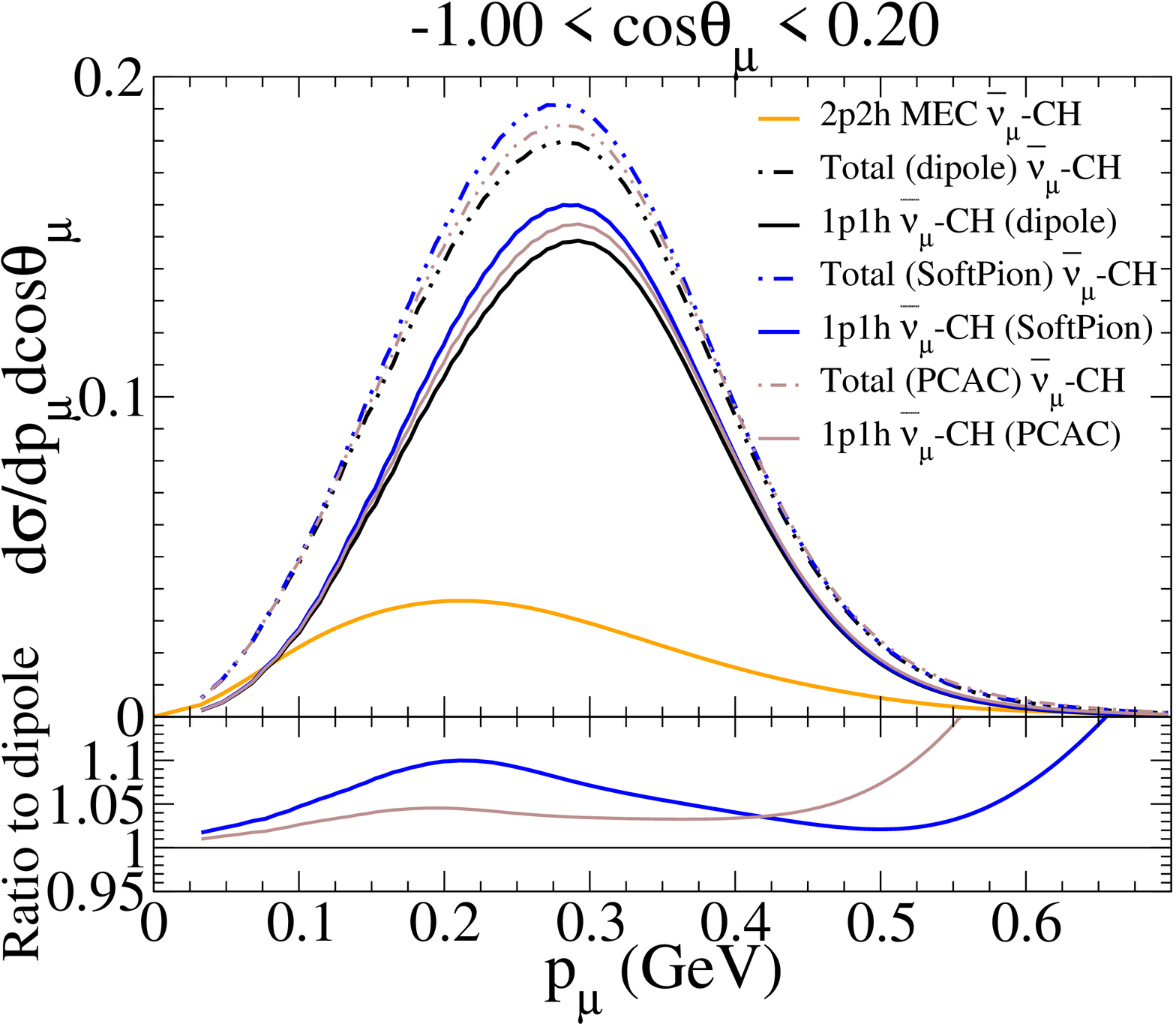}\hspace*{-0.25cm}		
				\includegraphics[scale=0.225, angle=270]{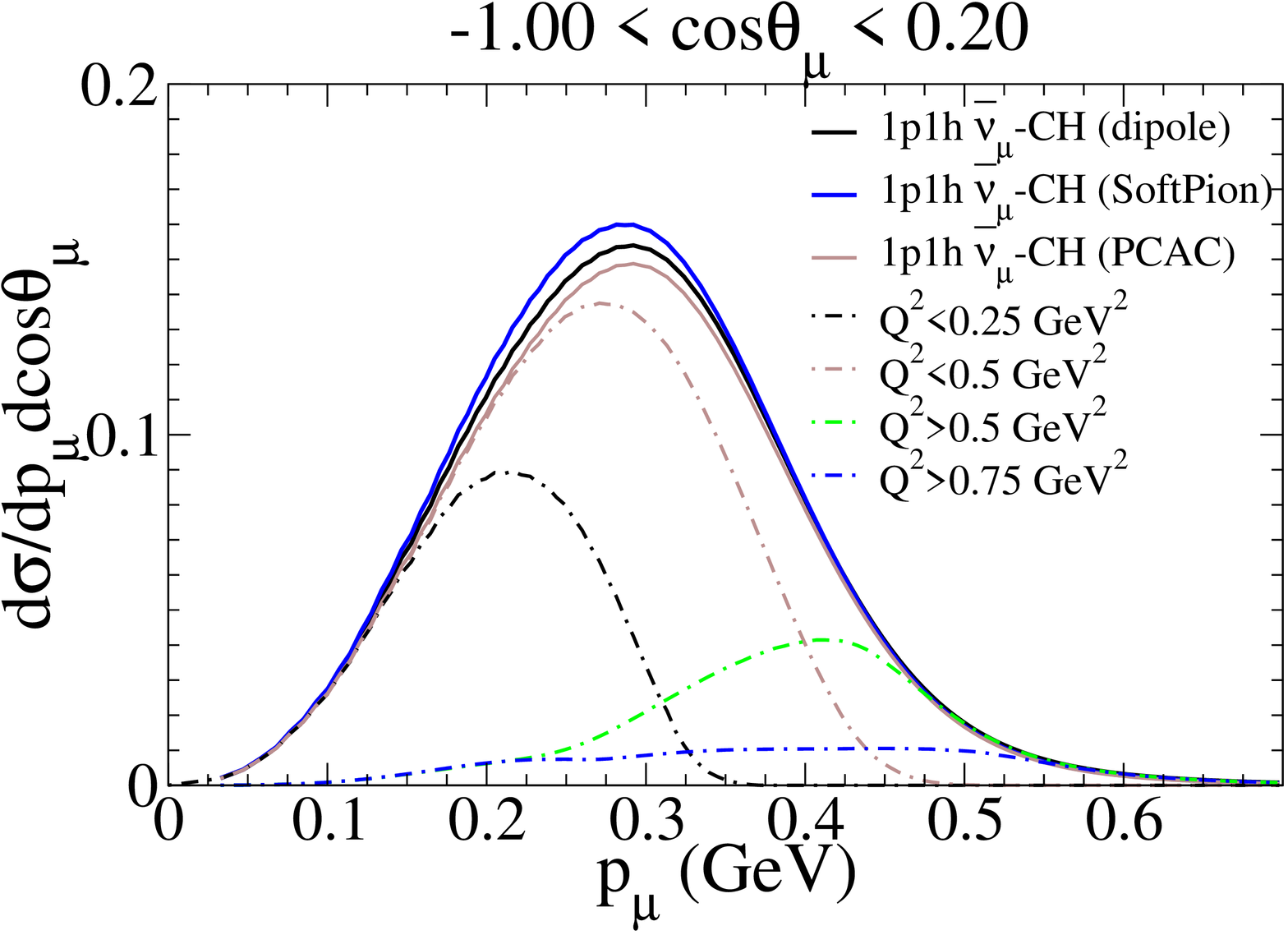}\hspace*{-0.25cm}
				\includegraphics[scale=0.225, angle=270]{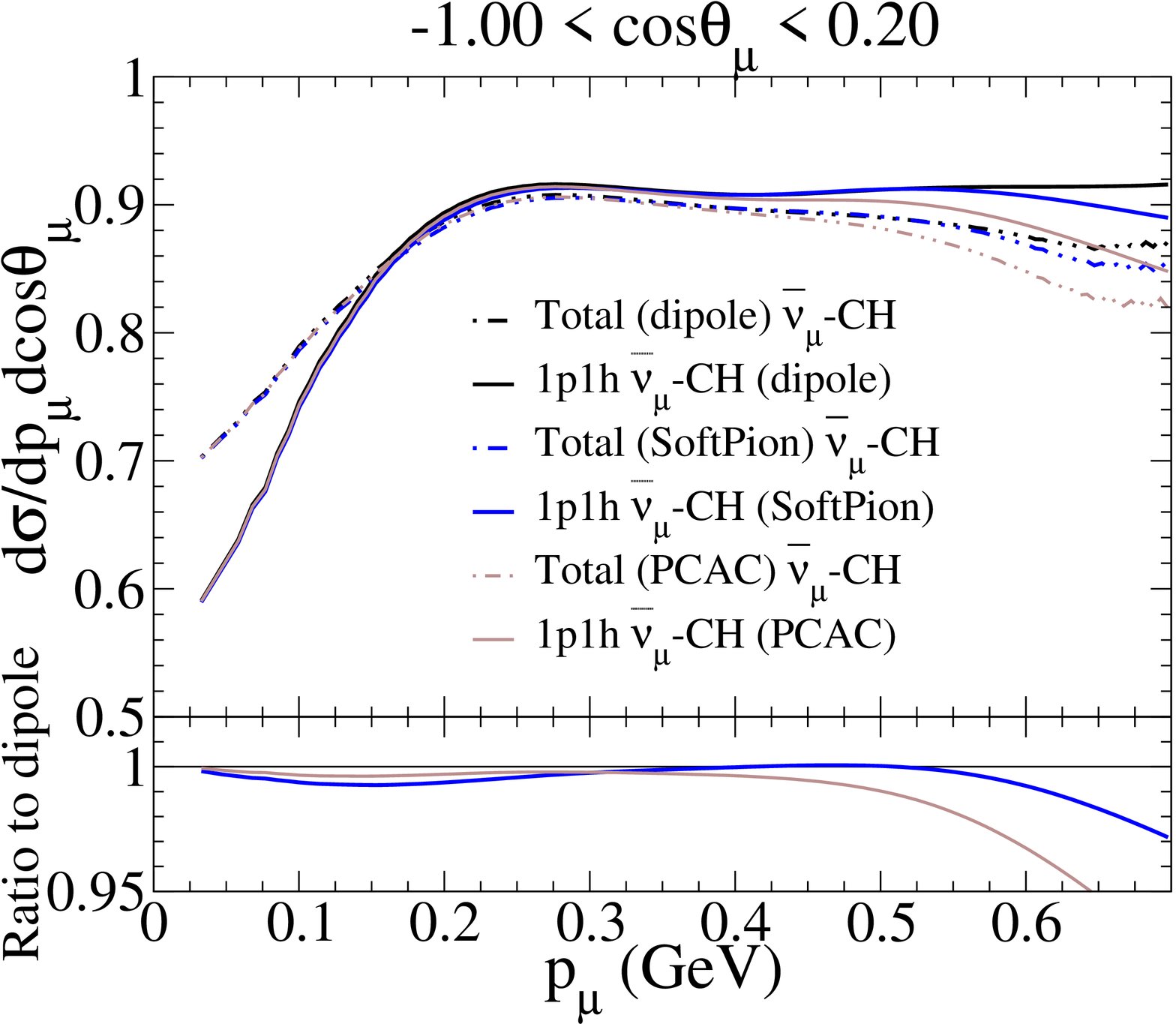}
	\end{center}
	\caption{T2K flux-integrated double-differential antineutrino cross-section (as in Fig.~\ref{fig:T2K_d2s}) for backward angle (left and center) and neutrino-antineutrino cross-section asymmetry (right) for different form factors. The different $Q^2$ contributions to the cross-section are shown in the center.
Double differential cross sections are shown in units of 10$^{-39}$ cm$^2$/GeV per nucleon.\label{fig:T2K_nubarcuts}}
\end{figure}

In Fig.~\ref{fig:T2K_d2s} the T2K neutrino and antineutrino double-differential cross-sections and the corresponding asymmetry
\begin{equation}
    \mbox{asymmetry} = \frac{d^2\sigma_{\nu}-d^2\sigma_{\bar\nu}}{d^2\sigma_\nu+d^2\sigma_{\bar\nu}}
    \label{eq:asym}
\end{equation}
are shown for different axial form factors.
The largest differences between neutrino cross-sections evaluated with different form factors is about 5\%. By mapping the muon kinematics ($p_\mu, \theta_\mu$) into $Q^2$ on the basis of Fig.~\ref{fig:weighted}, the largest difference with respect to the cross-section with dipole form factor, appears always in correspondence of $Q^2\approx$ 0.5~GeV$^2$ for SoftPion, as expected from Fig.~\ref{fig:FFfitNuPion2}. The impact of the different $Q^2$ regions is shown in Fig.~\ref{fig:T2K_nucuts}, where also the backward angle is analyzed. For forward angles, the $Q^2\approx 0.5$~GeV$^2$ region corresponds to the small tail at high momentum, so at these angles the cross-section is mostly unaffected by form factor differences. For backward angles, the $Q^2\approx$ 0.5~GeV$^2$ region corresponds instead exactly to the region of larger cross-section with intermediate muon momentum, thus the impact of form factors difference is larger. In the backward region, as shown in Fig.~\ref{fig:T2K_nubarcuts}, the form factor differences can reach 5\%. In such region the effect in the antineutrino case is even larger, up to 10\%. Still, in the neutrino-antineutrino asymmetry the effect is at percent level.   
The region $Q^2>1$~GeV, where the different axial form factors depart from each other sizeably, is negligible in T2K data. 

\subsubsection{Effects of axial form factors at MINERvA kinematics}
\label{sec:Minerva}

\begin{figure}\vspace{0.08cm}
	\begin{center}\vspace{0.80cm}
		\includegraphics[scale=0.23, angle=270]{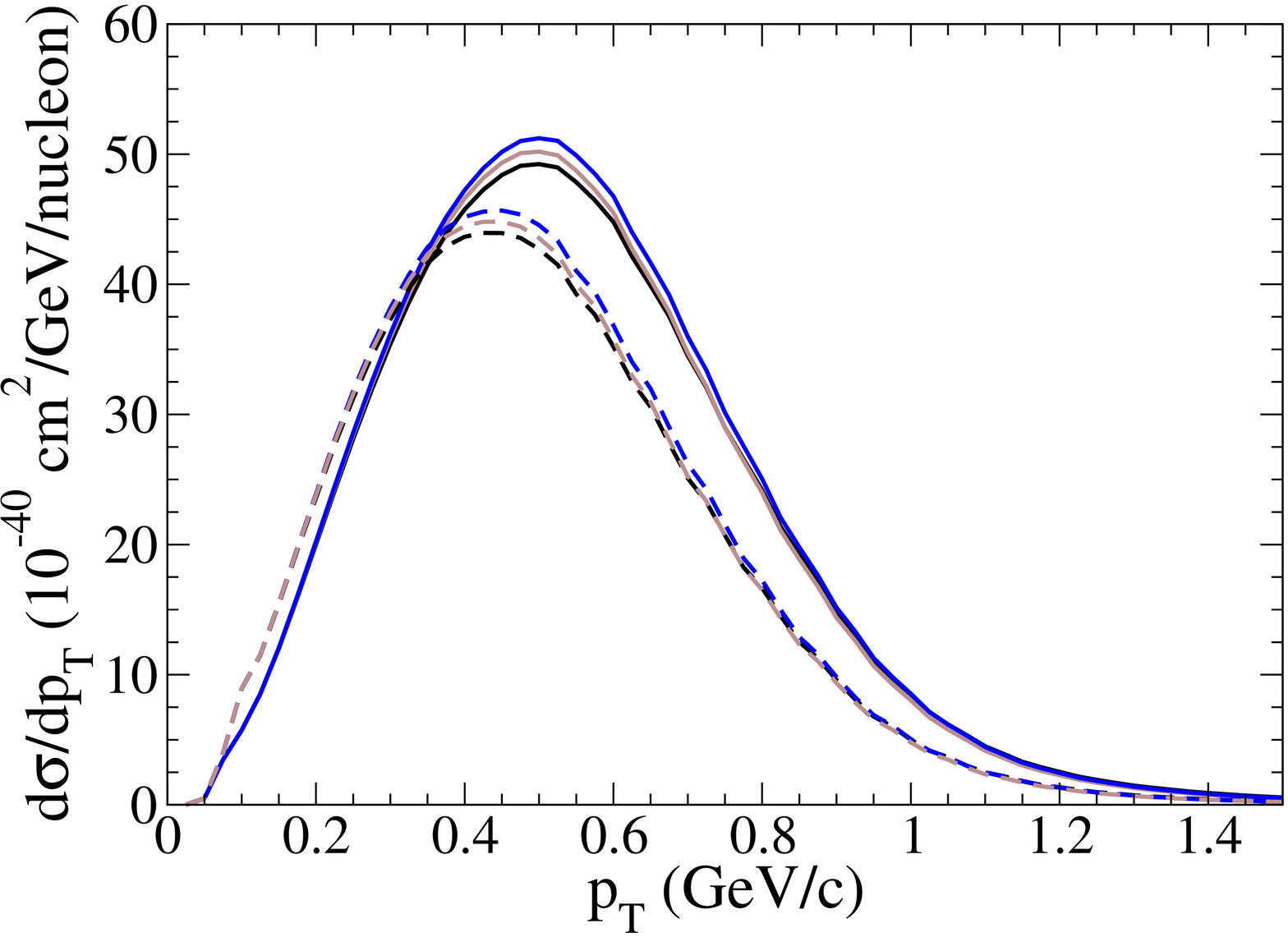}\hspace*{-0.15cm}%
		\includegraphics[scale=0.23, angle=270]{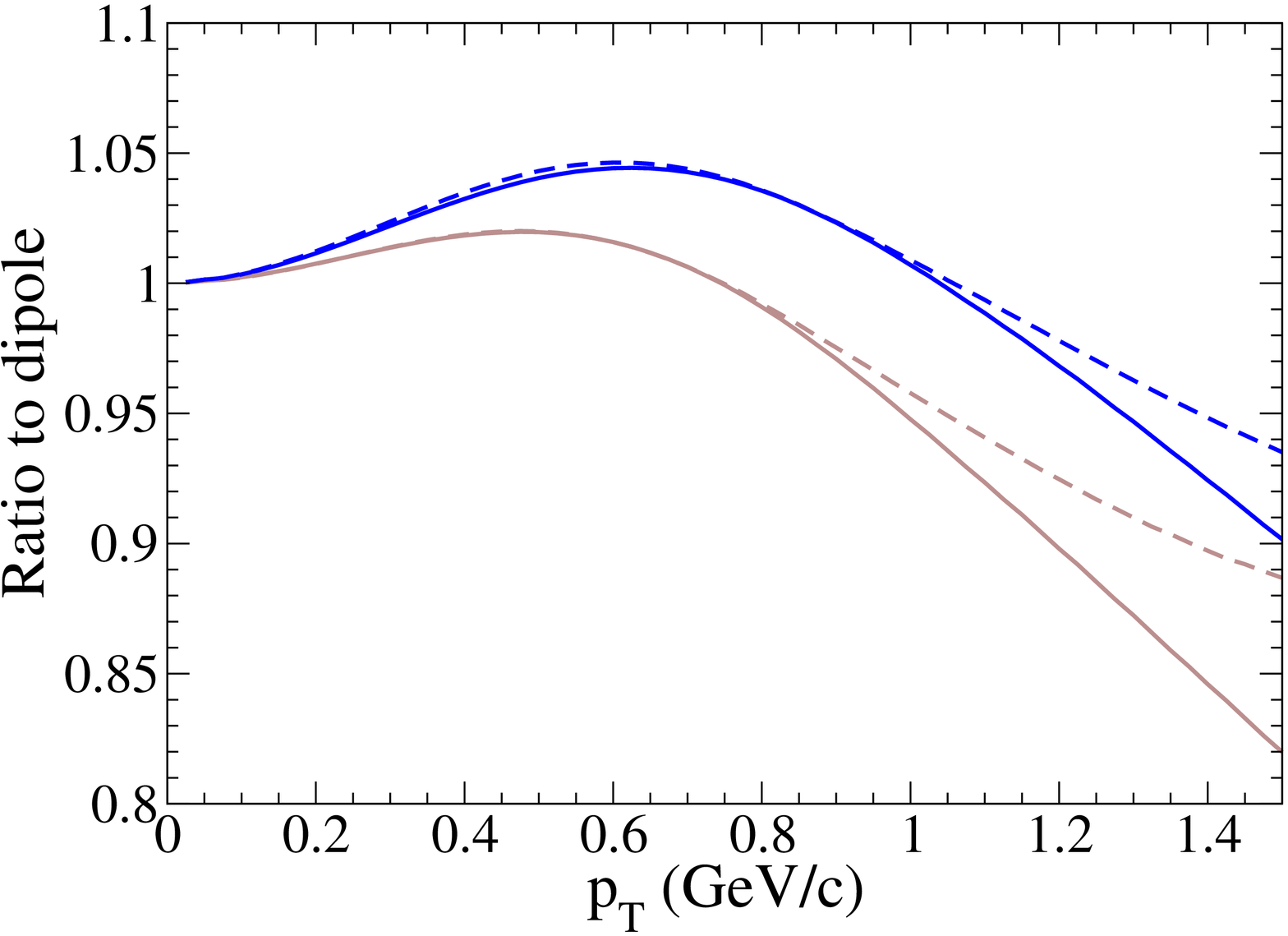}\\	\vspace*{-0.195cm}
		\includegraphics[scale=0.23, angle=270]{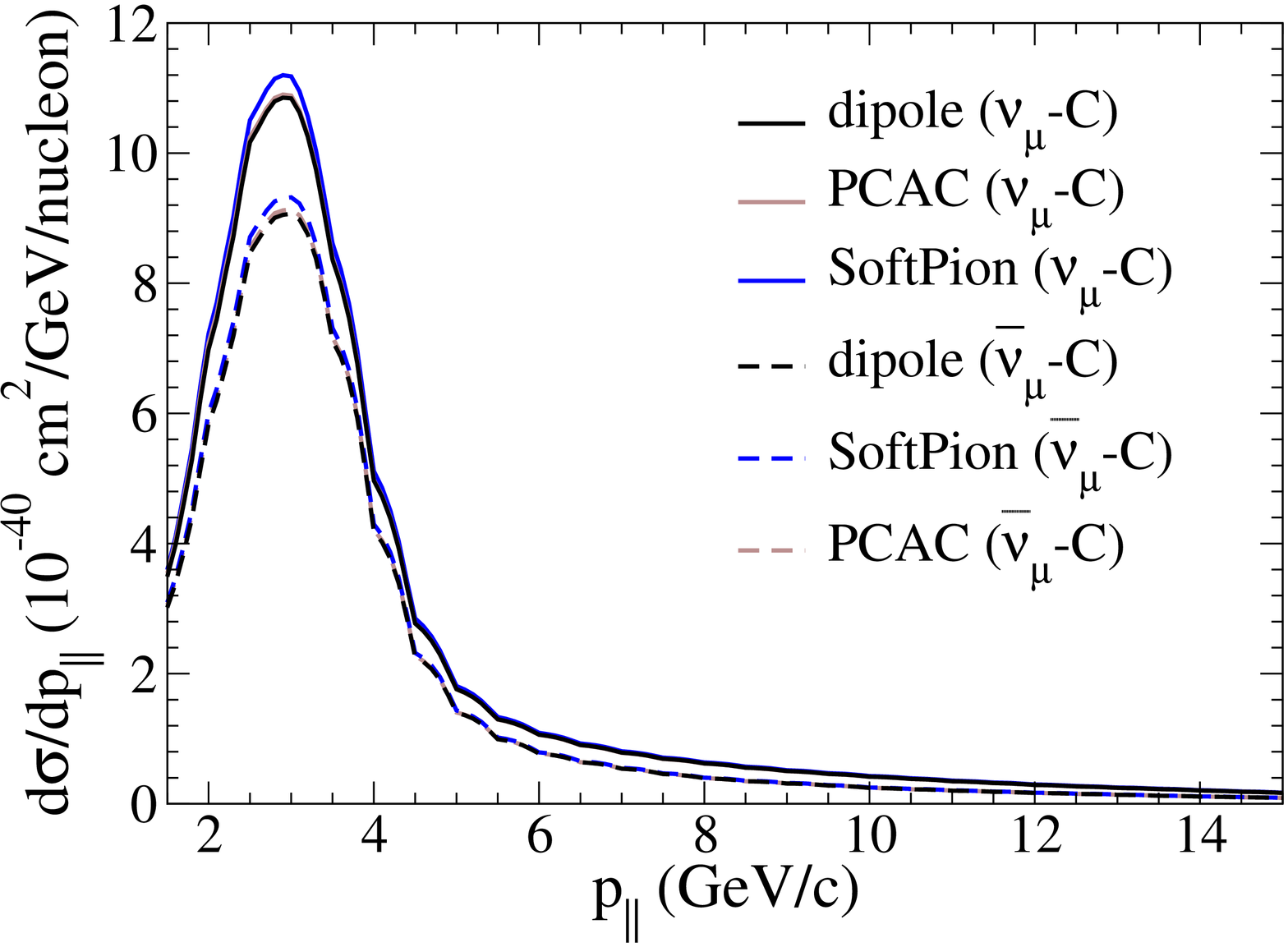}\hspace*{-0.15cm}
		\includegraphics[scale=0.23, angle=270]{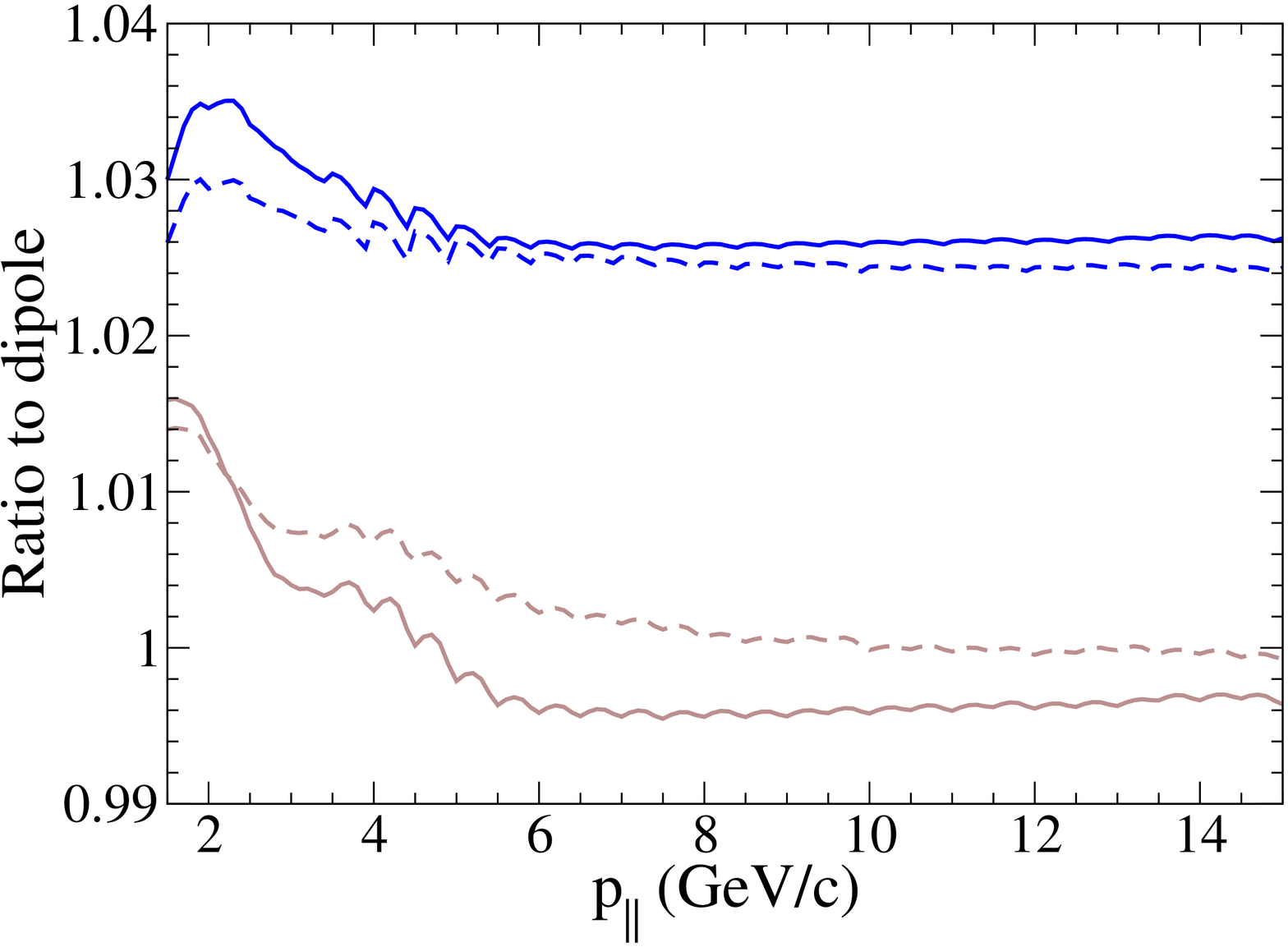}\\
	\end{center}
	\caption{
	Single differential cross-section (left) for neutrino and antineutrino as a function of $p_T$e (first row) and $p_{||}$ (second row) for different nucleon form factors at MINERvA kinematics. The ratio of the different form factors with respect to dipole is also shown (right). 
	\label{fig:Minerva_singlediff}}
\end{figure}

In Fig.~\ref{fig:Minerva_singlediff} the effect of the form factors on the single-differential cross-section as a function of muon transverse and longitudinal momentum ($p_T, p_L$) is shown. 

\begin{figure}\vspace{0.08cm}
	\begin{center}\vspace{0.80cm}
		\hspace*{-0.95cm}\includegraphics[scale=0.192, angle=270]{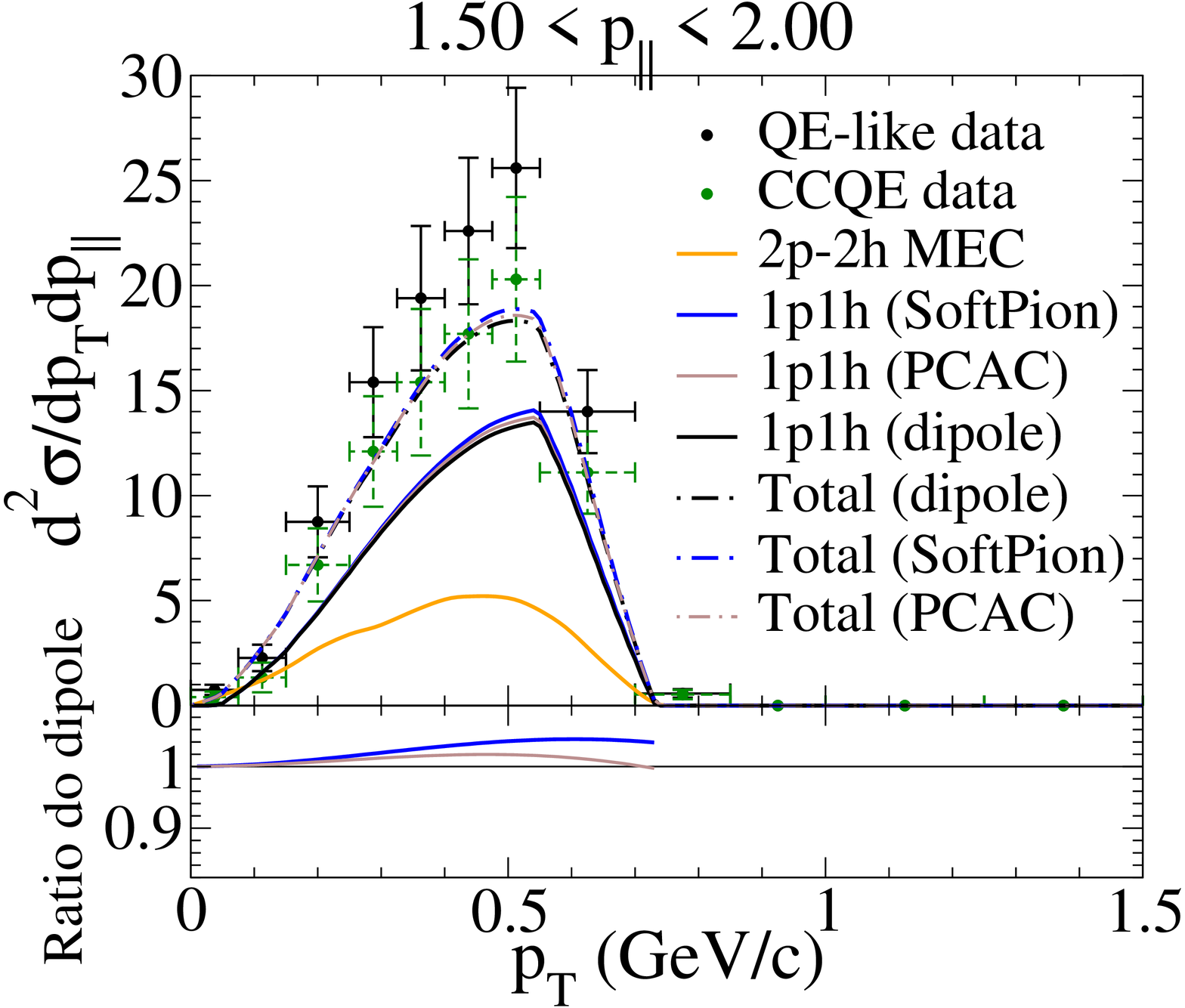}\hspace*{-0.495cm}%
		\includegraphics[scale=0.192, angle=270]{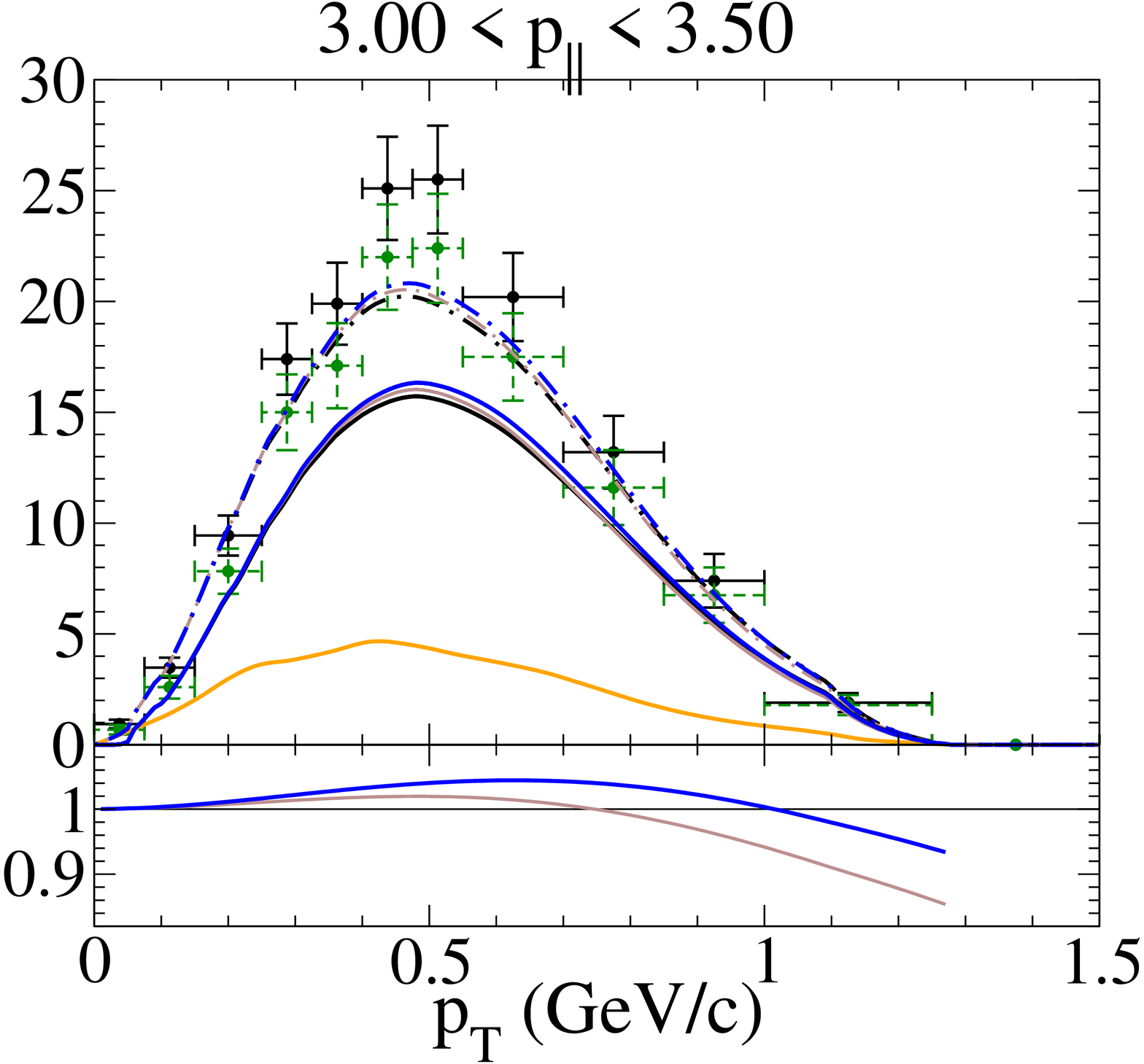}\hspace*{-0.495cm}%
		\includegraphics[scale=0.192, angle=270]{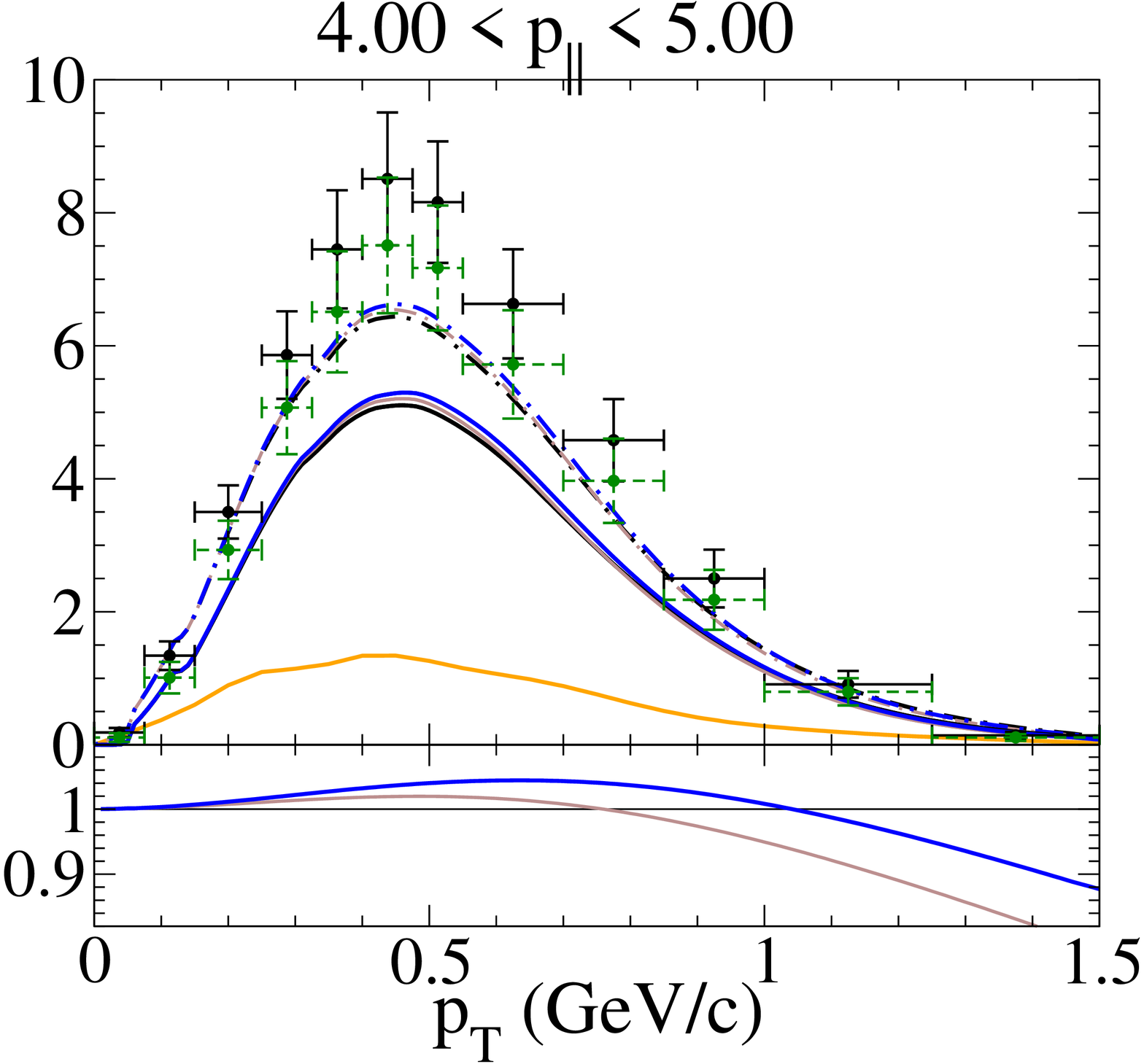}\hspace*{-0.495cm}%
		\includegraphics[scale=0.192, angle=270]{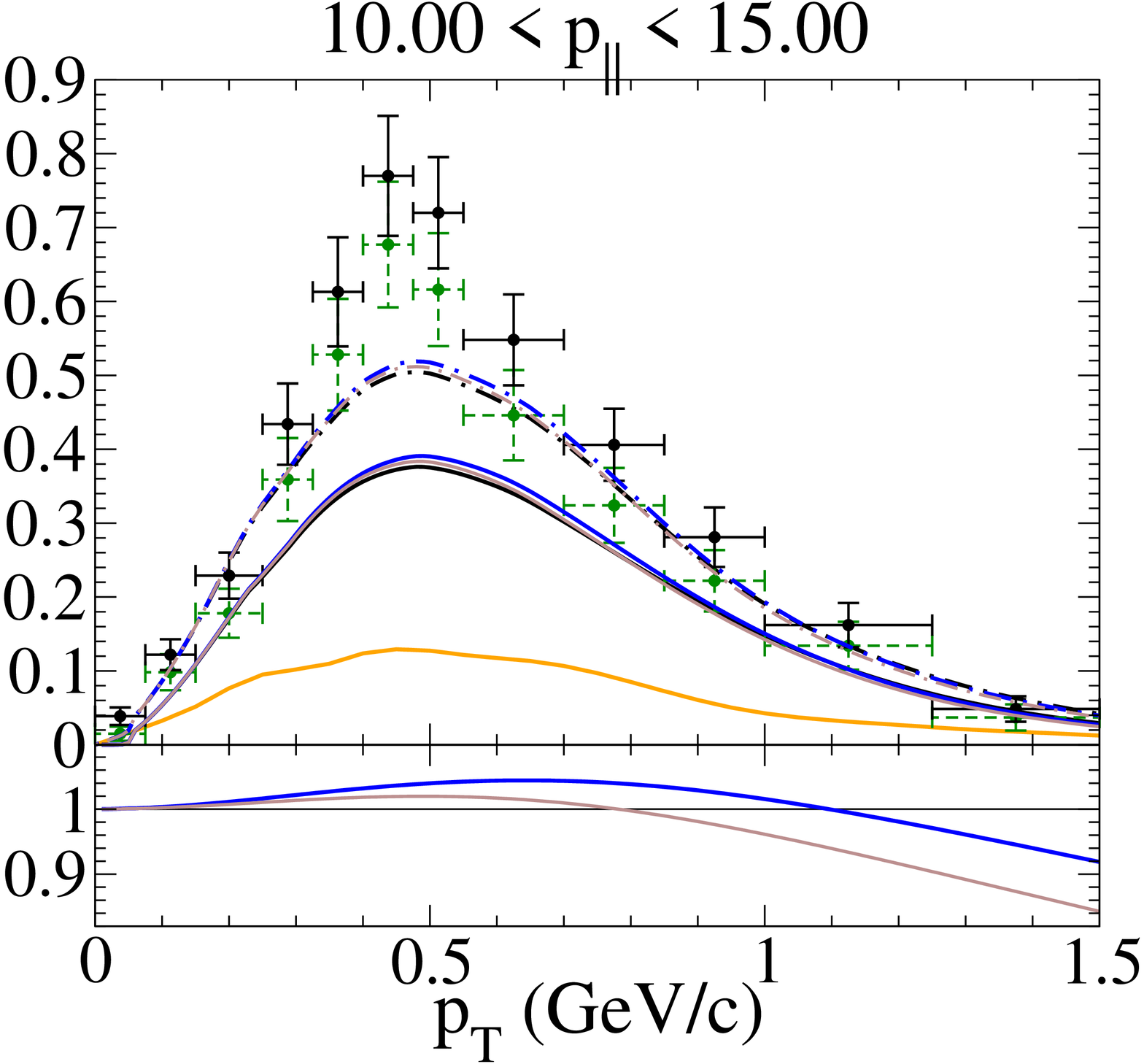}
		\\\vspace*{0.295cm}
		\hspace*{-0.95cm}\includegraphics[scale=0.192, angle=270]{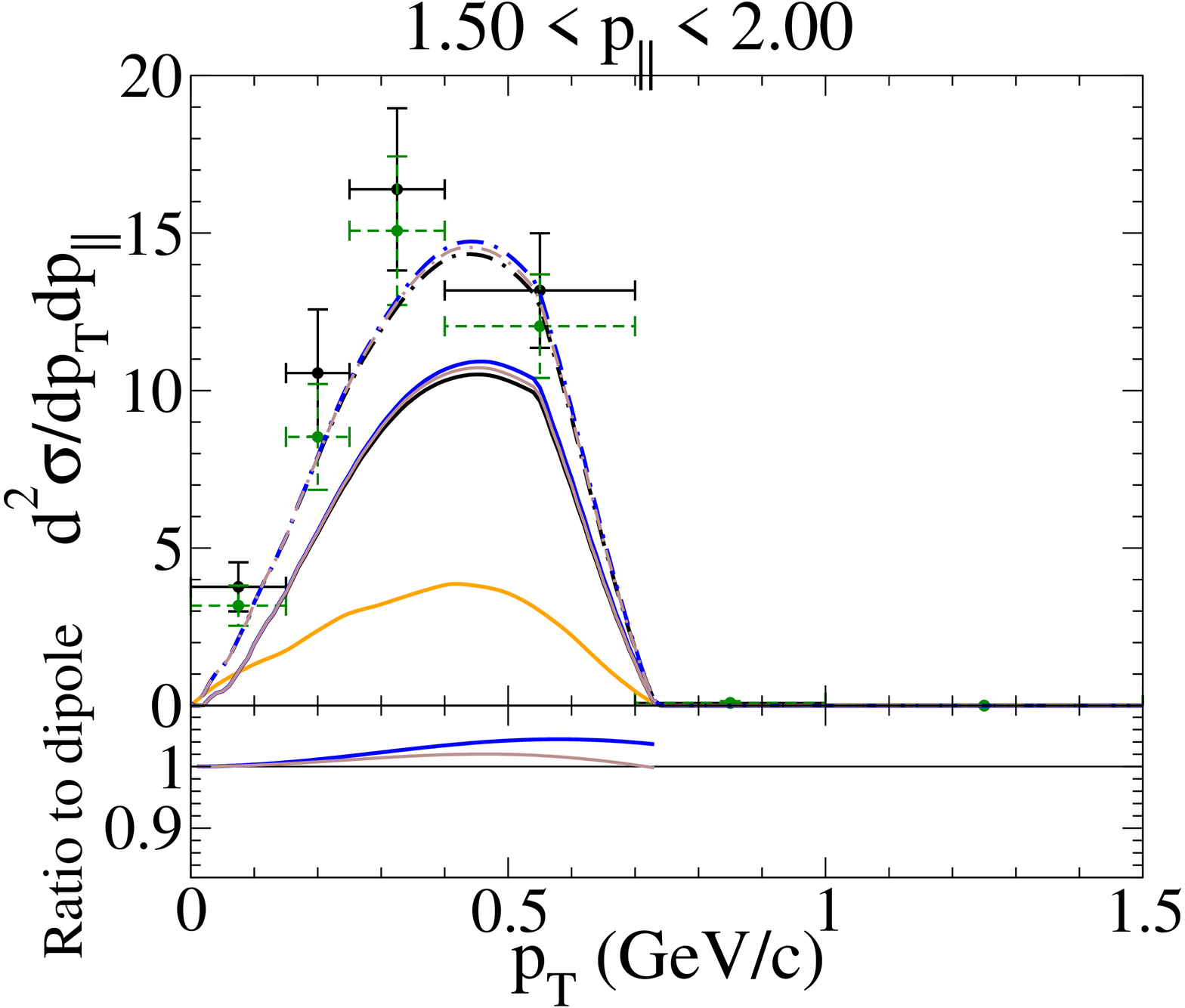}\hspace*{-0.495cm}%
		\includegraphics[scale=0.192, angle=270]{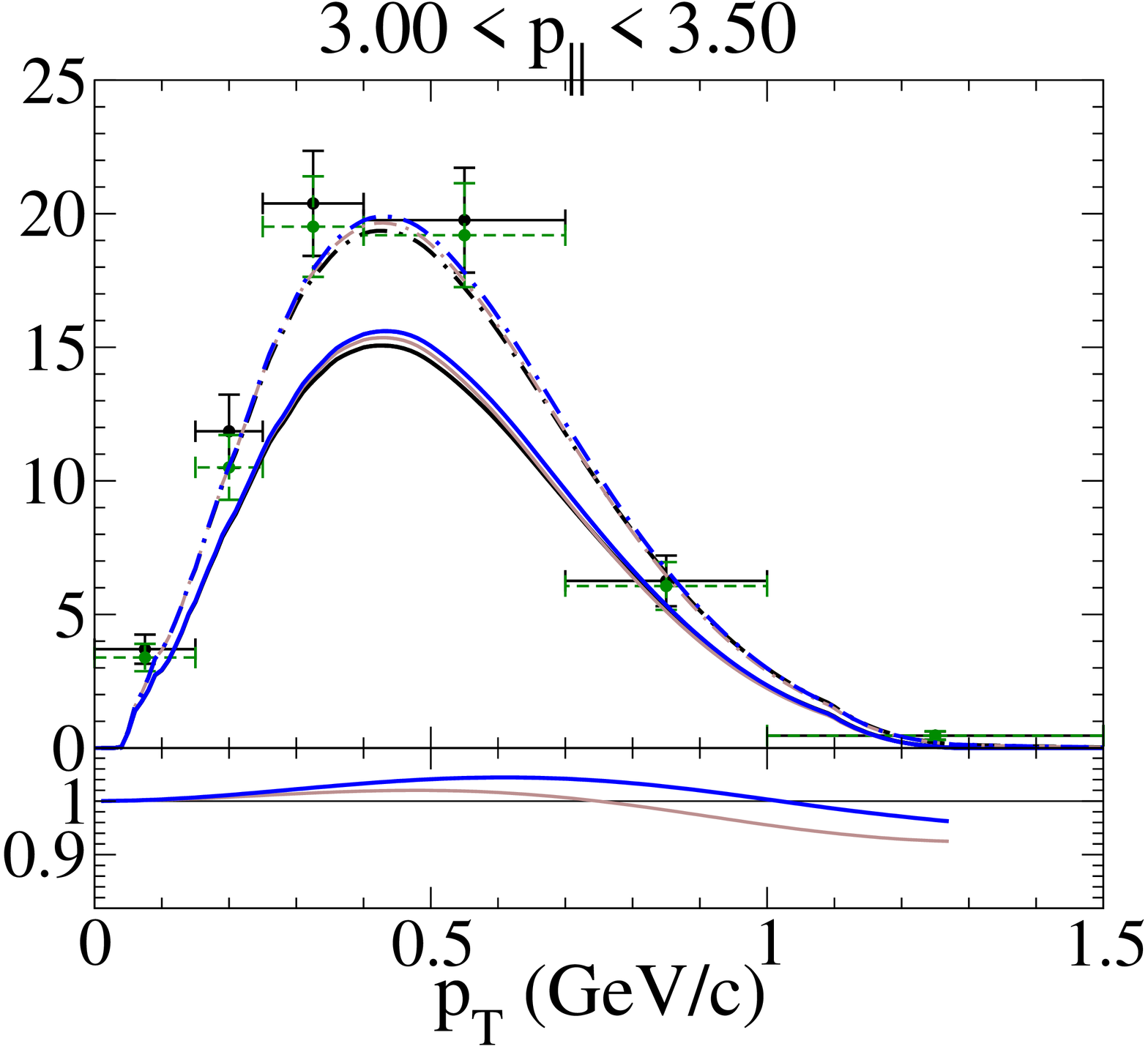}\hspace*{-0.495cm}%
		\includegraphics[scale=0.192, angle=270]{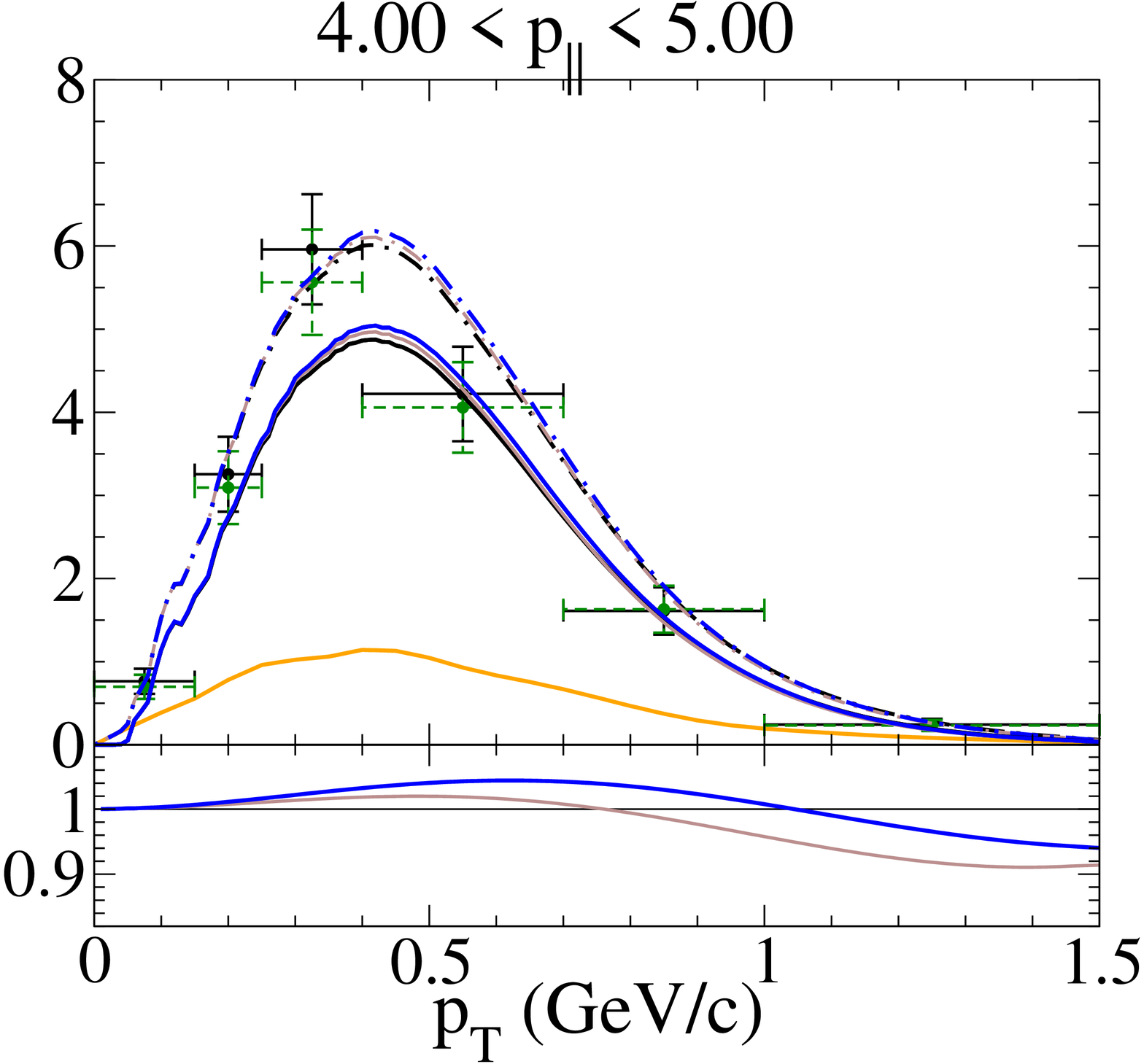}\hspace*{-0.495cm}%
		\includegraphics[scale=0.192, angle=270]{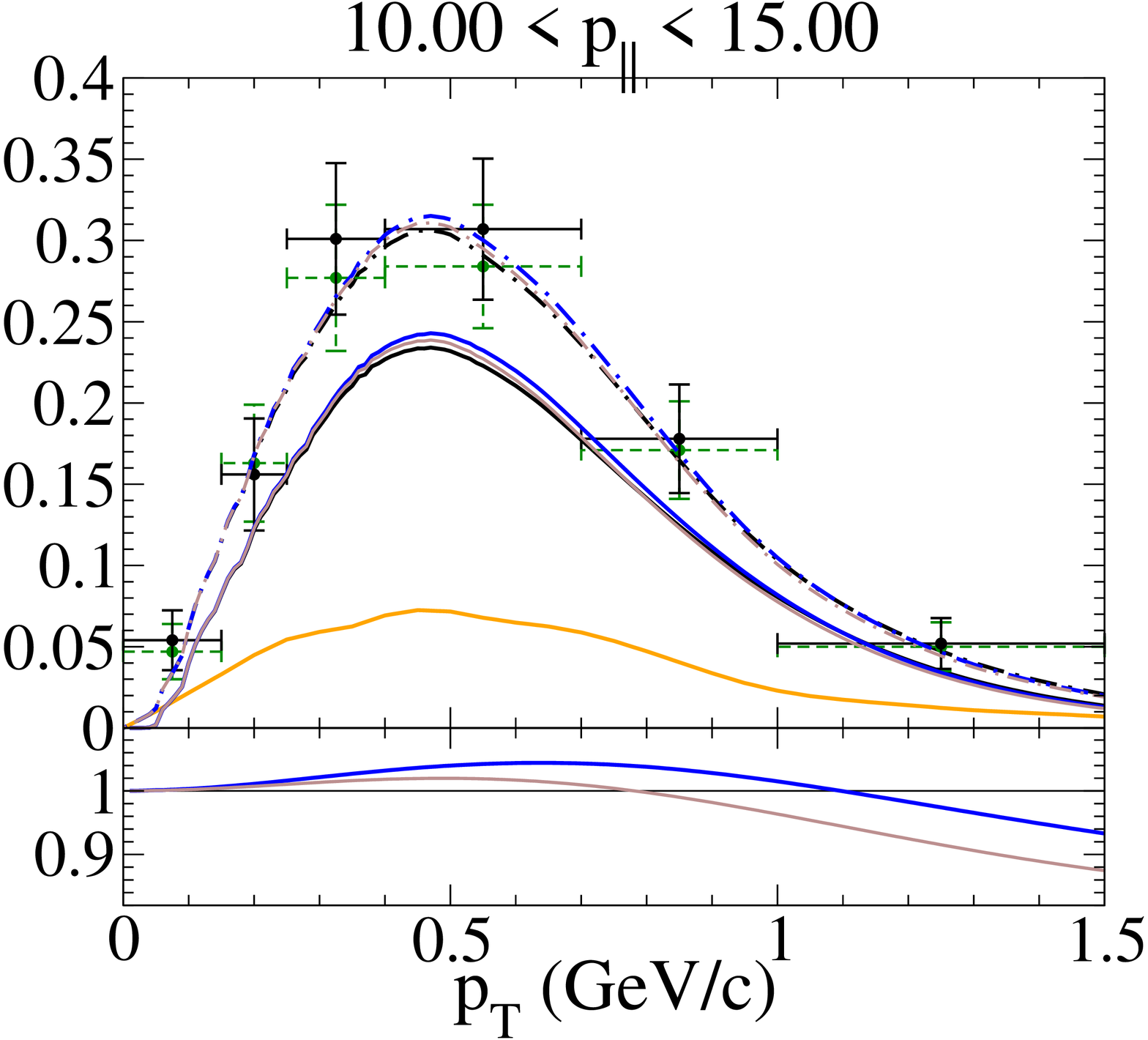}\\\vspace*{0.295cm}
		\hspace*{-0.95cm}\includegraphics[scale=0.192, angle=270]{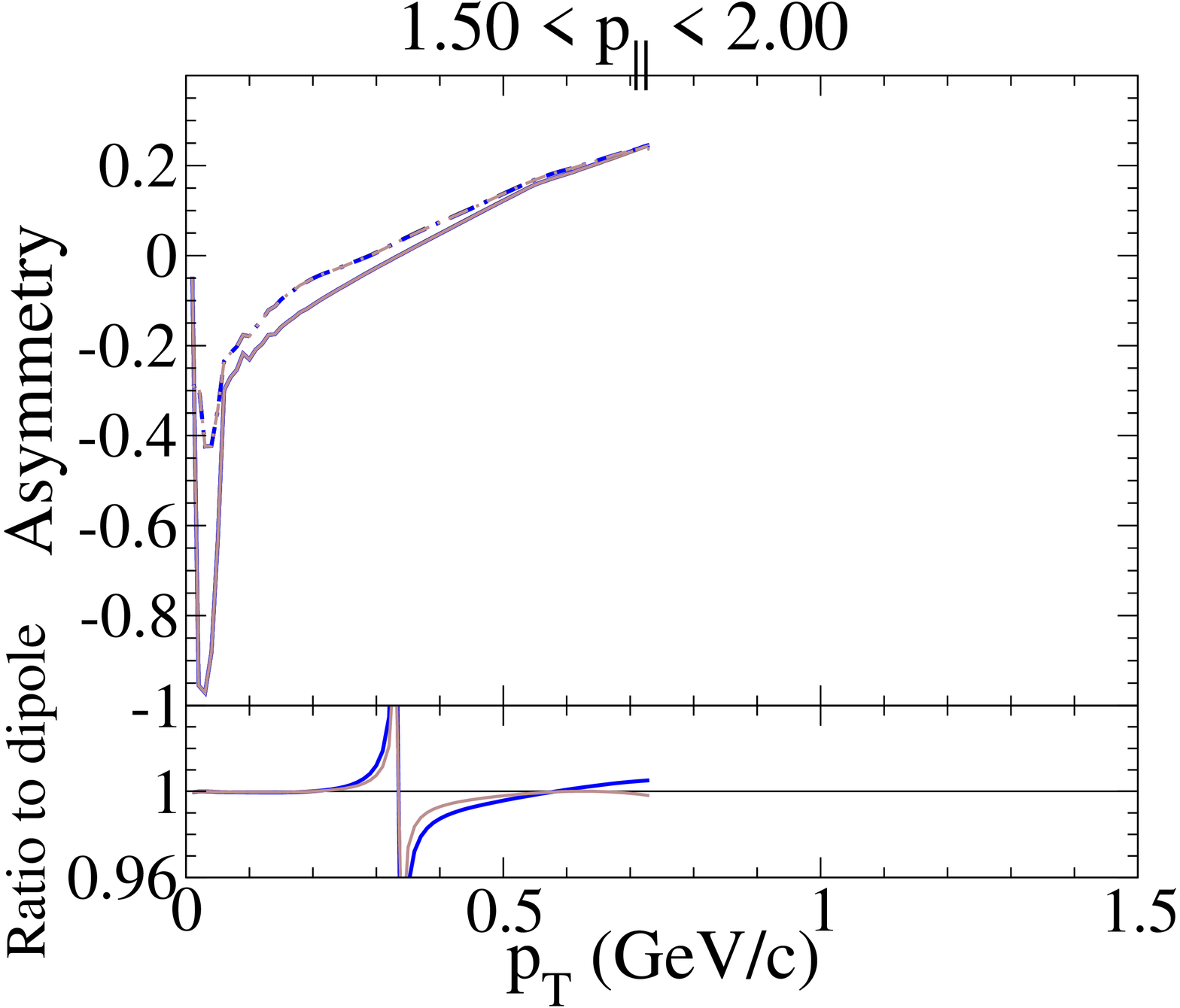}\hspace*{-0.495cm}%
		\includegraphics[scale=0.192, angle=270]{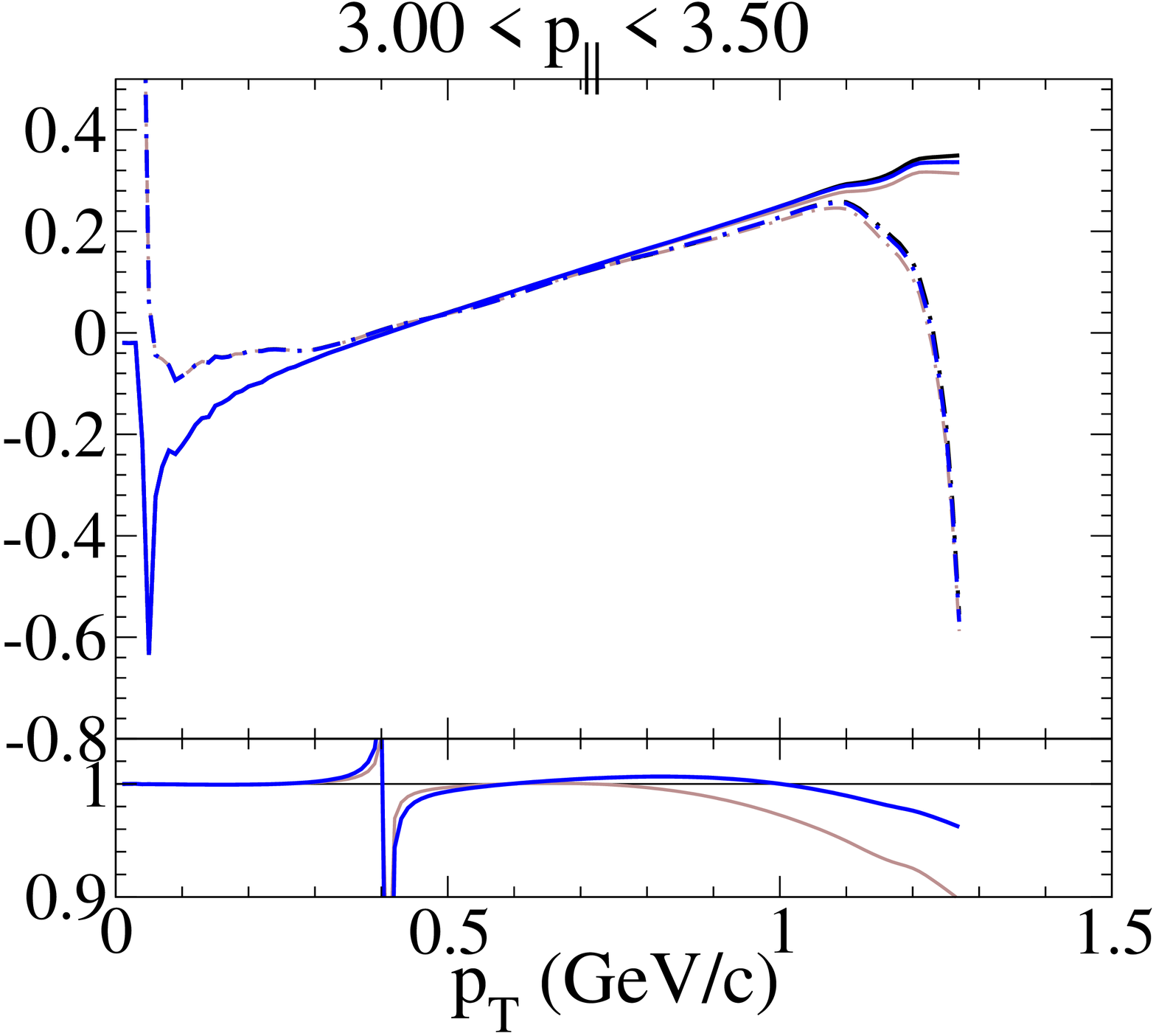}\hspace*{-0.495cm}%
		\includegraphics[scale=0.192, angle=270]{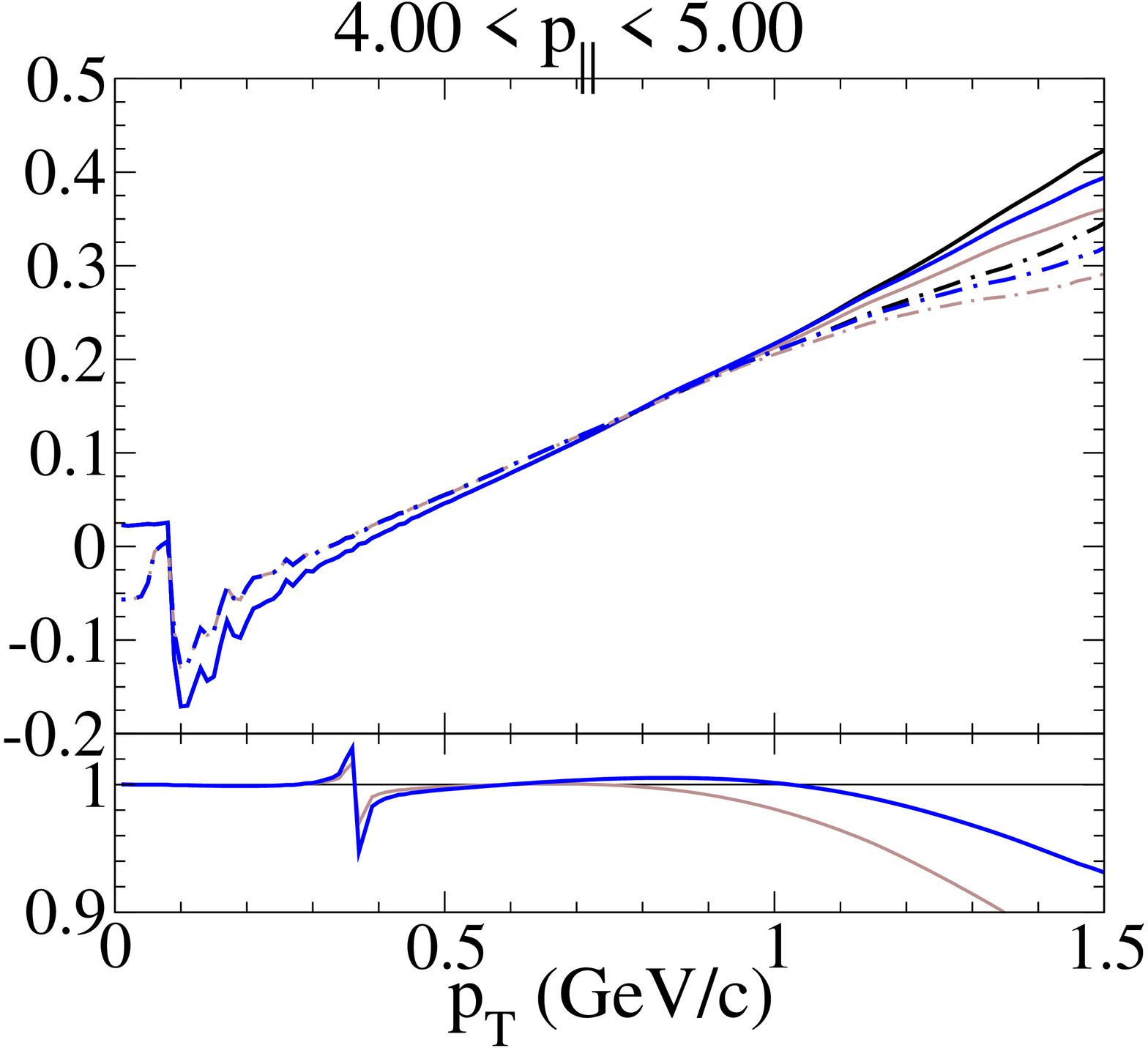}\hspace*{-0.495cm}%
		\includegraphics[scale=0.192, angle=270]{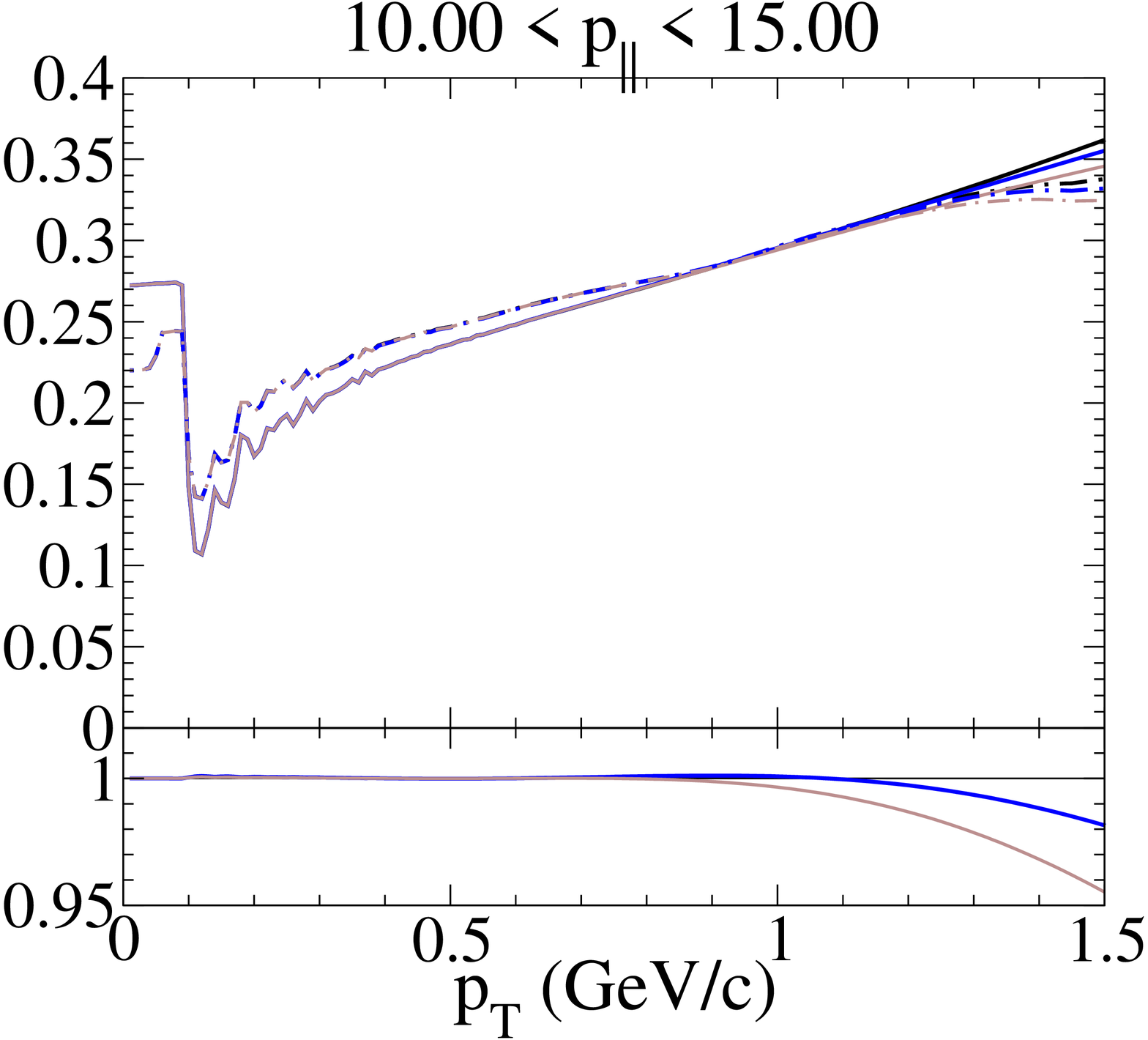}
	\end{center}
	\caption{MINERvA flux-integrated double-differential cross-section for different axial form factors, as a function of the muon transverse and longitudinal momentum, for neutrino (first row), antineutrino (second row) and the neutrino-antineutrino asymmetry (third row). Double differential cross sections are shown in units of 10$^{-39}$ cm$^2$/GeV$^2$ per nucleon.}
    \label{fig:Mnv_d2s}
\end{figure}

The double differential cross-section as a function of $p_T, p_L$ is shown in Fig~\ref{fig:Mnv_d2s}. The region of $Q^2 \approxeq 0.5$~GeV$^2$ shows differences of the order of 5\%, similarly to T2K, while in the region of high $p_T$ and lower cross-section effects up to 10\% and above can be observed.

\begin{figure}\vspace*{-1.298cm}
	\begin{center}\vspace{0.80cm}
		\includegraphics[scale=0.295, angle=270]{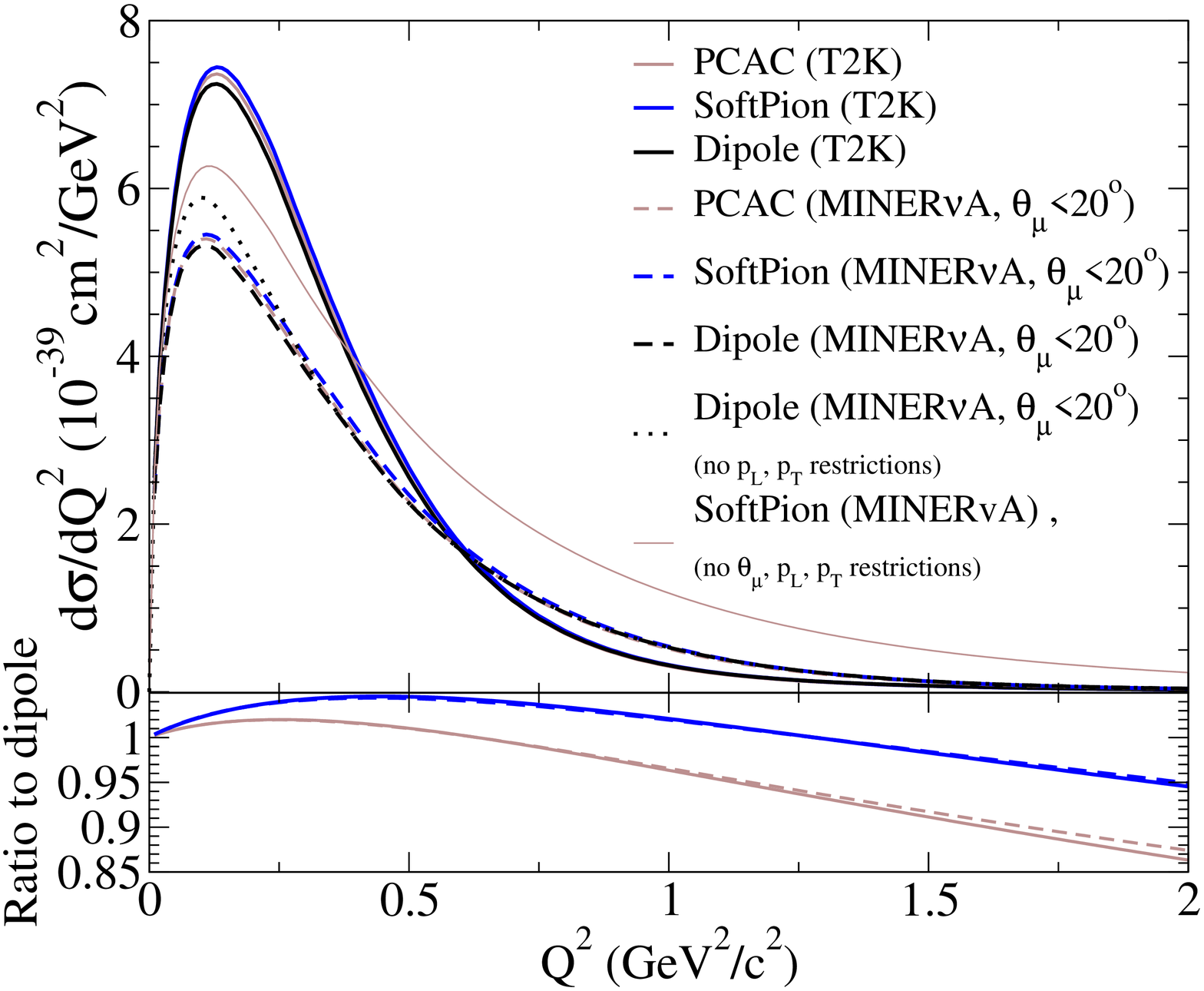}\hspace*{-0.15cm}%
		\includegraphics[scale=0.295, angle=270]{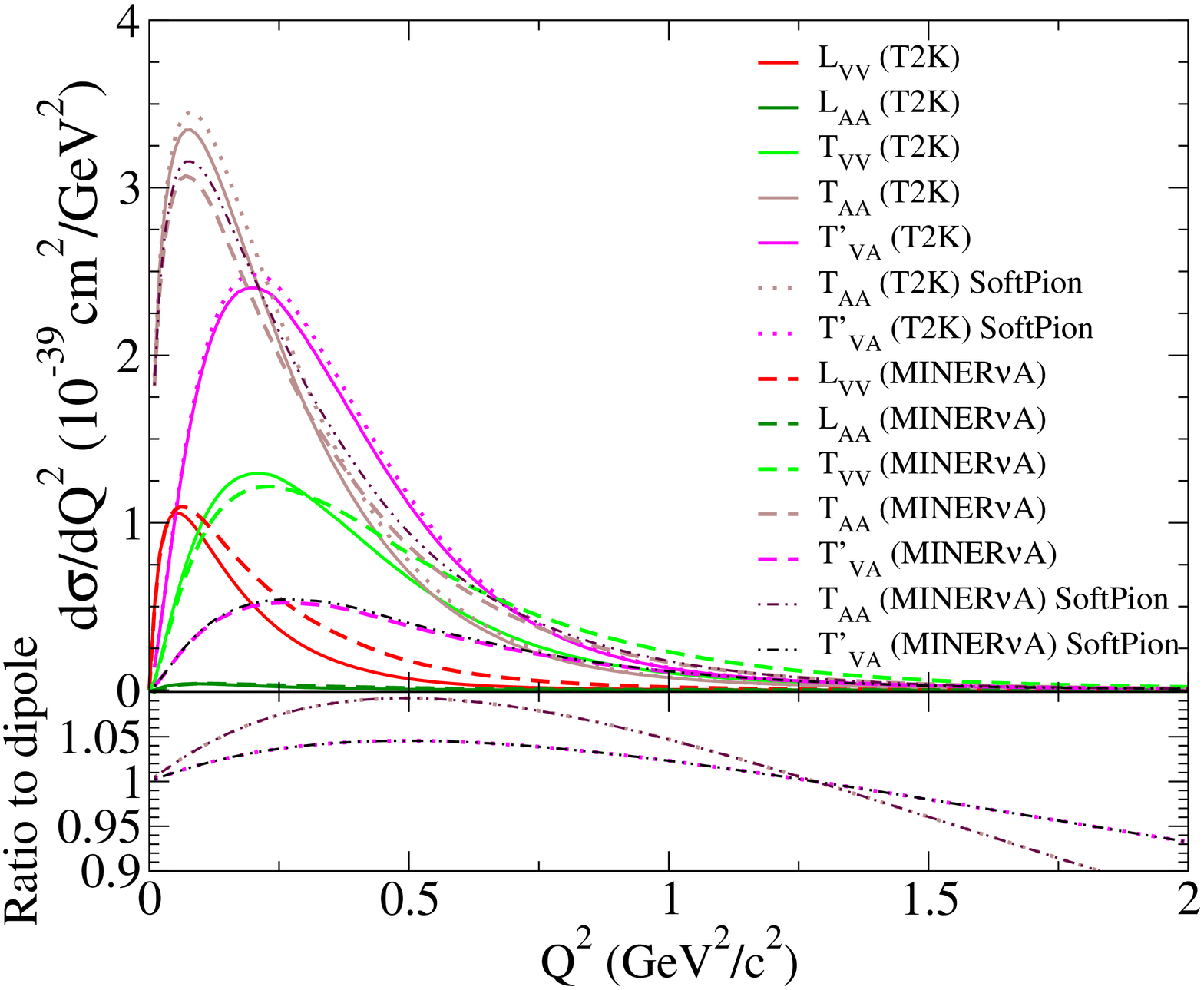}	
	\end{center}
	\caption{T2K and MINERvA cross-section as a function of $Q^2$ for different axial form factors (left), also shown separately for the different axial/vector contributions (right). 
	}\label{fig:Q2}
\end{figure}

In Fig.~\ref{fig:Q2} the cross-section as a function of $Q^2$ is compared for T2K and MINERvA and the role of the different axial and vector contributions is shown. The ratio between cross-sections evaluated with different form factors for the same $Q^2$ is almost identical for T2K and MINERvA, showing that the different axial/vector relative contributions in the two experiments play only a minor role in determining the effects of the axial form factor.

\section{Conclusions}
\label{sec:concl}

In this work we have re-evaluated the nucleon form factors in the two-component model and explored the impact of the uncertainties on the nucleon axial form factor, $G_A(Q^2)$, on the (anti)neutrino-nucleus cross-section. Firstly the electromagnetic form factors have been evaluated for  $Q^2\le 10$~GeV$^2$ by using the large set of electron scattering data available. The constraints obtained from such analysis have been used, then, to evaluate the axial form factor with a joint fit of pion electroproduction data and neutrino data. Updated values and proper uncertainties are reported for the parameters describing the electromagnetic and axial form factors and comparison to the dipole model are provided for the latter.

The evaluated axial form factors are then implemented in the CCQE cross-section using the SuSAv2 model to describe nuclear effects and compared to the cross-section obtained with the dipole form factor model. In general the agreement of the SuSAv2 model with data, using the different axial form factors, is satisfactory and the experimental uncertainties on T2K and MINERvA measurements do not allow yet to clearly discriminate between the various form factor evaluations. It is interesting to notice that, in the model considered here, the form factor effects have a different $Q^2$ dependence than the one of 2p2h, as well as a different neutrino/antineutrino dependence, making the disentagling of nucleon and nuclear effects feasible in future with higher statistics measurements. The feasibility of this approach relies on the capability of exploiting external data to drive the $Q^2$ dependence of the form factor. For this reason, the investigation of the earlier data of pion electro-production, as shown in Sec.~\ref{sec:fitaxial}, is of primary importance.

\begin{acknowledgments}
The authors are grateful to V. Bernard for interesting discussions, useful explanations and remarks. This work was partially supported by the Istituto Nazionale di Fisica Nucleare under project MANYBODY, by the University of Turin under contract BARM-RILO-17, by the Spanish Ministerio de Economia y Competitividad and ERDF (European Regional Development Fund) under contract FIS2017-88410-P, by the Junta de Andalucia (grant No. FQM160). M.B. Barbaro acknowledges support from the "Emilie du Ch\^atelet" programme of the P2IO LabEx (ANR-10-LABX-0038). G.D.M. acknowledges support from a P2IO-CNRS grant and from CEA, CNRS/IN2P3, France; and by the European Union’s Horizon 2020 research and innovation programme under the Marie Sklodowska-Curie grant agreement No. 839481. 
\end{acknowledgments}

\bibliography{biblio}

\end{document}